\documentclass[superscriptaddress,nobibnotes,amsmath,amssymb,notitlepage,twocolumn,prl,longbibliography]{revtex4-1}

\usepackage{bm,mathptmx,braket}

\usepackage{graphicx,color,hyperref}
\usepackage[caption=false]{subfig}
\hypersetup{colorlinks=true, linkcolor=blue, citecolor=blue, urlcolor=blue} 

\graphicspath{{./img/}}

\newcommand{\beq}[1]{\begin{equation}\label{#1}}
\newcommand{\eep}{\;.\end{equation}}
\newcommand{\eec}{\;,\end{equation}}
\newcommand{\eeq}{\end{equation}}

\newcommand*\dd{\mathop{}\!\mathrm{d}} 
\newcommand{\pd}[2]{\frac{\partial#1}{\partial#2}}

\newcommand{\lb}{\left(}
\newcommand{\rb}{\right)}

\newcommand*\chem[1]{\ensuremath{\mathrm{#1}}}

\renewcommand{\a}{\alpha}
\renewcommand{\b}{\beta}

\renewcommand{\d}{\delta}
\newcommand{\ep}{\epsilon}
\renewcommand{\k}{\kappa}
\newcommand{\la}{\lambda}
\renewcommand{\th}{\theta}

\newcommand{\p}{\phi}

\newcommand{\D}{\Delta}

\newcommand{\Om}{\Omega}

\DeclareMathAlphabet{\mathcal}{OMS}{cmsy}{m}{n}

\newcommand{\Df}{\mathcal{D}}
\newcommand{\Ef}{\mathcal{E}}

\newcommand{\bigO}{O}

\newcommand{\ointBZ}{\oint_{\text{BZ}}}

\newcommand{\BZ}{\text{BZ}}
\newcommand{\scBZ}{\text{scBZ}}
\newcommand{\Omsc}{\Om_{\text{sc}}}

\newcommand{\occ}{\text{occ}}

\newcommand{\Vc}{V_{\text{c}}} 

\newcommand{\Vsc}{V_{\text{sc}}} 
\newcommand{\Vdep}{V_{\text{dep},j}}

\renewcommand{\vec}[1]{{\bf #1}}

\newcommand{\x}{\vec{x}}
\newcommand{\kv}{\vec{k}}
\newcommand{\rv}{\vec{r}}
\newcommand{\R}{\vec{R}}

\newcommand{\A}{\vec{A}}
\newcommand{\w}{\vec{w}}

\renewcommand{\P}{\vec{P}} 
\newcommand{\Pvtot}{\vec{P}^{\text{tot}}}

\begin{document}

\title{Theory of polarization textures in crystal supercells}

\newcommand{\HarvardPhysics}{Department of Physics, Harvard University, Cambridge, Massachusetts 02138, United States}
\newcommand{\HarvardSeas}{John A.~Paulson School of Engineering and Applied Sciences, Harvard University, Cambridge, Massachusetts 02138, USA}
\newcommand{\TCM}{{Theory of Condensed Matter Group, Cavendish Laboratory, University of Cambridge, J.\,J.\,Thomson Avenue, Cambridge CB3 0HE, United Kingdom}}

\author{Daniel Bennett} \email{dbennett@seas.harvard.edu}
\affiliation{\HarvardSeas}

\author{Wojciech J. Jankowski}
\affiliation{\TCM}

\author{Gaurav Chaudhary}
\affiliation{\TCM}

\author{Efthimios Kaxiras}
\affiliation{\HarvardSeas}
\affiliation{\HarvardPhysics}

\author{Robert-Jan Slager}
\affiliation{\TCM}

\begin{abstract}
Recently, topologically non-trivial polarization textures have been predicted and observed in nanoscale systems. While these polarization textures are interesting and promising in terms of applications, their topology in general is yet to be fully understood. For example, the relation between topological polarization structures and band topology has not been explored, and polar domain structures are typically considered in topologically trivial systems. In particular, the local polarization in a crystal supercell is not well-defined, and typically calculated using approximations which do not satisfy gauge invariance. Furthermore, local polarization in supercells is typically approximated using calculations involving smaller unit cells, meaning the connection to the electronic structure of the supercell is lost. In this work, we propose a definition of local polarization which is gauge invariant and can be calculated directly from a supercell without approximations. We show using first-principles calculations for commensurate bilayer hexagonal boron nitride that our expressions for local polarization give the correct result at the unit cell level, which is a first approximation to the local polarization in a moir\'e superlattice. We also illustrate using an effective model that the local polarization can be directly calculated in real space. Finally, we discuss the relation between polarization and band topology, for which it is essential to have a correct definition of polarization textures.
\end{abstract}

\maketitle

\section{Introduction}

The formation of complex polar structures such as ferroelectric domains \cite{kmf,vacuum_1}, analogous to ferromagnetic domains \cite{kittel}, is a phenomenon intrinsic to ferroelectric materials with finite boundary conditions, and has been studied for many years. When going from bulk to a lower dimensional system, interfaces between a ferroelectric material and a non-polar material or vacuum lead to polar discontinuities which if not screened will lead to depolarizing fields that suppress ferroelectricity \cite{junquera2003critical}. Ferroelectric materials can form polydomain structures to mitigate depolarization effects, such as $180^{\circ}$ stripe domains, which for not too thin films can be described using a Landau-like theory \cite{bennett2020electrostatics}. These sharp domain structures eventually become unstable in very thin films, at which point it becomes more favourable to form softer domain walls, better described by a Ginzburg-Landau theory \cite{luk2009universal}; such domains have been observed in ferroelectric materials down to the monolayer limit \cite{monolayer_perovskite}. In some cases, even more complex structures involving polar vortices may form \cite{gomez2023kittel}, for example in ferroelectric/paraelectric (FE/PE) superlattices \cite{callori2012ferroelectric,zhang2017thermal,dawber2017balancing,park2018domain,susarla2021atomic} such as \chem{PbTiO_3/SrTiO_3} (PTO/STO), where the properties can be tuned through the use of different materials and by tuning the relative thicknesses of the layers. In FE/PE superlattices, both stripe domains and vortices can form \cite{dawber2017balancing}, depending on the strength of the coupling between the ferroelectric layers \cite{luk2009universal}, which can be determined by the ratio of the layers and the dielectric permittivities \cite{bennett2020electrostatics}.

Ferroelectric materials have been fabricated in many different geometries, from 2D thin films and FE/PE superlattices to 1D nanowires \cite{urban2002synthesis,yun2002ferroelectric} and nanotubes \cite{luo2003nanoshell,morrison2003high,mao2003hydrothermal}, and even 0D quantum dots \cite{chu2004impact,shin2005patterning}. Lower-dimensional ferroelectric systems typically exhibit size-dependent transitions where the polarization fields become more complex, before eventually becoming unstable and vanishing completely when the paraelectric phase is favored \cite{fu2003ferroelectricity,naumov2004unusual,geneste2006finite,morozovska2006ferroelectricity}. Soon after these complex polarization textures were discovered, they were proposed to be topologically non-trivial \cite{hong2010topology,junquera2023topologicaly}; skyrmion-like polarization structures were identified, for example in \chem{BaTiO_3} (BTO) nanowires embedded in a matrix of STO \cite{nahas2015discovery}. It has also been proposed that 3D skyrmions, i.e.~hopfions, may be created by controlling domains and domain walls in ferroelectrics, where at low temperatures the polarization rotates in-plane in the domain walls \cite{pereira2019theoretical}, resulting in a non-trivial winding. Ferroelectric skyrmions have been experimentally observed in PTO/STO superlattices \cite{das2019observation,han2022high}, and a skyrmion $\to$ meron transition with strain has recently been observed \cite{shao2023emergent}. Polar merons have also been observed in PTO under epitaxial strain from a \chem{SmScO_3} substrate \cite{wang2020polar}.

\begin{figure}[t!]
\centering
\includegraphics[width=\columnwidth]{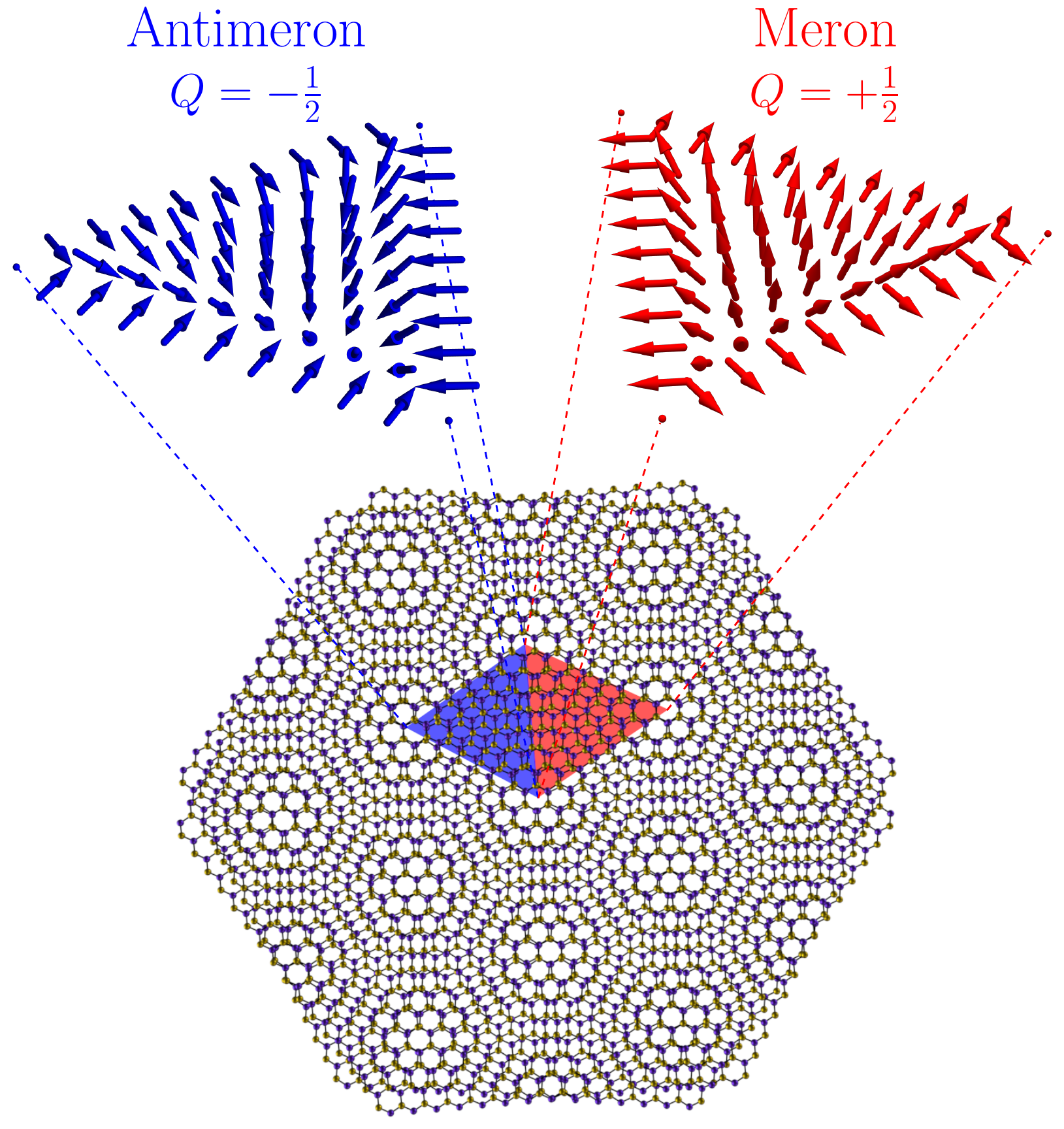}
\caption{Sketch of a twisted hBN bilayer. The MPDs in a single moir\'e cell are highlighted in blue and red, which have equal and opposite polarization. The normalized polarization field in the MPDs are sketched above, which form antimerons ($Q=-\frac{1}{2}$) and merons ($Q=+\frac{1}{2}$), respectively.}
\label{fig:meron}
\end{figure}

In recent years, a new type of ferroelectricity in layered van der Waals materials has been proposed \cite{li2017binary,bennett2022electrically,bennett2022theory} and experimentally observed \cite{zheng2020unconventional,yasuda2021stacking}. In layered systems such as 3R-stacked hexagonal boron nitride (hBN) or \chem{MoS_2}, an out-of-plane polarization occurs via an interlayer transfer of electronic charge when the relative stacking between, which can be switched by a relative sliding between the layers (van der Waals sliding \cite{stern2020interfacial}), resulting in ferroelectricity. When there is a relative twist or lattice mismatch between the layers, forming a moir\'e superlattice (see Fig.~\ref{fig:meron}), the interlayer charge transfer results in an out-of-plane polarization texture, and the stacking domains can be identified as moir\'e polar domains (MPDs), which have been experimentally shown to result in ferroelectricity \cite{yasuda2021stacking} via the growing and shrinking of the MPDs in response to an applied field \cite{bennett2022electrically,bennett2022theory}. Recently, we discovered that the MPDs also have an \textit{in-plane} component of polarization \cite{bennett2023polar}, resulting in possible real space non-trivial topological textures. For hBN and similar materials the topological charge $Q$ in each stacking domain,
\beq{eq:winding}
Q = \frac{1}{4\pi}\int \P \cdot \lb \partial_x\P \times \partial_y\P\rb\dd\rv
\eec
where the local polarization $\P(\rv)$ is normalized, integrates to $\pm\frac{1}{2}$, meaning the MPDs are polar merons and antimerons (see Fig.~\ref{fig:meron}). Moir\'e superlattices have also recently been fabricated with perovskites \cite{shen2022observation}, which for FE/PE interfaces such as BTO/STO also result in polar vortices \cite{sanchez20232d}.

While the complex structures and non-trivial topology of polar nanostructures are very interesting and promising for future applications in nanotechnology, some of their fundamental properties require a better understanding, namely:

\textbf{(i) The origin of topological polarization structures:} Polarization can be topological in the real space sense, i.e.~there may be non-trivial winding of the polarization vector field of a system. This winding must arise from both the geometry of the system and underlying crystal symmetries of the constituent materials. For example, in perovskites superlattices, the electrostatic boundary conditions at the interfaces between alternating layers make a uniform polarization in the ferroelectric layers energetically unfavourable. This promotes a polydomain structure, but does not guarantee non-trivial topology; a 180$^{\circ}$ stripe domain structure is preferable to a monodomain polarization, but is topologically trivial. The polar modes in each unit cell, related to the off-centering of the Ti atoms within the oxygen octahedra, lower the crystal symmetry, and allow coupling between polarization and strain such as piezoelectricity, which is forbidden by symmetry in the cubic phase. As a result, the strain across the domain walls and interfaces results in an additional component of polarization \cite{pereira2019theoretical}, which causes the polarization to wind and become topologically non-trivial. Rolling a thin film into a nanotube induces a radial polarization and hence a polar vortex via flexoelectricity \cite{artyukhov2020flexoelectricity,springolo2021direct,bennett2021flexoelectric}, which is a property of all insulators. In twisted or strained hBN, the MPDs are a result of the geometry of the moir\'e superlattice. However, the reason that the polarization winds and is topologically non-trivial is related to the underlying symmetry of the hBN bilayer \cite{bennett2023polar}. It was shown with a space group analysis that different mirror planes are broken by different local stackings, meaning that the polarization must point out-of-plane in the domain centers, and in-plane along the domain walls, leading to a network of merons and antimerons.

When considering the topology of polarization, one major problem is that the local polarization in a crystal in real space is not a well-defined quantity. From the modern theory of polarization \cite{vanderbilt1993electric,king1993theory, resta1994macroscopic,Wannier_Nakagawa, vanderbilt2018berry, vanMechelen_optical}, the change in \textit{total} polarization of a system can be measured from Berry phases, but the decomposition into individual contributions in real space is somewhat arbitrary. There are two types of methods for estimating the local polarization in a supercell: `configuration space' methods, where the local configurations in a each unit cell are emulated in a commensurate system, and real space methods, where the the local polarization is calculated directly from calculations involving the entire supercell. It is clear that calculating local polarization directly in real space would be preferable, although this would require expensive calculations, and defining local polarization in real space is difficult. One proposal is to measure the individual Wannier centers in each unit cell \cite{wu2006wannier}, which has been applied to slab-like systems to measure the out-of-plane polarization, averaged in the in-plane directions. However, this approximation does not necessarily satisfy gauge invariance. The more commonly used approximation relies on obtaining the local displacements in each cell and multiplying them by the Born effective charges obtained from the bulk system \cite{meyer2002ab,stengel2011band}. This was originally proposed for slab-like systems, and is the standard method for calculating the local polarization in perovskite systems. However, this relies on the assumption that the Born effective charges are uniform everywhere in space, which is not the case in twisted bilayers \cite{bennett2023polar}, for example. 

\textbf{(ii) The relation to electronic band topology:} Local polarization is not typically described at the electronic structure level, since it is defined in terms of local displacements in real space. Thus, the relation between the topology of polarization textures and band topology~\cite{Rmp1,Rmp2}, which has seen the development of various analytical diagnoses~\cite{Clas1,Clas2,Unal_quenched_Euler,Clas4, Clas5, Clas3, Shiozaki14, bouhon2019nonabelian, bouhon2020geometric, Song_2018}, is not clear. In the original description of polarization in terms of localized Wannier functions, the polarization is well-defined only for topologically trivial bands; for topologically non-trivial bands, the Wannier functions are not exponentially localized. The modern theory of polarization has since been generalized to Chern insulators by tracking the hybrid Wannier centers throughout closed loops in the  Brillouin zone (BZ) \cite{Sinisa2009,Song2021} (equivalent to calculating Wilson loops~\cite{Alex_BerryPhase,bouhon2019wilson}). However, the similarities and interplay between polar and band topology are not known. Recently there has been evidence that polarization can affect the topological properties of a system: in layered \chem{MnBi_2Te_4}, which is antiferromagnetic, ferroelectric, and exhibits quantum anomalous Hall (QAH) conductance, inverting the polarization via van der Waals sliding changes the sign of the Chern number and hence the QAH conductance \cite{liang2023ferroelectric}. Thus, a better understanding of the role that polarization can play in the properties of systems arising from band topology is needed. 

\textbf{(iii) Physical consequences:} Perhaps most importantly, the physical consequences of topological polarization are not well-understood. While it is has recently been shown that ferroelectric switching can change the Chern numbers in topologically non-trivial systems \cite{liang2023ferroelectric}, this is a result of uniform polarization; there may be additional phenomena which are unique to topologically non-trivial polar textures. Furthermore, finding physical consequences may also provide new ways to indirectly detect polar topology experimentally, which currently requires very careful microscopy measurements.

In this work, we address the problem of defining the local polarization in a crystal supercell, in the context of studying topological polarization. The most common approach for estimating local polarization, in twisted bilayers for example \cite{bennett2022electrically,bennett2022theory,bennett2023polar}, is using the configuration space mapping: the local polarization in real space is approximated by the total polarization in configuration space, i.e.~a commensurate system with a global distortion such as a relative shift between the layers. Estimating local polarization in this way requires the polarization to vary slowly, with a wavelength similar to the supercell period. Additionally, information about the electronic structure is lost, as the local polarization is not calculated from the bands of the supercell. While approximate expressions for the local polarization which can be calculated directly in real space using Wannier centers and Born effective charges have been proposed in the literature \cite{wu2006wannier,meyer2002ab,stengel2011band}, they are not necessarily well-defined. In the former case, the local polarization is given by a partial sum over Wannier centers, which in general is not gauge invariant. In the latter case, partial sums over the Born effective charges may not satisfy the acoustic sum rule, or charge neutrality. We discuss the different ways in which local polarization can be estimated, and propose definitions of local polarization in terms of Born effective charges or Wannier centers, which are well-defined (gauge invariant) and can be evaluated directly in real space. As a stepping stone to real space calculations, we show that these definitions yield the correct polarization in commensurate 3R-stacked hBN in configuration space. We illustrate using effective models in 1D (Aubry-Andr\'e model in the continuum limit \cite{Attila2018,qp1,qp2,aubry1980analyticity}) and 2D (Bistritzer-MacDonald \cite{Bistritzer2011,Balents2019}) that our proposed definitions can be used to calculate the local polarization in real space, without relying on the mapping to configuration space. Finally, we discuss the relation between polarization and band topology.

\section{Local polarization}

Our aim is to develop a definition of the local polarization field in a supercell comprised of a number of repeated unit cells, each with real space position $\rv_j$. Each unit cell may have local distortions with respect to a more symmetric reference configuration, and we define the local polarization as a result of these local distortions as a discrete vector field $\P(\rv_j)$, valued in each unit cell. When studying systems described by large supercells, in particular moir\'e superlattices, it is generally much more convenient to work in terms of the local distortions in each unit cell, which form a configuration space, rather than the full supercell in real space.

\subsection{Configuration Space}

\begin{figure}[t!]
\centering
\includegraphics[width=\columnwidth]{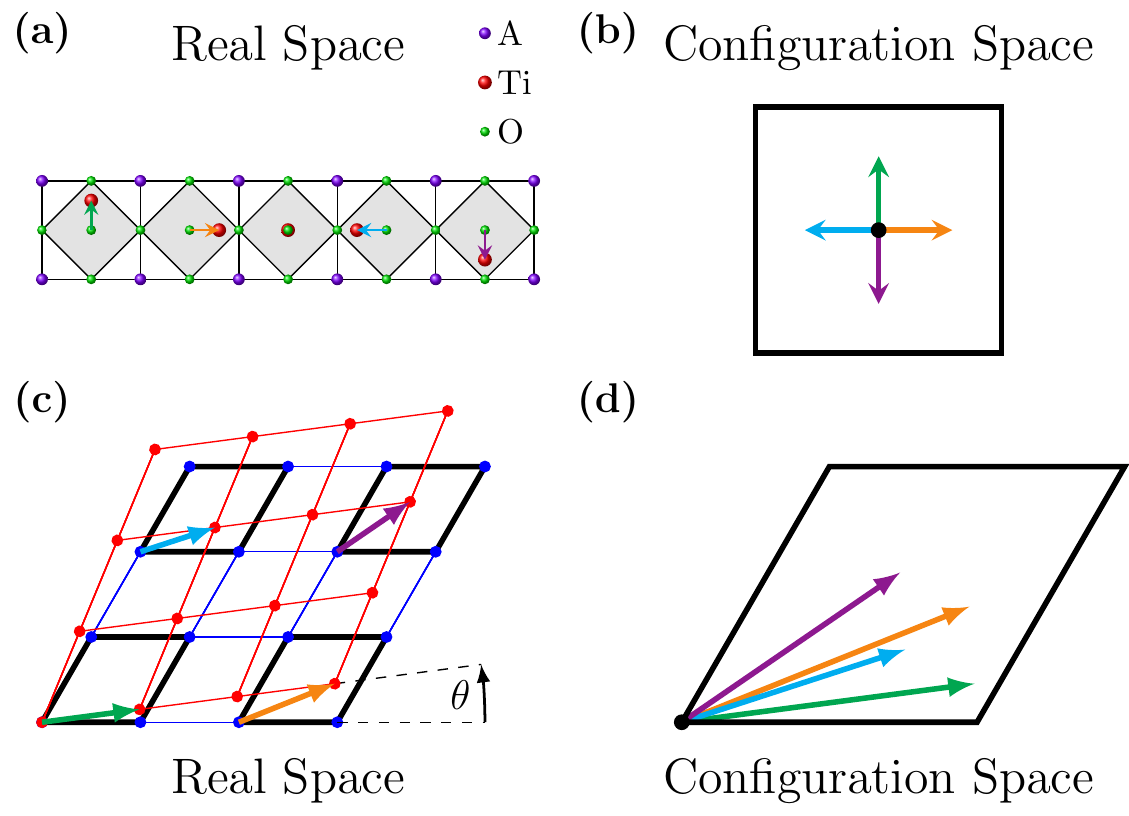}
\caption{\textbf{(a)}: Sketch of a $5\times 1\times 1$ supercell of a titanium oxide perovskite \chem{ATiO_3}. The arrows indicate the displacement of the Ti atom from the center of the oxygen octahedra in each unit cell. 
\textbf{(b)}: Mapping of the displacements in each unit cell from real space to configuration space, which has the dimensions of a single 5-atom unit cell. 
\textbf{(c)}: $3\times 3$ section of a bilayer with relative twist angle $\th$. 
\textbf{(d)}: Mapping of the relative displacements between the atoms in each layer from real space to configuration space, which has the dimensions of a single bilayer unit cell.}
\label{fig:config-space}
\end{figure}

The mapping between real space and configuration space is used to estimate local properties in periodic supercells and incommensurate systems \cite{carr2017twistronics,cazeaux2017analysis,massatt2017electronic,carr2018relaxation}. In the context of polar domains and textures, the two most common examples are oxide perovskite films (BTO, PTO, etc.), in which the supercell composed of five unit cells is repeated in the direction(s) normal to the interface in order to allow for the formation of polydomain structures, see Fig.~\ref{fig:config-space} (a), and a twisted/strained bilayer, in which two layers are twisted or strained relative to one another to form a moir\'e superlattice, see Fig.~\ref{fig:config-space} (c).

In an \chem{ABO_3} perovskite, the polarization arises from the off-centering between the B cations and the \chem{O_6} octahedra. In titanate oxide perovskites such as PTO and BTO, this has been attributed to the hybridisation of the Ti $3d$ and O $2p$ orbitals \cite{cohen1992origin}. For a perovskite supercell with a polydomain structure, the local polarization is a result of the polar mode displacements in each unit cell. Assuming for simplicity that the polar modes are dominated by the off-centering of the Ti atoms within each oxygen octahedron, the displacement in each unit cell is approximately ${\x(\rv_j) = \D\rv_{j}^{\chem{Ti}}}$, where $\rv_j$ is the position of unit cell $j$, and $\rv_j^{\chem{Ti}}$ is the position of the Ti atom in the cell. The displacements in each unit cell can be mapped to a single space known as configuration space, which is the size of a single 5-atom unit cell, see Fig.~\ref{fig:config-space} (b). Thus, local properties such polarization can be estimated in configuration space using a bulk 5-atom unit cell, which is much more efficient than performing calculations using the full supercell. For example, the polarization can be parameterized for every point in configuration space by calculating the Berry phases for a 5-atom bulk cell with various displacements of the Ti atom. Assuming the displacements in real space are known, the local polarization can then be parameterized in real space by using an inverse mapping from configuration space. While the mapping of displacements between real space and configuration space is exact, the parameterization of local properties relies on the approximation that the displacements vary slowly in real space and can therefore be neglected; we say that the local changes around a given cell are small, and we neglect them, which gives a commensurate cell which is much more easy to simulate. Of course, this approximation is not good for properties which modulate at the unit cell level, such as Peierls distortions, defects, dislocations, etc. However, for quantities which vary smoothly on the supercell scale, the approximation works well; in moir\'e materials for example, the separation of supercell and single-layer unit cell length scales justifies such approximate approach relying on the continuity of the mapping.

While the local polarization can be parameterized in configuration space by calculating Berry phases as a function of displacements, it is typically estimated as follows for perovskite systems: the displacements $\x(\rv_j)$ are measured in real space, either experimentally or from molecular dynamics simulations. The displacements in each cell are contracted with the Born effective charge tensor of the bulk cubic phase:
\beq{eq:P-approx-1}
P_{\b}(\rv_j) \approx \frac{1}{\Om_0}\sum_{\k\in\rv_j}\sum_{\a}Z^*_{\k,\a\b}x_{\k,\a}
\eec
where $\Om_{0}$ is the unit cell volume, contrary to the total supercell volume $\Om_{\text{sc}}$. The first sum is over the atoms $\k$ in unit cell $\rv_j$, and the second is over Cartesian directions $\a$. This is the generally accepted method to measure local polarization in oxide perovskites, and has been used to predict topologically non-trivial polarization. One problem with Eq.~\eqref{eq:P-approx-1} is that the individual unit cells are not guaranteed to satisfy the acoustic sum rule. In slab-like systems, the polarization in each layer is typically averaged over neighbouring layers \cite{stengel2011band}, and is therefore not truly localized to each unit cell. Eq.~\eqref{eq:P-approx-1} also requires the Born effective charges to be uniform in configuration space, which is reasonable when the polarization varies slowly in real space, such as in the centers of the polar domains, but not when the polarization varies sharply, such as across domain walls; in PTO/STO superlattices for example, the domain walls have been shown to be a single unit cell wide \cite{gomez2023kittel}. Additionally, it is possible for ferroelectric domain walls to be conducting, in which case it is not clear if the Born effective charges are well defined \cite{sluka2013free,bednyakov2015formation}.

One situation in which Eq.~\eqref{eq:P-approx-1} fails to correctly estimate the local polarization is in van der Waals materials, where the Born effective charges vary nonlinearly as one layer slides over the other \cite{bennett2023polar}. Although the charge transfer and modulation of the Born effective charges is small, they give rise to the unique polarization textures in twisted bilayers and thus cannot be neglected.

For a bilayer with relative twist angle $\th$ between the layers, see Fig.~\ref{fig:config-space} (c), the configuration space mapping is given by \cite{carr2018relaxation}
\beq{eq:config-space-twist}
\vec{x}(\vec{r}) = (\mathbb{I} - R_{\th}^{-1})\vec{r}
\eec
modulo any lattice vectors, where $\vec{r}$ is the real space position and $R_{\th}$ is a rotation matrix. While Eq.~\eqref{eq:config-space-twist} is exact, for small twist angles the local changes in environment around the black unit cells are small, and the local properties in each cell can be described by a commensurate bilayer with a relative translation ${\vec{x} \approx \theta\begin{bmatrix} 0 & -1 \\ 1 & 0 \end{bmatrix} \vec{r}}$ between the layers, see Fig.~\ref{fig:config-space} (d). Similarly, for a small homogeneous strain $\eta$, the equivalent mapping is ${\vec{x} = \eta\vec{r}}$. This allows the local properties in strained or small-angle twisted bilayers to be parameterized efficiently with first-principles calculations using a single commensurate cell of a bilayer, and sliding one layer over the other. Importantly, what we define as the unit cell $\rv_j$ for moir\'e systems, is a unit cell of one layer \textit{and} the atoms of the other layer which are contained in that unit cell. Such a general projective description is consistent with the configuration space picture, and provides an arena to define a local quantity, given its shortsightedness with respect to the neighbouring unit cells. However, within this picture, the configuration space calculations for simultaneously twisted and strained moir\'e bilayers performed on an elastically deformed unit cell might yield local polarization significantly deviating from the real values, on applying periodic boundary conditions in configuration space.  The main reason for this is that when both unit cell deformation and the stacking modulation due to the twist are present, the repetition of such approximate unit cells in configuration space for computational purposes, implicit in the imposed boundary conditions, does not account for the appreciable structural variations in the neighbourhood of the studied unit cell. Therefore, such possibilities also motivate pursuing the definitions of local polarization in crystal superlattices, beyond the notion of configuration space.

The local polarization can be calculated in the configuration space by sliding one layer over the other and directly calculating the Berry phases \cite{bennett2022electrically,bennett2022theory,bennett2023polar}. However, the Berry phase obtained for each point in configuration is not physically meaningful. In real space, the Berry phase is a global property of the system, which yields the \textit{total} polarization. A more natural way to define the local polarization is using the Born effective charges, since they are locally well-defined in real space. When the Born effective charges are not constant, Eq.\eqref{eq:P-approx-1} is not valid. Instead, the Born effective charges must be integrated:
\beq{eq:P-approx-2}
P_{\b}(\rv_j) \approx \frac{1}{\Om_{0}}\int_0^{\x(\rv_j)}Z^*_{\k,\a\b}(\x')dx'_{\k,\a}
\eec
where summation over repeated indices is assumed, and again only the atoms $\k \in \rv_j$ are displaced. When the Born effective charges are constant, Eq.~\eqref{eq:P-approx-2} simplifies to Eq.~\eqref{eq:P-approx-1}. The integration is performed from a reference state 0 to a general configuration $\x$, where in configuration space we can change smoothly from one to the other via a relative translation between the layers. 
In order to obtain the polarization, rather than an arbitrary change of polarization between two configurations, the reference state is chosen to be non-polar. 
For hBN and similar materials, the only non-polar configuration is when the layers are perfectly aligned (AA stacking) and unstrained, and therefore, while Eq.~\eqref{eq:P-approx-2} is naturally defined on a torus, the normalized polarization, used in Eq.~\eqref{eq:winding}, is defined on a \textit{punctured} torus, similarly to the strain fields \cite{engelke2023non}. Eq.~\eqref{eq:P-approx-2} yields a polarization identical to the one obtained from Berry phases \cite{bennett2023polar}.

While Eqs.~\eqref{eq:P-approx-1} and \eqref{eq:P-approx-2} have been successfully used to estimate polarization textures, they are both only valid for certain approximations such as large supercells, smoothly varying quantities and constant Born effective charges in the former case. Furthermore, we also lose any information about the electronic structure when using this approximation, because the electronic bands for each point in configuration space are not meaningful in real space. This is especially apparent when calculating the Berry phases; the Berry phase is a global property of a system, which we calculate from the bands for each point in configuration space. Although we get a good approximation to the local polarization, we do not have any information about the contribution from different bands of the supercell.

\subsection{Defining local polarization}
From the modern theory of polarization, the \textit{total} polarization $\Pvtot$ in a crystal is given by \cite{king1993theory,vanderbilt1993electric,vanderbilt2018berry}
\beq{eq:P-tot}
\Pvtot = \frac{-ief}{(2\pi)^3} \sum_{n}^{\occ} \oint_{\BZ} \braket{u_{n,\kv} | \nabla_{\kv} u_{n,\kv}}\dd\kv
\eec
where $\ket{u_{n,\vec{k}} }$ are the cell-periodic parts of the Bloch wavefunctions and $f$ is the occupation number of states in the valence bands (2 for spin-degenerate systems). Eq.~\eqref{eq:P-tot} can be rewritten as
\beq{eq:P-tot}
\Pvtot = \frac{-ef}{(2\pi)^3} \sum_{n}^{\occ} \ointBZ \A_n(\kv)\dd\kv = \frac{-ef}{\Om} \sum_{n}^{\occ} \frac{1}{2\pi}\p_{n,\a}\vec{a}_{\a}
\eec
where $\vec{a}_{\a}$ are the lattice vectors, $\Om$ is the system cell volume.
\beq{}
\A_n(\kv) = i\braket{u_{n,\kv} | \nabla_{\kv} u_{n,\kv}}
\eeq
is the Berry connection of band $n$, and
\beq{}
\p_{n,\a} = \frac{i\Om}{(2\pi)^3} \ointBZ \braket{u_{n,\kv} | \vec{b}_{\a}\cdot\nabla_{\kv} u_{n,\kv}} \dd\kv
\eeq
is the Berry phase of band $n$ in direction $\a$, defined up to a factor of $2\pi$, where $\vec{b}_{\a}$ are the reciprocal lattice vectors. For disordered systems, where $\kv$ is not a good quantum number, a generalization for the computation of macroscopic total polarization with single-point Berry phases has been proposed by Resta \cite{PhysRevLett.80.1800}.

It is well-known that the absolute polarization in a crystal is not well-defined. Only changes in polarization are well-defined, modulo any quanta of polarization (integer values of the Berry phases in different directions). Derivatives of the polarization with respect to perturbations (phonon, strain, electric field), which can be identified as the dielectric and electromechanical properties of a system, are well-defined and are routinely calculated from first-principles calculations, either using finite difference methods or density functional perturbation theory (DFPT) \cite{Wu2005}: the dielectric response of a system is related to the derivative of the total polarization with respect to electric field \cite{gonze1997dynamical}, and electromechanical properties can be measured by calculating the derivatives of the polarization with respect to strain (piezoelectricity \cite{Vanderbilt2000}), or strain and electric field (electrostriction \cite{bennett2022generalized}). For our purposes, the most relevant property is the derivative of the polarization with respect to phonon displacements, i.e.~the Born effective charge tensor \cite{gonze1997dynamical,ghosez1998dynamical}: 
\beq{eq:Z}
Z^{*}_{\k,\a\b} = \Om \pd{P_{\b}}{x_{\k,\a}} = \pd{F_{\k,\a}}{\Ef_{\b}}
\eec
where $x_{\k,\a}$ is the real space (phonon) displacement of atom $\k$ in direction $\a$, $F_{\k,\a}$ is the force on atom $\k$ in direction $\a$, and $\Ef_{\b}$ is an electric field in direction $\b$. Because the Born effective charge tensor is related to the mixed derivative of the free energy of the system with respect to phonon displacement and electric field, it can be interpreted as both the dipole generated by a phonon displacement, and the force generated on an atom by an electric field.

As mentioned previously, the change in polarization from one configuration to another can be obtained by integrating the Born effective charges using Eq.~\eqref{eq:P-approx-2}. For twisted/strained bilayers, this was done in configuration space in order to avoid expensive calculations involving large supercells \cite{bennett2023polar}. However, we propose that the local polarization may be calculated in a more well-defined way by calculating the Born effective charges in real space. While the polarization in configuration space is simply approximate, the Born effective charges are well-defined in real space because they are the derivatives of the well-defined \textit{total} polarization of the supercell with respect to the local atomic displacements in each unit cell. Expressed in this way, the Born effective charges are obtained directly from the electronic bands of the supercell, rather than the fictitious electronic bands in configuration space.

We define a unit cell $\rv_j$ as the smallest structural unit that can be mapped to configuration space, which captures the complete set of all possible configurations, i.e.~spanning the entire system. Writing the dynamical charges in each unit cell as \cite{ghosez2000band,sai2002theory}
\beq{eq:Z-mixed}
Z^{*}_{\k,\a\b}(\x(\rv_j)) = \frac{-2ief\Omsc}{(2\pi)^3}\sum_{n}^{\occ} \oint_{\scBZ} \braket{\partial_{x_{\k,\a}}u_{n,\kv}|\partial_{k_{\b}}u_{n,\kv}}\dd\kv
\eec
for all atoms $\k$ in cell $\rv_j$, the local polarization in each unit cell is given by
\beq{eq:P-local-modern}
P_{\b}(\rv_j) = \frac{-2ief}{(2\pi)^3} \int_0^{\x(\rv_j)} \sum_{n}^{\occ}\oint_{\scBZ} \braket{\partial_{x_{\k,\a}}u_{n,\kv}|\partial_{k_{\b}}u_{n,\kv}} \dd\kv \dd x'_{\k,\a}
\eep
It is important to stress that the momentum space integral is evaluated over the supercell Brillouin zone (scBZ) and the summation is performed over the bands of the supercell, while the integral with respect to $\x$ is performed over the relative displacements not in real space, but in configuration space, where the different configurations are connected by the phonon displacements $\x$. The limits of the integral are the nonpolar reference state 0 and the local configuration in each unit cell $\rv_j$.

The key difference between Eq.~\eqref{eq:P-local-modern} and Eq.~\eqref{eq:P-approx-2} is that the Born effective charges are calculated correctly: in real space, and using the electronic bands of the supercell. There are a few subtle details associated with the definition of local polarization in this way. First, the system must be a supercell comprised of a number of smaller unit cells, within each a local polarization is defined. The local polarization in each cell is really defined as a change in polarization with respect to a reference cell, but taking the reference cell to be non-polar, we write the polarization as $P$ rather than $\D P$. The integral over configurations in Eq.~\eqref{eq:P-local-modern} could be mapped to positions in real space. However, in real space the polarization is a discrete vector field, defined in each unit cell of the supercell, whereas configuration space is generally continuous and simply connected. A commensurate supercell is mapped to a finite subset of configuration space containing a discrete set of points, but for an incommensurate supercell, where the period goes to infinity, there is a one to one mapping between the two spaces. The integral over $\x$ in Eq.~\eqref{eq:P-local-modern} should be discretized over the unit cells in real space, but we can take advantage of the fact that configuration space is continuous and interpolate the Born charges (or any local quantity), and obtain the local polarization field in configuration space which is continuous and varies smoothly.

\subsection{Local polarization from a 2D continuum model}

In the configuration space method the polarization is computed as a global quantity in each configuration and then related to the local polarization via a mapping between configuration and the real space. 
This approach has its benefit as the global polarization is well defined for each point in configuration space. 
However, physically its relation to the local polarization becomes less transparent. 
The direct real-space picture described in the previous section is obtained at the unit cell level. 
However in moir\'e superlattices it is often more convenient to work with continuum models. While superlattices constructed from microscopic unit-cells are only well-defined at some commensurate twist angles or strains, the continuum model description accurately captures the low energy physics at all small angles and strains with smooth moir\'e periods, irrespective of microscopic periodicity~\cite{Bistritzer2011}. 

Here, we show that an expression for local polarization can be obtained in real space using a continuum field approach, derived in the context of deformation fields~\cite{Balents2019}.
For illustrative purposes we consider a moir\'e bilayer formed by a small strain or twist angle, although the generalization to the more complicated moir\'e patterns is straightforward. 

The continuum model describes the low energy physics near a band extremum that is located at the momentum $\vec{K}$ of the un-deformed monolayer.  
The electron field at $\vec{K}$ is given by
\beq{eq:elec_field_undeformed}
    c(\x) = \psi(\x) e^{i\vec{K}\cdot\x }
\eep
For concreteness, we assume that near this extremum the low energy physics of the monolayer is described by a 2D massive-Dirac model. 
This is appropriate for bilayer hBN, the main example considered in this work and in Ref.~\cite{bennett2023polar}, although generalization to other models and dimensions is straightforward. 
The monolayer Hamiltonian is given by
\beq{eq:Ham_ML}
    H_{\text{ML}} = \int \psi^{\dagger} \lb m\tau_3 - iv\tau^\mu\partial_{\x_{\mu}} \rb \psi \dd^2\vec{x}
\eec
where $m$ is the mass gap, $v$ is the Dirac-velocity and the $\tau$ Pauli-matrices act on some internal degrees, which for hBN are the two sublattices. Summation is assumed, with $\mu = 1,2$.

The effect of small strain and twist can be captured by a local deformation $\Df(\rv)$ field as
\beq{eq:deform}
    \rv = \x + \Df(\rv)
\eec
where $\rv$ is real space position in a ``laboratory frame",  $\x$ is the position in the monolayer and $\Df(\rv)$ is a deformation field as a result of strain and twist. 
The deformation field is assumed to be locally small, \textit{i.e.} $\partial_{\mu}\Df \ll 1$. 
It is crucial to define the ``small" deformation $\Df$ as a function of the variable $\rv$ instead of the variable $\x$, since when considering the multilayers, $\x$ are associated with the individual layers correspond to very different locations $\rv$ in the real space. 
Thus one cannot define a ``small local" deformation field in the variable $\x$. 
This is equivalent to the configuration-space consideration where configuration space vector is simply the local change in configurations of the unit cells from the top and the bottom layers, instead of overall global shift of the particular unit cell as one of the layer is twisted or strained.  

As a result of a small deformation in the layer, the electron field is locally modified as
\beq{eq:elec_field_deform}
    c(\rv) = \left|\text{det} \lb \frac{\partial x_{\mu}}{\partial r_{\nu}} \rb \right|^{1/2}  c(\x(\rv))
\eep
The $\psi$ field is correspondingly modified as
\beq{Eq:psi_deform}
    \psi(\rv) = \lb 1-\bm{\nabla}\cdot\Df(\rv) \rb^{1/2} \psi(\x(\rv))e^{-i\vec{K}\cdot \Df(\rv)}
\eec
and the integral measure is modified as
\beq{eq:int_measure_deform}
    \dd^2\x = \text{det} \lb \frac{\partial x_{\mu}}{\partial r_{\nu}} \rb \dd^2\rv
    \sim (1-\bm{\nabla}\cdot \Df(\rv))  \dd^2\rv
\eep
The continuum Hamiltonian of the decoupled bilayers can be obtained when each layer experiences an independent deformation field $\Df_l$, where $l= \text{t,b}$ is the layer index:
\beq{eq:Ham_BL}
\begin{split}
    H_{\text{BL}} &= \sum_{l= \text{t,b}}\int \psi^{\dagger}_l \biggl[ m\tau_3 - iv \lb( \tau^{\mu} + \frac{\partial \Df_{l,\mu}}{\partial r_\nu}\tau^{\nu} \rb \partial_{r_{\mu}} \\
    &\hspace{2cm} + v (\vec{K}\cdot \partial_{r_\mu} \Df_{l}) \tau^{\mu}\biggr] \psi_{l} \dd^2\rv
\end{split}
\eec
where we have kept only the terms linear in the deformation field. 
For example, if the two layers are twisted rigidly by angles $\pm \theta/2$, the deformation field is given by
\beq{eq:twist_deform}
    \Df_{\text{t}}(\rv) = -\Df_{\text{b}}(\rv) = \frac{\th}{2}\hat{z}\times \rv 
\eep
The twist deformation is simply the continuum field analog of the real space to configuration space mapping for a twisted moir\'e bilayer in Eq.~\eqref{eq:config-space-twist}. The inter-layer tunneling Hamiltonian is also modified under the twist deformation and takes a general form
\beq{eq:Ham_tunnel}
    H_{\text{tun}} = \int  \psi^{\dagger}_2 T(\Df_{\text{t}} - \Df_{\text{b}}) \psi_1 \dd^2\rv \ + \ \text{h.c.} 
\eec
Here, we have assumed that the interlayer tunneling $T$ is purely local and does not depend on the gradients of the deformation fields. 
Furthermore, if the two layers are deformed identically, the local interlayer tunneling must remain unchanged. 
Thus we take the tunneling Hamiltonian to only depend on the relative deformation of the two layers. 
The tunneling Hamiltonian can further be constrained by the relevant symmetries near $\vec{K}$. 
A number of those symmetry-related constraints are system dependent, however, the discrete lattice translation symmetry ${T(\Df(\rv+\a)) = T(\Df) }$ is a common feature. Thus,
\beq{eq:tunneling_periodicity}
    T(\Df) = \sum_{\vec{G}} T_{\vec{G}} e^{i\vec{G}\cdot \Df }
\eec
where $\vec{G}$ are the reciprocal lattice vectors of the undeformed monolayer.
Finally the Hamiltonian of the moir\'e bilayer can be represented as
\beq{eq:Ham_moire_continuum}
    H_{\text{moir\'e}} = H_{\text{BL}} + H_{\text{tun}} \equiv 
    H_{0} + H[\Df,\bm{\nabla}\Df],
\eeq
where $H_{0}$ is the Hamiltonian of the undeformed bilayer, which we take to be non-polar, i.e.~a commensurate bilayer with AA stacking.
Thus, the local polarization can be characterized by the evolution of the total polarization as the deformation is turned on adiabatically:
\beq{eq:polarization_cont}
\P(\Df(\rv)) = \int^{\Df(\rv)}_0 \frac{\partial \P}{\partial \Df } d\Df
\eec
where,
\beq{Eq:dpolarization_cont}
    \frac{\partial \P}{\partial \Df} = \frac{-2ief}{(2\pi)^2} \sum^{\occ}_{n} \oint_{\text{mBZ}} \braket{ \partial_{\Df} u^{n}_{\Df,\vec{G}}(\kv) | \partial_{\kv} u^{n}_{\Df,\vec{G}}(\kv)}
\eec
denoting mBZ as the moir\'e BZ, a specific case of the more general scBZ. Here, $u^{n}_{\Df,\vec{G}} (\kv)$ are the Bloch wavefunctions, obtained by solving the continuum model Hamiltonian in Eq.~\eqref{eq:Ham_moire_continuum}.

\subsection{Wannier functions and gauge invariance}

When considering polarization in a crystal, it is natural to work in terms of localized states such as Wannier functions \cite{marzari2012maximally}:
\beq{}
\begin{split}
\ket{w_{n,\R}} &= \frac{\Om}{(2\pi)^3}\oint_{\BZ}e^{-i\kv\cdot\R}\ket{\psi_{n,\kv}}\dd\kv\\
\ket{\psi_{n,\kv}} & = \sum_{\vec{R}} e^{i\kv\cdot\R} \ket{w_{n,\R}}
\end{split}
\eec
which are the Fourier transforms of the Bloch states. We have one for each band $n$ and each lattice vector $\R$. In the case of a supercell, $\R$ represents a supercell vector. The Wannier functions are orthonormal, have translational invariance, and are exponentially localized for a system with topologically trivial electronic bands. In seminal works by King-Smith and Vanderbilt \cite{vanderbilt1993electric,king1993theory}, it was shown that the Wannier centers ${\bar{\w}_n \equiv \braket{w_{n,0}|\rv|w_{n,0}}}$, the expectation values of the position operator in the Wannier basis, can be identified as the integral of the Berry phases, with units of length (see Appendix A):
\beq{W-center-final}
\bar{\w}_{n} \equiv \bra{w_{n,0}}\rv\ket{w_{n,0}} = \frac{i\Om}{(2\pi)^3}\oint_{\BZ}\braket{u_{n,\kv}|\nabla_{\kv} u_{n,\kv}}\dd\kv 
\eep
A well-known property of the Wannier functions is that their centers are invariant, modulo a lattice vector, with respect to single-band gauge transformations of the Bloch states:
\beq{}
\ket{u_{n,\kv}}\to \ket{u'_{n,\kv}} = e^{-i\b_n(\kv)}\ket{u_{n,\kv}}
\eec
where $\beta(\kv) = \beta(\kv+\vec{G}) + \vec{G} \cdot \R$, and $\vec{G}$ is a reciprocal lattice vector. For a so-called small gauge transformation defined by $\R = 0$, each Wannier center is invariant: $\bar{\w}'_{n} = \bar{\w}_{n}$, whereas for a large transformation with $\R \neq 0$: $\bar{\w}'_{n} = \bar{\w}_{n} + \R$, which contributes a quantum of polarization to the total polarization, but does not affect the physical observables. 

All such transformations provide automorphisms of isolated bands. However, when there are crossings between bands, the identity of single bands within the band subspaces is lost, which can lead to potential problems with the smoothness of integrands when computing Wannier centers \cite{vanderbilt2018berry}. In this case, on isolating an occupied band subspace from a Bloch bundle, more general multi-band gauge transformations apply
\beq{}
    \ket{u_{n,\kv}}\to \ket{u'_{n,\kv}} = \sum^{\occ}_m U_{nm}(\kv) \ket{u_{m,\kv}}
\eec
with the single-band gauge transformations ${U_{nm}(\kv) = \delta_{n,m}e^{-i\beta_n(\kv)}}$ constituting only a subset of the non-Abelian matrix transformations. Under such transformations, the trace of the matrix-valued non-Abelian Berry connection,
\beq{}
\A_{nm}(\kv) = i\braket{u_{n,\kv} | \nabla_{\kv} u_{m,\kv}}
\eec
is preserved, rather than the individual components. Therefore in general, the gauge invariant quantity is not the individual Wannier centers, but the sum of Wannier centers of occupied bands \cite{gresch2017z2pack}:
\beq{}
    \sum^{\occ}_n \bar{\w}'_n = \sum^{\occ}_n \bar{\w}_n
\eep
This gauge freedom is typically used to change the representation of the states to obtain maximally localized Wannier functions \cite{marzari2012maximally}, using the {\sc Wannier90} code \cite{pizzi2020wannier90}, for example. The Wannier centers can be calculated in configuration space using first-principles calculations and {\sc Wannier90}:
\beq{}
\bar{w}_{n}(\x) = \braket{w_{n,0}(\x)|\rv|w_{n,0}(\x)}
\eep
As mentioned previously, the Wannier centers have been used to estimate the local polarization in real space using \cite{wu2006wannier}:
\beq{eq:P-wannier-approx}
\P(\rv_j) = -\frac{ef}{\Om_0}\sum^{\occ}_{ \bar{\w}_n \in \rv_j} \bar{\w}_n
\eec
where the local polarization in cell $\rv_j$ is related to the sum over Wannier centers in cell $\rv_j$. However, Eq.~\eqref{eq:P-wannier-approx} is not gauge invariant in general, since individual sums of Wannier centers are not gauge invariant in the multi-band case. Going from a unit cell to a supercell shrinks the BZ and introduces significant band folding, in which case it might not always be possible to disentangle the Wannier functions into isolated unit cells.

We propose that the correct way to define the local polarization in terms of Wannier centers is to calculate the changes of \textit{all} the Wannier centers of occupied bands in a supercell with respect to local perturbations:
\beq{eq:P-wannier-modern}
\P(\rv_j) = -\frac{ef}{\Om_0}\sum_{n}^{\occ} \int_0^{\x(\rv_j)} \partial_{x'_{\k,\a}}\bar{\w}_n \dd x'_{\k,\a}
\eec
where $\k\in \rv_j$.
Manifestly, this shows that the local polarization is defined mod $e\R/\Omega_0$ under large gauge transformations of the Bloch bands over the supercell BZ. As with Eq.~\eqref{eq:P-local-modern}, Eq.~\eqref{eq:P-wannier-modern} calculates the change in a global property of the system, which is well-defined, with respect to local perturbations. If the perturbation in a given cell only affects the Wannier centers attributed to that cell, then Eq.~\eqref{eq:P-wannier-modern} reduces to Eq.~\eqref{eq:P-wannier-approx}.

The expression for local polarization given by Eq.~\eqref{eq:P-wannier-modern} is exact, but evaluation of the Wannier centers and their variation in a supercell is computationally and technically demanding.~In practice, Wannier functions are typically obtained in configuration space, which works well in many circumstances; for example, tight-binding models of twisted bilayers parametrized using Wannier functions have been shown to give accurate descriptions of the electronic bands for a range of supercell sizes \cite{carr2019exact,carr2019derivation}. 

Nonetheless, it is interesting to assess how large a discrepancy there can be between the local polarization obtained from configuration space and real space Wannier centers. Consider a unit cell, repeated to form a supercell, but with an identical configuration in each cell. Now, we switch on an additional supercell potential correction, $\la \D V_{\text{sl}}(\x(\rv))$, to the potential imposed by configuration space, where $\la = 0 \rightarrow 1$ parameterizes the switching on/off of the correction. For $\la = 0$, the Wannier functions should be the same in each unit cell by translational invariance, and are thus equivalent to the Wannier functions at a given point in configuration space: ${\ket{w^{\text{cs}}_{n,\R}} \equiv \ket{w^{\la = 0}_{n,\R}}}$. For $\la = 1$, the potential in each unit cell is different, and the Wannier functions are no longer equivalent by translational invariance. We denote these as the supercell Wannier functions: ${\ket{w^{\text{sl}}_{n,\R}} \equiv \ket{w^{\la = 1}_{n,\R}}}$, i.e.~those obtained directly in real space. In both $\ket{w^{\text{cs}}_{n,\R}}$ and $\ket{w^{\text{sl}}_{n,\R}}$, $\R$ is a supercell vector. The Wannier functions are given by ${w^{\la}_{n,\R}(\rv-\R) = \braket{\rv|w^{\la}_{n,\R}}}$ in the position representation, and for any $\la$, and $\R = 0$ we have
\beq{eq:wannier-lambda}
\bar{\w}^{\la}_{n} = \bra{w^{\la}_{n,0}} \rv \ket{w^{\la}_{n,0}}
\eec
which can be used to track the evolution of the Wannier centres after switching on the supercell potential $\D V_{\text{sl}}(\x(\rv))$. The cumulative change on applying the supercell potential correction can be written as
\beq{eq:w-switch}
   \ket{w^{\text{sl}}_{n,0}} = \ket{w^{\text{cs}}_{n,0}} + \ket{\Delta w_{n,0}}
\eec
where
\beq{}
\ket{\Delta w_{n,0}} = \int^1_0  \ket{\partial_\la w_{n,0}} \dd\la
\eep
The change in the Wannier functions can be determined from the corresponding changes in the Bloch states:
\beq{}
    \ket{\partial_\la w_{n, 0}} = \frac{\Omsc}{(2\pi)^3} \int_{\scBZ}  e^{i\kv \cdot \rv}  \ket{\partial_\la u_{n\kv}} \dd\kv
\eec
where the changes in the Bloch states are determined by the Sternheimer equation \cite{sternheimer1954electronic,gonze1995adiabatic,baroni2001phonons},
\beq{}
\ket{\partial_{\la}u_{n\kv}} = - A_n(\kv)\ket{u_{n\kv}} + \sum_{m\neq n}\frac{\ket{u_{m\kv}}\bra{u_{m\kv}}}{E_n-E_m} \lb \partial_{\la} H\rb\ket{u_{n\kv}}
\eec
which can be evaluated using DFPT. The difference between the local polarization obtained in real space and configuration space is determined by the error in the Wannier centers:
\beq{}
\begin{split}
    \D\bar{\w}_{n} = \bar{\w}^{\text{sl}}_{n} - \bar{\w}^{\text{cs}}_{n} = \bra{\Delta w_{n,0}} \rv \ket{w^{\text{cs}}_{n,0}} + \text{c.c.}
    + \bigO(\D \bar{\w}_n ^2)
\end{split}
\eec
to first order in $\D\bar{\w}_{n}$. 
The correction to local polarization is then given by
\beq{eq:P-loc-err}
\begin{split}
\P^{\text{err}}(\rv_j) &= \P^{\text{sl}}(\rv_j) - \P^{\text{cs}}(\rv_j) \\
&= -\frac{ef}{\Om_0}\sum_{n}^{\occ} \int_0^{\x(\rv_j)} \partial_{x'_{\k,\a}} \Delta \bar{\w}_n \dd x'_{\k,\a}
 \end{split}
\eec
where $\P^{\text{sl/cs}}(\rv_j)$ represent the local polarization calculated using Wannier centers in real space / configuration space:
\beq{eq:P-loc-err}
\P^{\text{sl/cs}}(\rv_j)
 = -\frac{ef}{\Om_0}\sum_{n}^{\occ} \int_0^{\x(\rv_j)} \partial_{x'_{\k,\a}} \bar{\w}^{\text{sl/cs}}_n \dd x'_{\k,\a}.
\eeq

\section{Results}
\subsection{First-principles calculations}

\begin{figure*}[t!]
\centering
\includegraphics[width=\linewidth]{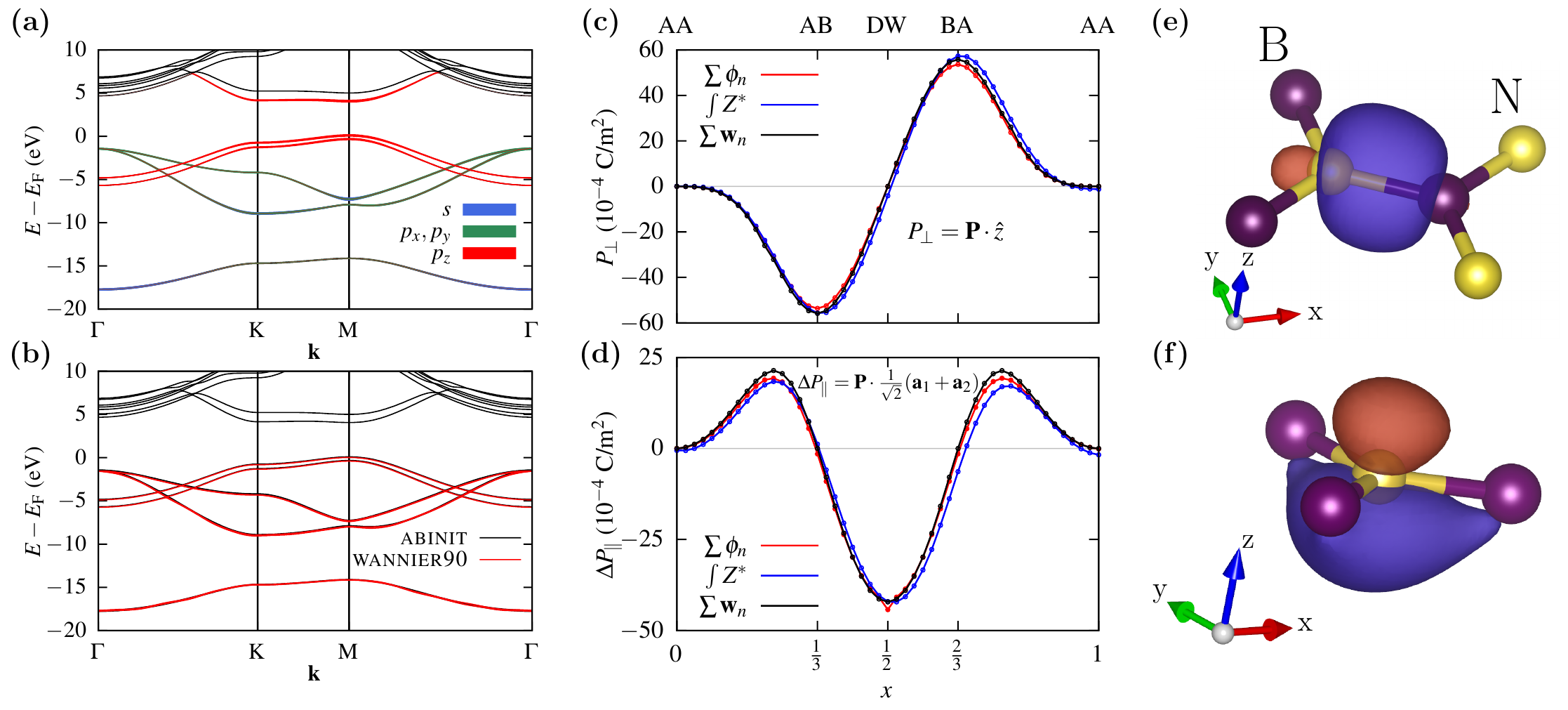}
\caption{ \textbf{(a)} Orbital character of the 8 valence bands and the bottom of the conduction band in bilayer hBN (AA stacking). \textbf{(b)} electronic band structure obtained from first-principles calculations using {\sc abinit} (black) and Wannier interpolation using {\sc Wannier90} (red). \textbf{(c)} Out-of-plane and \textbf{(d)} in-plane polarization in fractional coordinates along the configuration space diagonal, calculated from the Berry phases (red), integrating the Born effective charges (blue) and the Wannier centers (black). The positions high symmetry stackings AA ($x=0$), AB ($x=\frac{1}{3}$), DW ($x=\frac{1}{2}$) and BA ($x=\frac{2}{3}$) are indicated by the ticks. \textbf{(e-f)}: Maximally localized Wannier functions obtained in bilayer hBN: \textbf{(e)} an $sp$ orbital aligned with the B-N bond (three per layer) and \textbf{(f)} an $sp$ orbital normal to the layers (one per layer).}
\label{fig:bands-pol-wannier}
\end{figure*}

While we propose that the most correct way to calculate local polarization is from the Wannier centers / Born effective charges in real space, this is a very computationally heavy and technically demanding task for large supercells, which deserves its own dedicated study and is beyond the scope of this work. As a stepping stone, we first show that our proposed expressions for local polarization, Eqs.~\eqref{eq:P-local-modern} and \eqref{eq:P-wannier-modern}, give the correct expression in configuration space, i.e.~they are in agreement with the result obtained from Berry phases. As an example, we consider 3R-stacked bilayer hBN, following the methodology in Refs.~\cite{bennett2022electrically,bennett2022theory,bennett2023polar}.

First-principles density functional theory (DFT) calculations were performed using the {\sc abinit} \cite{gonze2009abinit} code, using {\sc psml} \cite{psml} norm-conserving pseudopotentials \cite{norm_conserving}, obtained from Pseudo-Dojo \cite{pseudodojo}. {\sc abinit} employs a plane wave basis set, which was determined using a kinetic energy cutoff of $1000$ eV, and a Monkhorst-Pack $\kv$-point grid \cite{mp} of $12 \times 12 \times 1$ was used. Calculations were converged until the relative changes in the total energy were less than $10^{-10}$ Ha. The revPBE exchange-correlation functional was used \cite{zhang1998comment}, and the vdw-DFT-D3(BJ) \cite{becke2006simple} correction was used to treat the long-range interactions between the layers.

The top hBN layer was translated along the unit cell diagonal over the bottom layer, which was held fixed. At each point a geometry relaxation was performed to obtain the equilibrium layer separation, while keeping the in-plane lattice vectors fixed. The out-of-plane and in-plane polarization were then obtained by calculating the Berry phases of the Bloch states. The out-of-plane polarization $P_{\perp}$ was found to be odd with respect to the relative stacking, and the in-plane polarization, $\D\P_{\parallel}$, was found to be even, as reported in Ref.~\cite{bennett2023polar}. At each point along the unit cell diagonal, the Born effective charges were calculated by calculating the change in the Hamiltonian and Bloch states in response to phonon and electric field perturbations, using the DFPT routines in {\sc abinit}. At each point in configuration space, the Wannier functions were also calculated using {\sc Wannier90}, which interfaces with {\sc abinit}. Maximally localized Wannier functions \cite{marzari2012maximally} were obtained by using the gauge freedom to minimize the spread. The orbital characters of the 8 valence bands and the lowest lying conduction band, see Fig.~\ref{fig:bands-pol-wannier}, were used to determine the initial projections onto atomic orbitals. The lowest valence bands are comprised of in-plane $sp$-like states, with three in each layer. The highest valence bands and lowest conduction bands have $p_z$ character, originating from the N and B atoms, respectively. The spread was minimized until the relative change was less than $10^{-10}$ \AA$^2$ and the bands obtained from Wannier interpolation were in agreement with the bands obtained from {\sc abinit}, see Fig.~\ref{fig:bands-pol-wannier} (b). The bands are well-reproduced with three in-plane $sp$-like states (Fig.~\ref{fig:bands-pol-wannier} (e)) and one $sp_z$-like state (Fig.~\ref{fig:bands-pol-wannier} (f)) for each layer.

The out-of-plane and in-plane polarization of bilayer hBN in configuration space are shown in Fig.~\ref{fig:bands-pol-wannier} (c) and (d), respectively. The red data show the polarization obtained from Berry phases, i.e. calculating the polarization from the electronic bands of a commensurate bilayer plus a relative translation between the layers. The blue data show the polarization obtained by integrating the Born effective charges along the unit cell diagonal, i.e.~Eq.~\eqref{eq:P-local-modern}. As shown in Ref.~\cite{bennett2023polar}, Eq.~\eqref{eq:P-approx-1} is not appropriate in layered ferroelectrics, but integrating the Born effective charges yields a polarization identical to the one obtained from the Berry phases. The black data show the polarization obtained by measuring the sum of Wannier centers in configuration space, which is in exact agreement with the other two methods. In configuration space, the 8 Wannier functions describe all of the occupied bands, so the sum of Wannier centers for each configuration is always gauge invariant, and Eq.~\eqref{eq:P-wannier-modern} reduces to Eq.~\eqref{eq:P-wannier-approx}.

\subsection{Effective model}

\begin{figure}[h!]
\centering
\includegraphics[width=\columnwidth]{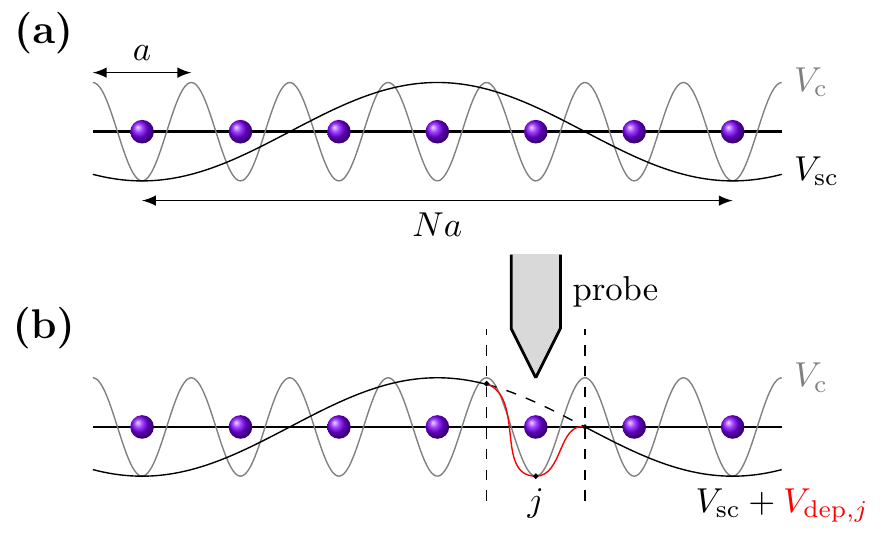}
\caption{\textbf{(a)} Sketch of the two-cosine model for a one-dimensional supercell: a periodic chain of atoms with spacing $a$. The potential from the ions $\Vc(x)$ is shown in gray. The supercell potential $\Vsc(x)$, with period $Na$, is shown in black. \textbf{(b)} Sketch of the system with a modified potential due to a depolarizing perturbation from a probe. The action of the perturbation brings the cell $j$ to a non-polar configuration and the response of the Wannier centers of occupied bands can be determined to measure the local polarization.}
\label{fig:sketch-1d}
\end{figure}

The calculation of polarization from first-principles calculations in configuration space are a valid approximation to the local polarization in real space for large supercells. As a proof of concept, we show using a one-dimensional effective model that our proposed definition of local polarization, Eq.~\eqref{eq:P-wannier-modern}, can be calculated directly in real space without relying on this approximation. We consider a model inspired by the Aubry-Andr\'e (AA) model \cite{aubry1980analyticity} which consists of a one-dimensional chain of atoms with an atomic spacing $a$ and a supercell potential with period $Na$, where ${N \in \mathbb{N} }$ is not necessarily large, see Fig.~\ref{fig:sketch-1d} (a). The model is described by the one-body Hamiltonian
\beq{eq:H-1d}
    H = -\frac{1}{2} \frac{d^2}{dx^2} + \Vc(x) + \Vsc(x)
\eec
in terms of one-dimensional position $x$, where the first term represents the kinetic energy of an electron, ${\Vc(x) = 2 V_0 \cos(Gx)}$ is the ionic potential with $G=\frac{2\pi}{a}$,  and ${V_{\text{sc}}(x) = 2 \la V_0\cos(G_N x)}$ is the supercell potential, which can be interpreted as a contribution due to the local displacement of the cores, where ${ G_N = \frac{G}{N} }$. As in the previous section, we introduce a parameter ${\la}$ which determines the relative strength between the core and supercell potentials and can be used to switch on the supercell potential. Contrary to the previous case, $\lambda$ can be larger than $\la = 1$ if the supercell potential is stronger than the core potential. 

\begin{figure}[h!]
\centering
\includegraphics[width=\columnwidth]{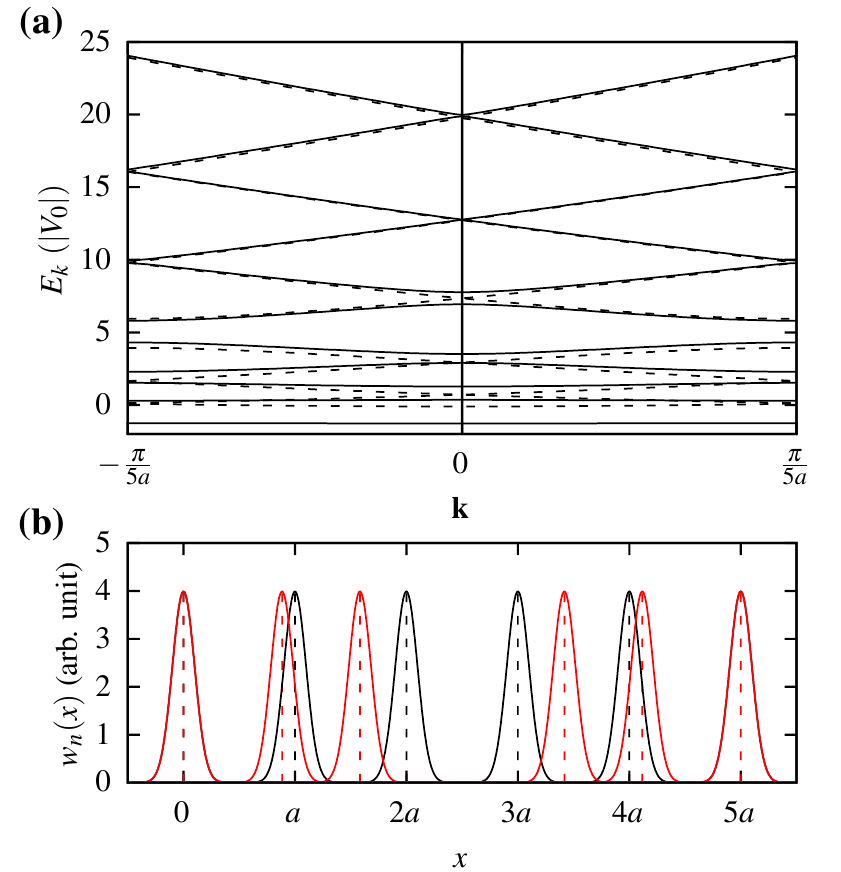}
\caption{\textbf{(a)} $N=5$ two-cosine model bands for $\la = 0$ (dashed) and for $\la = 1$ (solid), with $V_0 = -1$. As the supercell potential strength is switched on, additional gaps open. \textbf{(b)} Wannier centers for the five occupied supercell bands, using two different multi-band gauges. The first gauge results in atom-centered Wannier functions (black), and the second gauge results in Wannier functions with cell-dependent off-centering (red). The Wannier centers are indicated by the vertical dashed lines in both cases. The total sum of Wannier centers is the same for both gauges. As the Wannier functions in arbitrary gauges are not well-localized, their representations around different Wannier centers are purely illustrative.}
\label{fig:bands-1d}
\end{figure}

For weak core and superlattice potentials, we can perturbatively obtain the eigenstates of \eqref{eq:H-1d} using the nearly-free electron gas approach via nearly-degenerate perturbation theory, using a basis of plane-wave states $\ket{k}$ with energies ${\ep_k = \frac{1}{2}k^2 }$ from the free-particle problem. The potential mixes different plane wave states $\{\ket{k+mG_N}\}$, for ${\{m \in\mathbb{Z}: |m|\leq N\}}$, yielding a set of secular equations. Evaluating the corresponding matrix elements of the effective Hamiltonian, 
\beq{}
\begin{split}
\bra{k+mG_N} H \ket{k+nG_N} &= \ep_{k+mG_N} \d_{m,n} \\
&+ V_0 (\d_{m,n+N} + \d_{m,n-N}) \\
&+ \la V_0 (\d_{m,n+1} + \d_{m,n-1})
\end{split}
\eec
we obtain the following matrix equation:
\beq{eq:matrix}\resizebox{0.85\columnwidth}{!}{$
\begin{pmatrix}

\ep_{k} & \la V_0 & \la V_0 & \cdots & V_0 & V_0 \\
\la V^*_0 & \ep_{k+G_N} & 0 & \cdots & 0 & 0 \\
\la V^*_0 & 0 & \ep_{k-G_N} & \ddots & 0 & 0 \\
\vdots & \vdots & \ddots & \ddots & \vdots & \vdots \\

 V^*_0 & 0 & 0 & \cdots & \ep_{k+G}  & 0 \\
 V^*_0 & 0 & 0 & \cdots & 0 & \ep_{k-G} 

\end{pmatrix} 
\begin{pmatrix}
c_{k}\\ 
c_{k+G_N}\\ 
c_{k-G_N}\\
\vdots \\
c_{k+G} \\
c_{k-G}
\end{pmatrix}
= E_k
\begin{pmatrix}
c_{k}\\ 
c_{k+G_N}\\ 
c_{k-G_N}\\ 
\vdots \\
c_{k+G} \\
c_{k-G}
\end{pmatrix}
$}
\eep
The approximate (unnormalized) eigenstates are given by
\beq{}
\ket{\psi} = \sum^{N}_{m={-N}} c_{k+mG_N}\ket{k+mG_N}
\eec
with position representation 
\beq{eq:psi-position}
\psi(x) = e^{ikx} \sum^{N}_{m={-N}} c_{k+mG_N} e^{imG_Nx}
\eep
The approximate eigenstates are Bloch states, ${\psi(x) = e^{ikx} u_{n\kv}(x)}$, with cell-periodic parts which are periodic over a supercell period: ${ u_{n\kv}(x) = u_{n\kv}(x+Na) }$. Solving the secular equations yields $2N+1$ coefficients $c_{k+mG_N}$ for each $k$, and a set of energy bands $E_k$ which are sensitive to the parameters $V_0$ and $\la$. The basis states corresponding to larger reciprocal lattice vectors can also be included to increase the accuracy of the approximate solution to the model. Eq.~\eqref{eq:matrix} was solved numerically for $N=5$, and the resulting bands are shown in Fig.~\ref{fig:bands-1d} (a). For $\la = 0$, the folding of the bands into the scBZ introduces many band crossings. Switching on the supercell potential with $\la\neq 0$ opens up several gaps where the bands cross.

Having found the approximate bands and eigenstates, we can proceed to obtain the Wannier functions $w^{\la}_{n,X}$ for a given $\la$, where $n$ is the band index and $X$ is a lattice vector. For $\la = 0$, the Wannier functions are given by
\beq{}
w^{0}_{n,X} (x - ja) \propto  \int_{\BZ}  e^{ik(x-ja)} \lb c_k + c_{k+G}e^{iGx}\rb \dd k
\eec
where $X$ is a unit cell vector and the integral is over the BZ, ${\left[-\frac{\pi}{a},\frac{\pi}{a} \right] }$, because the periodicity reduces to $a$ in the absence of a supercell potential. For $\la \neq 0$, the Wannier functions are given by
\beq{}
w^{\la}_{n,X} (x - jNa) \propto \int_{\scBZ} e^{ik(x-jNa)} \sum^{N}_{m={-N}} c_{k+mG_N} e^{imG_Nx} \dd k
\eec
where $X$ is a supercell vector and the integral is over the scBZ, ${[-\frac{\pi}{Na}, \frac{\pi}{Na}]}$. The Wannier functions were obtained and their centers are plotted in Fig.~\ref{fig:bands-1d} (b) for two different gauges. The first gauge projected the Wannier centers onto the atomic sites, as indicated by the black lines. For the second gauge, there is a displacement between the Wannier centers and the atomic sites in each cell, as indicated by the red lines. However, the sum of Wannier centers is the same in each case, and both sets of Wannier functions can be used as a localized basis. This illustrates the arbitrariness of describing local polarization in a supercell using partial sums over Wannier centers with Eq.~\eqref{eq:P-wannier-approx}, as the partial sums are not gauge invariant.

\begin{figure}[h!]
\centering
\includegraphics[width=\columnwidth]{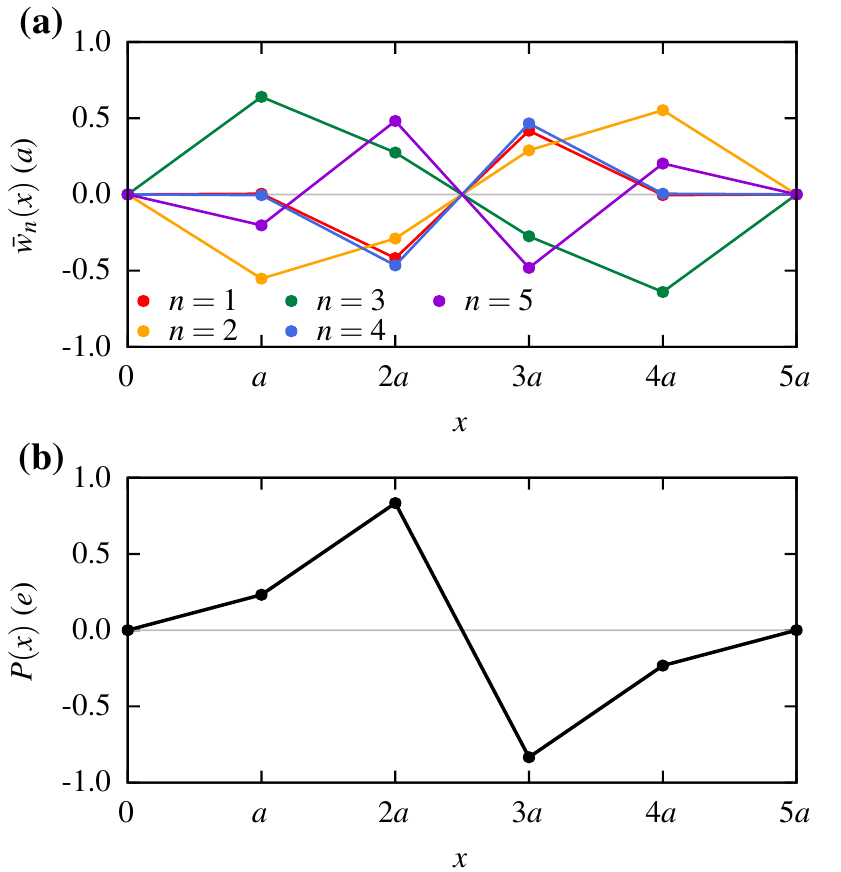}
\caption{\textbf{(a)} Shift of Wannier centers in response to local perturbations $\Vdep$ in each cell, for $\la = 2$ and $V_0 = -2$. \textbf{(b)} Local polarization calculated from the displacement of all Wannier centers in response to the perturbation potential $\Vdep$.}
\label{fig:pol-1d}
\end{figure}

In the spirit of Eq.~\eqref{eq:P-wannier-modern}, we introduce an additional local potential to Eq.~\eqref{eq:H-1d} in order to calculate the displacements of \textit{all} the Wannier centers in response to local perturbations. This can be achieved by switching a cell $j$ to a non-polar configuration using a depolarizing potential $\Vdep(x)$:
\beq{}\resizebox{0.87\columnwidth}{!}{$
    \Vdep(x) = \left[\th\lb x - ja + \frac{a}{2}\rb - \th\lb x - j a -\frac{a}{2}\rb\right] \left[V_{\text{sc}}(x-ja) - V_{\text{sc}}(x)\right] $}
\eec
where $\theta(x)$ is the Heaviside step function. Such a potential could be achieved using a tip to locally probe the system, see Fig.~\ref{fig:sketch-1d} (a). The supercell potential in cell $j$ is removed and replaced with the potential of the non-polar configuration. The matrix elements of $\Vdep$ are given by
\beq{}\resizebox{0.87\columnwidth}{!}{$
\begin{split}
    (V_{\text{dep},j})_{nm} &=  \Lambda_j\int^{ja+\frac{a}{2}}_{ja-\frac{a}{2}} e^{iG_N(m-n)x} \sin(G_N(x - ja/2)) \dd x \\
    \Lambda_j &= \frac{4 \lambda V_0}{Na} \sin{\lb \frac{\pi j}{N}\rb}
\end{split} $}
\eep
Adding these matrix elements to the secular equations, Eq.~\eqref{eq:matrix}, the change in the Wannier centers of the occupied bands in response to local perturbations in each cell can be obtained, see Fig.~\ref{fig:pol-1d} (a), which allows the local polarization to be calculated, see Fig.~\ref{fig:pol-1d} (b). We note that the polarization is odd, and reminiscent of the out-of-plane polarization in bilayer hBN along the unit cell diagonal, see Fig.~\ref{fig:bands-pol-wannier} (c). By the symmetry of the potential, the local polarization is odd about $x=0$, and sums to zero.

\section{Discussions and Conclusions}

In this work, we propose a definition of the local polarization in a crystal supercell, obtained either by integrating the Born effective charges, Eq.~\eqref{eq:P-local-modern}, or the changes in the Wannier centers of all occupied bands with respect to local perturbations, Eq.~\eqref{eq:P-wannier-modern}. In this way, the polarization can be calculated directly in real space, and is gauge invariant. We show that Eqs.~\eqref{eq:P-local-modern} and \eqref{eq:P-wannier-modern} are in agreement with the polarization obtained from Berry phases when evaluated in configuration space. Although we propose that it is more correct to calculate the polarization directly in real space, calculations involving DFPT or Wannierization with large supercells deserve their own dedicated study and we do not pursue this here. However, this is an important subject of future research: in many systems where topologically non-trivial polarization textures are observed, it should be checked that the shape and topology are the same when calculating the polarization in configuration space and in real space. For example, in twisted bilayers, the configuration space approximation is only valid for small angles, and while it is clear there should be a polar--nonpolar transition somewhere between ${0^{\circ} < \th  < 60^{\circ}}$, it is not immediately clear where this transition occurs, or what the polarization field looks like beyond the small angle regimes \cite{bennett2022theory}. Such large-scale calculations may not be unrealistically expensive: fortunately, larger twist angles result in smaller supercells, meaning the length scales where the configuration space approximation breaks down naturally contain fewer atoms. Furthermore, because the Born effective charge tensor is a mixed derivative of the free energy, it can be obtained using either type of perturbation (phonon or electric field). Thus, instead of $3N$ phonon perturbations, where $N$ is very large, the effective charges can be obtained with only 3 electric field perturbations. This is known as the `interchange theorem' \cite{gonze1997dynamical}. 

We also demonstrate as a proof of concept that the local polarization can be calculated directly in real space in a one-dimensional effective model. The polarization is calculated in a small supercell, for which the configuration space approximation is not valid, by calculating the change in the Wannier centers of all occupied bands in response to a depolarizing potential from a local probe. The calculated local polarization is consistent with the form of the potential experienced by electrons. 

A better definition and understanding of local polarization is essential for considering topological polarization. As mentioned previously, because polarization is topological in the real space sense, the topological properties are solely determined by the geometry of the supercell and the underlying crystal symmetry of the lattice. Having a correct definition of the local polarization such as Eq.~\eqref{eq:P-local-modern} shows that topological polarization can indeed be calculated from the bands of a supercell, and that this real space topology can be described at the electronic level. This is important when considering the relation between polarization and band topology. Recently, it was predicted that ferroelectric switching via van der Waals sliding can lead to a change in the QAH conductance in a topological insulator \cite{liang2023ferroelectric}. Another recent theory proposes that that for a material in a moir\'e potential, which results in electronic bands with nonzero Chern numbers, the Chern numbers can be altered by applying an electric field \cite{ghorashi2022topological}. In this example, the moir\'e superlattice potential was achieved using a non-polar material. Substituting a ferroelectric material may lead to Chern bands which can be switched at zero field. Although in both Refs.~\cite{liang2023ferroelectric} and \cite{ghorashi2022topological} the polarization was uniform, they both suggest that polarization can influence the topological properties of a system. There may be additional phenomena which are unique to topologically non-trivial polarization textures, although currently no such phenomena have been considered or proposed.

Similar to the effect of polarization on the band topology, one may consider the effect of the band topology on the local polarization textures. 
When considering local polarization, we can define a higher-dimensional space spanned by the BZ and configuration space, which for the case of a bilayer would be a 4-dimensional torus. 
In this higher dimensional space, the local polarization resembles the phase-space Berry curvature (see Eq.~\ref{eq:P-local-modern}). 
Thus the local polarization can be linked to the evolution of the phase-space Berry curvature and Chern numbers as a function of relative configurations. However, the local polarization is defined as an integral from the nonpolar reference state which does not form a closed loop in configuration space, and thus the Chern theorem does not apply unless considering a translation by a unit cell. For example, the most important shift in a bilayer is the one associated to the ferroelectric switching of the polarization, achieved by a relative sliding of a third of a unit cell diagonal: ${\x=\frac{1}{3}(\vec{a}_1+\vec{a}_2)}$. The change in polarization associated to this switching process does not form a closed loop in configuration space, and thus cannot result in a nonzero Chern number. Furthermore, in bilayer hBN the electronic bands are topologically trivial for every point in configuration space, and sliding one layer over the other does not change this. Because polarization alone does not break time-reversal symmetry, a polarization texture cannot by itself lead to any non-trivial Chern band topology. For this to occur, time-reversal symmetry must be broken by some other means. In order to see the interplay between topological polarization and band topology, we must consider a system which is already topologically non-trivial, but is also polar, such as those considered in Refs.~\cite{liang2023ferroelectric} and \cite{ghorashi2022topological}. In this case, the description of local polarization developed here would need to be generalized to the case of Chern insulators, by tracking the evolution of hybrid Wannier centers \cite{Sinisa2009,Song2021}. Here, bulk-boundary relations~\cite{Rmp1,Rmp2,Slager_bbc,Slager2018_unifiedbbc,Hatsugai93} also pose an intriguing question for future pursuits.

\section{Acknowledgements}
D.B.~and E.~K.~acknowledge funding from the US Army Research Office (ARO) MURI project under grant No.~W911NF-21-0147 and by the National Science Foundation DMREF program under Award No.~DMR-1922172.
W.~J.~J. acknowledges funding from the Rod Smallwood Studentship at Trinity College, Cambridge. 
R.~J.~S. and G.~C.~acknowledge funding from a New Investigator Award, EPSRC grant EP/W00187X/1. 
R.~J.~S. also acknowledges funding from Trinity College, University of Cambridge.

\section{Appendix A: Relation between Berry phases and Wannier centers}

For illustrative purposes, we briefly review the derivation for the important relation between the Berry phases and the Wannier centers \cite{vanderbilt2018berry}. First, we apply the position operator to a Wannier state $\ket{w_{n,\R}}$:
\beq{}
\begin{split}
\rv\ket{w_{n,\R}} 
&= \frac{\Om}{(2\pi)^3}\oint_{\BZ}\rv e^{-i\kv\cdot\R}\ket{\psi_{n,\kv}}\dd\kv \\
&= \frac{\Om}{(2\pi)^3}\oint_{\BZ}\rv e^{i\kv\cdot(\rv-\R)}\ket{u_{n,\kv}}\dd\kv
\end{split}
\eec
where we write the Bloch state as $\ket{\psi_{n,\kv}} = e^{i\kv\cdot\rv} \ket{u_{n,\kv}}$. Next we rewrite the product of the position operator and the exponential:
\beq{}
\rv e^{i\kv\cdot(\rv-\R)} = (\R-i\nabla_{\kv})e^{i\kv\cdot(\rv-\R)}
\eec
giving
\beq{}
\rv\ket{w_{n,\R}}
= \frac{\Om}{(2\pi)^3}\oint_{\BZ} (\R-i\nabla_{\kv})e^{i\kv\cdot(\rv-\R)}\ket{u_{n,\kv}}\dd\kv
\eep
For the term proportional to $\nabla_{\kv}$, we perform integration by parts. The boundary term vanishes because the integral is around a closed loop in the BZ and $\ket{u_{n,\kv}}$ has translational invariance. This gives
\beq{}
\rv\ket{w_{n,\R}}
= \frac{\Om}{(2\pi)^3}\oint_{\BZ}e^{i\kv\cdot(\rv-\R)}(\R+i\nabla_{\kv})\ket{u_{n,\kv}}\dd\kv
\eep
Now we multiply by
\beq{}
\bra{w_{n,0}} =  \frac{\Om}{(2\pi)^3}\oint_{\BZ}e^{-i\kv'\cdot \rv}\bra{u_{n,\kv'}}\dd\kv'
\eeq
which gives
\beq{}
\bra{w_{n0}}\rv\ket{w_{n\R}} = \frac{\Om}{(2\pi)^3}\oint_{\BZ}e^{-i\kv\cdot\R} \bra{u_{n\kv}}\R+i\nabla_{\kv}\ket{u_{n\kv}} \dd\kv 
\eep
The term proportional to $\R$ gives $\R\d_{0\R}$, which vanishes on setting $\R\to0$, as we get the expected expression for the Wannier centers:
\beq{W-center-final}
\bar{\w}_{n} \equiv \bra{w_{n,0}}\rv\ket{w_{n,0}} = \frac{i\Om}{(2\pi)^3}\oint_{\BZ}\braket{u_{n,\kv}|\nabla_{\kv} u_{n,\kv}}\dd\kv 
\eep


\begin{thebibliography}{111}%
\makeatletter
\providecommand \@ifxundefined [1]{%
 \@ifx{#1\undefined}
}%
\providecommand \@ifnum [1]{%
 \ifnum #1\expandafter \@firstoftwo
 \else \expandafter \@secondoftwo
 \fi
}%
\providecommand \@ifx [1]{%
 \ifx #1\expandafter \@firstoftwo
 \else \expandafter \@secondoftwo
 \fi
}%
\providecommand \natexlab [1]{#1}%
\providecommand \enquote  [1]{``#1''}%
\providecommand \bibnamefont  [1]{#1}%
\providecommand \bibfnamefont [1]{#1}%
\providecommand \citenamefont [1]{#1}%
\providecommand \href@noop [0]{\@secondoftwo}%
\providecommand \href [0]{\begingroup \@sanitize@url \@href}%
\providecommand \@href[1]{\@@startlink{#1}\@@href}%
\providecommand \@@href[1]{\endgroup#1\@@endlink}%
\providecommand \@sanitize@url [0]{\catcode `\\12\catcode `\$12\catcode
  `\&12\catcode `\#12\catcode `\^12\catcode `\_12\catcode `\%12\relax}%
\providecommand \@@startlink[1]{}%
\providecommand \@@endlink[0]{}%
\providecommand \url  [0]{\begingroup\@sanitize@url \@url }%
\providecommand \@url [1]{\endgroup\@href {#1}{\urlprefix }}%
\providecommand \urlprefix  [0]{URL }%
\providecommand \Eprint [0]{\href }%
\providecommand \doibase [0]{http://dx.doi.org/}%
\providecommand \selectlanguage [0]{\@gobble}%
\providecommand \bibinfo  [0]{\@secondoftwo}%
\providecommand \bibfield  [0]{\@secondoftwo}%
\providecommand \translation [1]{[#1]}%
\providecommand \BibitemOpen [0]{}%
\providecommand \bibitemStop [0]{}%
\providecommand \bibitemNoStop [0]{.\EOS\space}%
\providecommand \EOS [0]{\spacefactor3000\relax}%
\providecommand \BibitemShut  [1]{\csname bibitem#1\endcsname}%
\let\auto@bib@innerbib\@empty
\bibitem [{\citenamefont {Mitsui}\ and\ \citenamefont {Furuichi}(1953)}]{kmf}%
  \BibitemOpen
  \bibfield  {author} {\bibinfo {author} {\bibfnamefont {Toshito}\ \bibnamefont
  {Mitsui}}\ and\ \bibinfo {author} {\bibfnamefont {Jiro}\ \bibnamefont
  {Furuichi}},\ }\bibfield  {title} {\enquote {\bibinfo {title} {{Domain
  structure of {R}ochelle salt and KH$_2$PO$_4$}},}\ }\href {\doibase
  10.1103/PhysRev.90.193} {\bibfield  {journal} {\bibinfo  {journal} {Phys.
  Rev.}\ }\textbf {\bibinfo {volume} {90}},\ \bibinfo {pages} {193} (\bibinfo
  {year} {1953})}\BibitemShut {NoStop}%
\bibitem [{\citenamefont {Kopal}\ \emph {et~al.}(1997)\citenamefont {Kopal},
  \citenamefont {Bahnik},\ and\ \citenamefont {Fousek}}]{vacuum_1}%
  \BibitemOpen
  \bibfield  {author} {\bibinfo {author} {\bibfnamefont {A.}~\bibnamefont
  {Kopal}}, \bibinfo {author} {\bibfnamefont {T.}~\bibnamefont {Bahnik}}, \
  and\ \bibinfo {author} {\bibfnamefont {J.}~\bibnamefont {Fousek}},\
  }\bibfield  {title} {\enquote {\bibinfo {title} {{Domain formation in thin
  ferroelectric films: the role of depolarization energy}},}\ }\href {\doibase
  10.1080/00150199708213485} {\bibfield  {journal} {\bibinfo  {journal}
  {Ferroelectrics}\ }\textbf {\bibinfo {volume} {202}},\ \bibinfo {pages}
  {267--274} (\bibinfo {year} {1997})}\BibitemShut {NoStop}%
\bibitem [{\citenamefont {Kittel}(1946)}]{kittel}%
  \BibitemOpen
  \bibfield  {author} {\bibinfo {author} {\bibfnamefont {Charles}\ \bibnamefont
  {Kittel}},\ }\bibfield  {title} {\enquote {\bibinfo {title} {Theory of the
  structure of ferromagnetic domains in films and small particles},}\ }\href
  {\doibase 10.1103/PhysRev.70.965} {\bibfield  {journal} {\bibinfo  {journal}
  {Phys. Rev.}\ }\textbf {\bibinfo {volume} {70}},\ \bibinfo {pages} {965--971}
  (\bibinfo {year} {1946})}\BibitemShut {NoStop}%
\bibitem [{\citenamefont {Junquera}\ and\ \citenamefont
  {Ghosez}(2003)}]{junquera2003critical}%
  \BibitemOpen
  \bibfield  {author} {\bibinfo {author} {\bibfnamefont {Javier}\ \bibnamefont
  {Junquera}}\ and\ \bibinfo {author} {\bibfnamefont {Philippe}\ \bibnamefont
  {Ghosez}},\ }\bibfield  {title} {\enquote {\bibinfo {title} {Critical
  thickness for ferroelectricity in perovskite ultrathin films},}\ }\href
  {\doibase https://doi.org/10.1038/nature01501} {\bibfield  {journal}
  {\bibinfo  {journal} {Nature}\ }\textbf {\bibinfo {volume} {422}},\ \bibinfo
  {pages} {506--509} (\bibinfo {year} {2003})}\BibitemShut {NoStop}%
\bibitem [{\citenamefont {Bennett}\ \emph {et~al.}(2020)\citenamefont
  {Bennett}, \citenamefont {Mu{\~n}oz~Basagoiti},\ and\ \citenamefont
  {Artacho}}]{bennett2020electrostatics}%
  \BibitemOpen
  \bibfield  {author} {\bibinfo {author} {\bibfnamefont {Daniel}\ \bibnamefont
  {Bennett}}, \bibinfo {author} {\bibfnamefont {Maitane}\ \bibnamefont
  {Mu{\~n}oz~Basagoiti}}, \ and\ \bibinfo {author} {\bibfnamefont {Emilio}\
  \bibnamefont {Artacho}},\ }\bibfield  {title} {\enquote {\bibinfo {title}
  {Electrostatics and domains in ferroelectric superlattices},}\ }\href
  {\doibase https://doi.org/10.1098/rsos.201270} {\bibfield  {journal}
  {\bibinfo  {journal} {R. Soc. Open Sci.}\ }\textbf {\bibinfo {volume} {7}},\
  \bibinfo {pages} {201270} (\bibinfo {year} {2020})}\BibitemShut {NoStop}%
\bibitem [{\citenamefont {Luk'yanchuk}\ \emph {et~al.}(2009)\citenamefont
  {Luk'yanchuk}, \citenamefont {Lahoche},\ and\ \citenamefont
  {Sen\'e}}]{luk2009universal}%
  \BibitemOpen
  \bibfield  {author} {\bibinfo {author} {\bibfnamefont {Igor~A.}\ \bibnamefont
  {Luk'yanchuk}}, \bibinfo {author} {\bibfnamefont {Laurent}\ \bibnamefont
  {Lahoche}}, \ and\ \bibinfo {author} {\bibfnamefont {Ana\"{\i}s}\
  \bibnamefont {Sen\'e}},\ }\bibfield  {title} {\enquote {\bibinfo {title}
  {Universal properties of ferroelectric domains},}\ }\href {\doibase
  10.1103/PhysRevLett.102.147601} {\bibfield  {journal} {\bibinfo  {journal}
  {Phys. Rev. Lett.}\ }\textbf {\bibinfo {volume} {102}},\ \bibinfo {pages}
  {147601} (\bibinfo {year} {2009})}\BibitemShut {NoStop}%
\bibitem [{\citenamefont {Ji}\ \emph {et~al.}(2019)\citenamefont {Ji},
  \citenamefont {Cai}, \citenamefont {Paudel}, \citenamefont {Sun},
  \citenamefont {Zhang}, \citenamefont {Han}, \citenamefont {Wei},
  \citenamefont {Zang}, \citenamefont {Gu}, \citenamefont {Zhang} \emph
  {et~al.}}]{monolayer_perovskite}%
  \BibitemOpen
  \bibfield  {author} {\bibinfo {author} {\bibfnamefont {Dianxiang}\
  \bibnamefont {Ji}}, \bibinfo {author} {\bibfnamefont {Songhua}\ \bibnamefont
  {Cai}}, \bibinfo {author} {\bibfnamefont {Tula~R}\ \bibnamefont {Paudel}},
  \bibinfo {author} {\bibfnamefont {Haoying}\ \bibnamefont {Sun}}, \bibinfo
  {author} {\bibfnamefont {Chunchen}\ \bibnamefont {Zhang}}, \bibinfo {author}
  {\bibfnamefont {Lu}~\bibnamefont {Han}}, \bibinfo {author} {\bibfnamefont
  {Yifan}\ \bibnamefont {Wei}}, \bibinfo {author} {\bibfnamefont {Yipeng}\
  \bibnamefont {Zang}}, \bibinfo {author} {\bibfnamefont {Min}\ \bibnamefont
  {Gu}}, \bibinfo {author} {\bibfnamefont {Yi}~\bibnamefont {Zhang}},  \emph
  {et~al.},\ }\bibfield  {title} {\enquote {\bibinfo {title} {Freestanding
  crystalline oxide perovskites down to the monolayer limit},}\ }\href
  {\doibase https://doi.org/10.1038/s41586-019-1255-7} {\bibfield  {journal}
  {\bibinfo  {journal} {Nature}\ }\textbf {\bibinfo {volume} {570}},\ \bibinfo
  {pages} {87--90} (\bibinfo {year} {2019})}\BibitemShut {NoStop}%
\bibitem [{\citenamefont {G{\'o}mez-Ortiz}\ \emph {et~al.}(2023)\citenamefont
  {G{\'o}mez-Ortiz}, \citenamefont {Aramberri}, \citenamefont {L{\'o}pez},
  \citenamefont {Garc{\'\i}a-Fern{\'a}ndez}, \citenamefont {{\'I}{\~n}iguez},\
  and\ \citenamefont {Junquera}}]{gomez2023kittel}%
  \BibitemOpen
  \bibfield  {author} {\bibinfo {author} {\bibfnamefont {Fernando}\
  \bibnamefont {G{\'o}mez-Ortiz}}, \bibinfo {author} {\bibfnamefont {Hugo}\
  \bibnamefont {Aramberri}}, \bibinfo {author} {\bibfnamefont {Juan~M}\
  \bibnamefont {L{\'o}pez}}, \bibinfo {author} {\bibfnamefont {Pablo}\
  \bibnamefont {Garc{\'\i}a-Fern{\'a}ndez}}, \bibinfo {author} {\bibfnamefont
  {Jorge}\ \bibnamefont {{\'I}{\~n}iguez}}, \ and\ \bibinfo {author}
  {\bibfnamefont {Javier}\ \bibnamefont {Junquera}},\ }\bibfield  {title}
  {\enquote {\bibinfo {title} {{Kittel law and domain formation mechanism in
  PbTiO$_3$/SrTiO$_3$ superlattices}},}\ }\href
  {https://doi.org/10.48550/arXiv.2303.01755} {\bibfield  {journal} {\bibinfo
  {journal} {arXiv:2303.01755}\ } (\bibinfo {year} {2023})}\BibitemShut
  {NoStop}%
\bibitem [{\citenamefont {Callori}\ \emph {et~al.}(2012)\citenamefont
  {Callori}, \citenamefont {Gabel}, \citenamefont {Su}, \citenamefont
  {Sinsheimer}, \citenamefont {Fernandez-Serra},\ and\ \citenamefont
  {Dawber}}]{callori2012ferroelectric}%
  \BibitemOpen
  \bibfield  {author} {\bibinfo {author} {\bibfnamefont {S.~J.}\ \bibnamefont
  {Callori}}, \bibinfo {author} {\bibfnamefont {J.}~\bibnamefont {Gabel}},
  \bibinfo {author} {\bibfnamefont {D.}~\bibnamefont {Su}}, \bibinfo {author}
  {\bibfnamefont {J.}~\bibnamefont {Sinsheimer}}, \bibinfo {author}
  {\bibfnamefont {M.~V.}\ \bibnamefont {Fernandez-Serra}}, \ and\ \bibinfo
  {author} {\bibfnamefont {M.}~\bibnamefont {Dawber}},\ }\bibfield  {title}
  {\enquote {\bibinfo {title} {{Ferroelectric PbTiO$_3$/SrRuO$_3$ superlattices
  with broken inversion symmetry}},}\ }\href {\doibase
  10.1103/PhysRevLett.109.067601} {\bibfield  {journal} {\bibinfo  {journal}
  {Phys. Rev. Lett.}\ }\textbf {\bibinfo {volume} {109}},\ \bibinfo {pages}
  {067601} (\bibinfo {year} {2012})}\BibitemShut {NoStop}%
\bibitem [{\citenamefont {Zhang}\ \emph {et~al.}(2017)\citenamefont {Zhang},
  \citenamefont {Dufresne}, \citenamefont {Chen}, \citenamefont {Park},
  \citenamefont {Cosgriff}, \citenamefont {Yusuf}, \citenamefont {Dong},
  \citenamefont {Fong}, \citenamefont {Zhou}, \citenamefont {Cai},
  \citenamefont {Harder}, \citenamefont {Callori}, \citenamefont {Dawber},
  \citenamefont {Evans},\ and\ \citenamefont {Sandy}}]{zhang2017thermal}%
  \BibitemOpen
  \bibfield  {author} {\bibinfo {author} {\bibfnamefont {Qingteng}\
  \bibnamefont {Zhang}}, \bibinfo {author} {\bibfnamefont {Eric~M.}\
  \bibnamefont {Dufresne}}, \bibinfo {author} {\bibfnamefont {Pice}\
  \bibnamefont {Chen}}, \bibinfo {author} {\bibfnamefont {Joonkyu}\
  \bibnamefont {Park}}, \bibinfo {author} {\bibfnamefont {Margaret~P.}\
  \bibnamefont {Cosgriff}}, \bibinfo {author} {\bibfnamefont {Mohammed}\
  \bibnamefont {Yusuf}}, \bibinfo {author} {\bibfnamefont {Yongqi}\
  \bibnamefont {Dong}}, \bibinfo {author} {\bibfnamefont {Dillon~D.}\
  \bibnamefont {Fong}}, \bibinfo {author} {\bibfnamefont {Hua}\ \bibnamefont
  {Zhou}}, \bibinfo {author} {\bibfnamefont {Zhonghou}\ \bibnamefont {Cai}},
  \bibinfo {author} {\bibfnamefont {Ross~J.}\ \bibnamefont {Harder}}, \bibinfo
  {author} {\bibfnamefont {Sara~J.}\ \bibnamefont {Callori}}, \bibinfo {author}
  {\bibfnamefont {Matthew}\ \bibnamefont {Dawber}}, \bibinfo {author}
  {\bibfnamefont {Paul~G.}\ \bibnamefont {Evans}}, \ and\ \bibinfo {author}
  {\bibfnamefont {Alec~R.}\ \bibnamefont {Sandy}},\ }\bibfield  {title}
  {\enquote {\bibinfo {title} {{Thermal fluctuations of ferroelectric
  nanodomains in a ferroelectric-dielectric PbTiO$_3$/SrTiO$_3$
  superlattice}},}\ }\href {\doibase 10.1103/PhysRevLett.118.097601} {\bibfield
   {journal} {\bibinfo  {journal} {Phys. Rev. Lett.}\ }\textbf {\bibinfo
  {volume} {118}},\ \bibinfo {pages} {097601} (\bibinfo {year}
  {2017})}\BibitemShut {NoStop}%
\bibitem [{\citenamefont {Dawber}(2017)}]{dawber2017balancing}%
  \BibitemOpen
  \bibfield  {author} {\bibinfo {author} {\bibfnamefont {Matthew}\ \bibnamefont
  {Dawber}},\ }\bibfield  {title} {\enquote {\bibinfo {title} {Balancing polar
  vortices and stripes},}\ }\href {\doibase https://doi.org/10.1038/nmat4962}
  {\bibfield  {journal} {\bibinfo  {journal} {Nat. Mater.}\ }\textbf {\bibinfo
  {volume} {16}},\ \bibinfo {pages} {971--972} (\bibinfo {year}
  {2017})}\BibitemShut {NoStop}%
\bibitem [{\citenamefont {Park}\ \emph {et~al.}(2018)\citenamefont {Park},
  \citenamefont {Mangeri}, \citenamefont {Zhang}, \citenamefont {Yusuf},
  \citenamefont {Pateras}, \citenamefont {Dawber}, \citenamefont {Holt},
  \citenamefont {Heinonen}, \citenamefont {Nakhmanson},\ and\ \citenamefont
  {Evans}}]{park2018domain}%
  \BibitemOpen
  \bibfield  {author} {\bibinfo {author} {\bibfnamefont {Joonkyu}\ \bibnamefont
  {Park}}, \bibinfo {author} {\bibfnamefont {John}\ \bibnamefont {Mangeri}},
  \bibinfo {author} {\bibfnamefont {Qingteng}\ \bibnamefont {Zhang}}, \bibinfo
  {author} {\bibfnamefont {M~Humed}\ \bibnamefont {Yusuf}}, \bibinfo {author}
  {\bibfnamefont {Anastasios}\ \bibnamefont {Pateras}}, \bibinfo {author}
  {\bibfnamefont {Matthew}\ \bibnamefont {Dawber}}, \bibinfo {author}
  {\bibfnamefont {Martin~V}\ \bibnamefont {Holt}}, \bibinfo {author}
  {\bibfnamefont {Olle~G}\ \bibnamefont {Heinonen}}, \bibinfo {author}
  {\bibfnamefont {Serge}\ \bibnamefont {Nakhmanson}}, \ and\ \bibinfo {author}
  {\bibfnamefont {Paul~G}\ \bibnamefont {Evans}},\ }\bibfield  {title}
  {\enquote {\bibinfo {title} {{Domain alignment within
  ferroelectric/dielectric PbTiO$_3$/SrTiO$_3$ superlattice nanostructures}},}\
  }\href {\doibase https://doi.org/10.1039/C7NR07203A} {\bibfield  {journal}
  {\bibinfo  {journal} {Nanoscale}\ }\textbf {\bibinfo {volume} {10}},\
  \bibinfo {pages} {3262--3271} (\bibinfo {year} {2018})}\BibitemShut {NoStop}%
\bibitem [{\citenamefont {Susarla}\ \emph {et~al.}(2021)\citenamefont
  {Susarla}, \citenamefont {Garc{\'\i}a-Fern{\'a}ndez}, \citenamefont {Ophus},
  \citenamefont {Das}, \citenamefont {Aguado-Puente}, \citenamefont {McCarter},
  \citenamefont {Ercius}, \citenamefont {Martin}, \citenamefont {Ramesh},\ and\
  \citenamefont {Junquera}}]{susarla2021atomic}%
  \BibitemOpen
  \bibfield  {author} {\bibinfo {author} {\bibfnamefont {Sandhya}\ \bibnamefont
  {Susarla}}, \bibinfo {author} {\bibfnamefont {Pablo}\ \bibnamefont
  {Garc{\'\i}a-Fern{\'a}ndez}}, \bibinfo {author} {\bibfnamefont {Colin}\
  \bibnamefont {Ophus}}, \bibinfo {author} {\bibfnamefont {Sujit}\ \bibnamefont
  {Das}}, \bibinfo {author} {\bibfnamefont {Pablo}\ \bibnamefont
  {Aguado-Puente}}, \bibinfo {author} {\bibfnamefont {Margaret}\ \bibnamefont
  {McCarter}}, \bibinfo {author} {\bibfnamefont {Peter}\ \bibnamefont
  {Ercius}}, \bibinfo {author} {\bibfnamefont {Lane~W}\ \bibnamefont {Martin}},
  \bibinfo {author} {\bibfnamefont {Ramamoorthy}\ \bibnamefont {Ramesh}}, \
  and\ \bibinfo {author} {\bibfnamefont {Javier}\ \bibnamefont {Junquera}},\
  }\bibfield  {title} {\enquote {\bibinfo {title} {Atomic scale crystal field
  mapping of polar vortices in oxide superlattices},}\ }\href {\doibase
  https://doi.org/10.1038/s41467-021-26476-5} {\bibfield  {journal} {\bibinfo
  {journal} {Nat. Commun.}\ }\textbf {\bibinfo {volume} {12}},\ \bibinfo
  {pages} {6273} (\bibinfo {year} {2021})}\BibitemShut {NoStop}%
\bibitem [{\citenamefont {Urban}\ \emph {et~al.}(2002)\citenamefont {Urban},
  \citenamefont {Yun}, \citenamefont {Gu},\ and\ \citenamefont
  {Park}}]{urban2002synthesis}%
  \BibitemOpen
  \bibfield  {author} {\bibinfo {author} {\bibfnamefont {Jeffrey~J}\
  \bibnamefont {Urban}}, \bibinfo {author} {\bibfnamefont {Wan~Soo}\
  \bibnamefont {Yun}}, \bibinfo {author} {\bibfnamefont {Qian}\ \bibnamefont
  {Gu}}, \ and\ \bibinfo {author} {\bibfnamefont {Hongkun}\ \bibnamefont
  {Park}},\ }\bibfield  {title} {\enquote {\bibinfo {title} {Synthesis of
  single-crystalline perovskite nanorods composed of barium titanate and
  strontium titanate},}\ }\href {\doibase https://doi.org/10.1021/ja017694b}
  {\bibfield  {journal} {\bibinfo  {journal} {J. Am. Chem. Soc.}\ }\textbf
  {\bibinfo {volume} {124}},\ \bibinfo {pages} {1186--1187} (\bibinfo {year}
  {2002})}\BibitemShut {NoStop}%
\bibitem [{\citenamefont {Yun}\ \emph {et~al.}(2002)\citenamefont {Yun},
  \citenamefont {Urban}, \citenamefont {Gu},\ and\ \citenamefont
  {Park}}]{yun2002ferroelectric}%
  \BibitemOpen
  \bibfield  {author} {\bibinfo {author} {\bibfnamefont {Wan~Soo}\ \bibnamefont
  {Yun}}, \bibinfo {author} {\bibfnamefont {Jeffrey~J}\ \bibnamefont {Urban}},
  \bibinfo {author} {\bibfnamefont {Qian}\ \bibnamefont {Gu}}, \ and\ \bibinfo
  {author} {\bibfnamefont {Hongkun}\ \bibnamefont {Park}},\ }\bibfield  {title}
  {\enquote {\bibinfo {title} {Ferroelectric properties of individual barium
  titanate nanowires investigated by scanned probe microscopy},}\ }\href
  {\doibase https://doi.org/10.1021/nl015702g} {\bibfield  {journal} {\bibinfo
  {journal} {Nano Lett.}\ }\textbf {\bibinfo {volume} {2}},\ \bibinfo {pages}
  {447--450} (\bibinfo {year} {2002})}\BibitemShut {NoStop}%
\bibitem [{\citenamefont {Luo}\ \emph {et~al.}(2003)\citenamefont {Luo},
  \citenamefont {Szafraniak}, \citenamefont {Zakharov}, \citenamefont
  {Nagarajan}, \citenamefont {Steinhart}, \citenamefont {Wehrspohn},
  \citenamefont {Wendorff}, \citenamefont {Ramesh},\ and\ \citenamefont
  {Alexe}}]{luo2003nanoshell}%
  \BibitemOpen
  \bibfield  {author} {\bibinfo {author} {\bibfnamefont {Yun}\ \bibnamefont
  {Luo}}, \bibinfo {author} {\bibfnamefont {Izabela}\ \bibnamefont
  {Szafraniak}}, \bibinfo {author} {\bibfnamefont {Nikolai~D}\ \bibnamefont
  {Zakharov}}, \bibinfo {author} {\bibfnamefont {Valanoor}\ \bibnamefont
  {Nagarajan}}, \bibinfo {author} {\bibfnamefont {Martin}\ \bibnamefont
  {Steinhart}}, \bibinfo {author} {\bibfnamefont {Ralf~B}\ \bibnamefont
  {Wehrspohn}}, \bibinfo {author} {\bibfnamefont {Joachim~H}\ \bibnamefont
  {Wendorff}}, \bibinfo {author} {\bibfnamefont {Ramamoorthy}\ \bibnamefont
  {Ramesh}}, \ and\ \bibinfo {author} {\bibfnamefont {Marin}\ \bibnamefont
  {Alexe}},\ }\bibfield  {title} {\enquote {\bibinfo {title} {Nanoshell tubes
  of ferroelectric lead zirconate titanate and barium titanate},}\ }\href
  {\doibase https://doi.org/10.1063/1.1592013} {\bibfield  {journal} {\bibinfo
  {journal} {Appl. Phys. Lett.}\ }\textbf {\bibinfo {volume} {83}},\ \bibinfo
  {pages} {440--442} (\bibinfo {year} {2003})}\BibitemShut {NoStop}%
\bibitem [{\citenamefont {Morrison}\ \emph {et~al.}(2003)\citenamefont
  {Morrison}, \citenamefont {Ramsay},\ and\ \citenamefont
  {Scott}}]{morrison2003high}%
  \BibitemOpen
  \bibfield  {author} {\bibinfo {author} {\bibfnamefont {Finlay~D}\
  \bibnamefont {Morrison}}, \bibinfo {author} {\bibfnamefont {Laura}\
  \bibnamefont {Ramsay}}, \ and\ \bibinfo {author} {\bibfnamefont {James~F}\
  \bibnamefont {Scott}},\ }\bibfield  {title} {\enquote {\bibinfo {title} {High
  aspect ratio piezoelectric strontium--bismuth--tantalate nanotubes},}\ }\href
  {\doibase https://doi.org/10.1088/0953-8984/15/33/103} {\bibfield  {journal}
  {\bibinfo  {journal} {J. Phys.: Condens. Matter}\ }\textbf {\bibinfo {volume}
  {15}},\ \bibinfo {pages} {L527} (\bibinfo {year} {2003})}\BibitemShut
  {NoStop}%
\bibitem [{\citenamefont {Mao}\ \emph {et~al.}(2003)\citenamefont {Mao},
  \citenamefont {Banerjee},\ and\ \citenamefont {Wong}}]{mao2003hydrothermal}%
  \BibitemOpen
  \bibfield  {author} {\bibinfo {author} {\bibfnamefont {Yuanbing}\
  \bibnamefont {Mao}}, \bibinfo {author} {\bibfnamefont {Sarbajit}\
  \bibnamefont {Banerjee}}, \ and\ \bibinfo {author} {\bibfnamefont
  {Stanislaus~S}\ \bibnamefont {Wong}},\ }\bibfield  {title} {\enquote
  {\bibinfo {title} {Hydrothermal synthesis of perovskite nanotubes},}\ }\href
  {\doibase https://doi.org/10.1039/B210633G} {\bibfield  {journal} {\bibinfo
  {journal} {Chem. Commun.}\ ,\ \bibinfo {pages} {408--409}} (\bibinfo {year}
  {2003})}\BibitemShut {NoStop}%
\bibitem [{\citenamefont {Chu}\ \emph {et~al.}(2004)\citenamefont {Chu},
  \citenamefont {Szafraniak}, \citenamefont {Scholz}, \citenamefont {Harnagea},
  \citenamefont {Hesse}, \citenamefont {Alexe},\ and\ \citenamefont
  {G{\"o}sele}}]{chu2004impact}%
  \BibitemOpen
  \bibfield  {author} {\bibinfo {author} {\bibfnamefont {Ming-Wen}\
  \bibnamefont {Chu}}, \bibinfo {author} {\bibfnamefont {Izabela}\ \bibnamefont
  {Szafraniak}}, \bibinfo {author} {\bibfnamefont {Roland}\ \bibnamefont
  {Scholz}}, \bibinfo {author} {\bibfnamefont {Catalin}\ \bibnamefont
  {Harnagea}}, \bibinfo {author} {\bibfnamefont {Dietrich}\ \bibnamefont
  {Hesse}}, \bibinfo {author} {\bibfnamefont {Marin}\ \bibnamefont {Alexe}}, \
  and\ \bibinfo {author} {\bibfnamefont {Ulrich}\ \bibnamefont {G{\"o}sele}},\
  }\bibfield  {title} {\enquote {\bibinfo {title} {Impact of misfit
  dislocations on the polarization instability of epitaxial nanostructured
  ferroelectric perovskites},}\ }\href {\doibase
  https://doi.org/10.1038/nmat1057} {\bibfield  {journal} {\bibinfo  {journal}
  {Nat. Mater.}\ }\textbf {\bibinfo {volume} {3}},\ \bibinfo {pages} {87--90}
  (\bibinfo {year} {2004})}\BibitemShut {NoStop}%
\bibitem [{\citenamefont {Shin}\ \emph {et~al.}(2005)\citenamefont {Shin},
  \citenamefont {Choi}, \citenamefont {Yang}, \citenamefont {Park},
  \citenamefont {Kuk},\ and\ \citenamefont {Kang}}]{shin2005patterning}%
  \BibitemOpen
  \bibfield  {author} {\bibinfo {author} {\bibfnamefont {Hyung-Joon}\
  \bibnamefont {Shin}}, \bibinfo {author} {\bibfnamefont {Je~Hyuk}\
  \bibnamefont {Choi}}, \bibinfo {author} {\bibfnamefont {Hee~Jun}\
  \bibnamefont {Yang}}, \bibinfo {author} {\bibfnamefont {Young~Dae}\
  \bibnamefont {Park}}, \bibinfo {author} {\bibfnamefont {Young}\ \bibnamefont
  {Kuk}}, \ and\ \bibinfo {author} {\bibfnamefont {Chi-Jung}\ \bibnamefont
  {Kang}},\ }\bibfield  {title} {\enquote {\bibinfo {title} {Patterning of
  ferroelectric nanodot arrays using a silicon nitride shadow mask},}\ }\href
  {\doibase https://doi.org/10.1063/1.2048818} {\bibfield  {journal} {\bibinfo
  {journal} {Appl. Phys. Lett.}\ }\textbf {\bibinfo {volume} {87}},\ \bibinfo
  {pages} {113114} (\bibinfo {year} {2005})}\BibitemShut {NoStop}%
\bibitem [{\citenamefont {Fu}\ and\ \citenamefont
  {Bellaiche}(2003)}]{fu2003ferroelectricity}%
  \BibitemOpen
  \bibfield  {author} {\bibinfo {author} {\bibfnamefont {Huaxiang}\
  \bibnamefont {Fu}}\ and\ \bibinfo {author} {\bibfnamefont {L}~\bibnamefont
  {Bellaiche}},\ }\bibfield  {title} {\enquote {\bibinfo {title}
  {Ferroelectricity in barium titanate quantum dots and wires},}\ }\href
  {\doibase https://doi.org/10.1103/PhysRevLett.91.257601} {\bibfield
  {journal} {\bibinfo  {journal} {Phys. Rev. Lett.}\ }\textbf {\bibinfo
  {volume} {91}},\ \bibinfo {pages} {257601} (\bibinfo {year}
  {2003})}\BibitemShut {NoStop}%
\bibitem [{\citenamefont {Naumov}\ \emph {et~al.}(2004)\citenamefont {Naumov},
  \citenamefont {Bellaiche},\ and\ \citenamefont {Fu}}]{naumov2004unusual}%
  \BibitemOpen
  \bibfield  {author} {\bibinfo {author} {\bibfnamefont {Ivan~I}\ \bibnamefont
  {Naumov}}, \bibinfo {author} {\bibfnamefont {L}~\bibnamefont {Bellaiche}}, \
  and\ \bibinfo {author} {\bibfnamefont {Huaxiang}\ \bibnamefont {Fu}},\
  }\bibfield  {title} {\enquote {\bibinfo {title} {Unusual phase transitions in
  ferroelectric nanodisks and nanorods},}\ }\href {\doibase
  https://doi.org/10.1038/nature03107} {\bibfield  {journal} {\bibinfo
  {journal} {Nature}\ }\textbf {\bibinfo {volume} {432}},\ \bibinfo {pages}
  {737--740} (\bibinfo {year} {2004})}\BibitemShut {NoStop}%
\bibitem [{\citenamefont {Geneste}\ \emph {et~al.}(2006)\citenamefont
  {Geneste}, \citenamefont {Bousquet}, \citenamefont {Junquera},\ and\
  \citenamefont {Ghosez}}]{geneste2006finite}%
  \BibitemOpen
  \bibfield  {author} {\bibinfo {author} {\bibfnamefont {Gregory}\ \bibnamefont
  {Geneste}}, \bibinfo {author} {\bibfnamefont {Eric}\ \bibnamefont
  {Bousquet}}, \bibinfo {author} {\bibfnamefont {Javier}\ \bibnamefont
  {Junquera}}, \ and\ \bibinfo {author} {\bibfnamefont {Philippe}\ \bibnamefont
  {Ghosez}},\ }\bibfield  {title} {\enquote {\bibinfo {title} {{Finite-size
  effects in BaTiO$_3$ nanowires}},}\ }\href {\doibase
  https://doi.org/10.1063/1.2186104} {\bibfield  {journal} {\bibinfo  {journal}
  {Appl. Phys. Lett.}\ }\textbf {\bibinfo {volume} {88}},\ \bibinfo {pages}
  {112906} (\bibinfo {year} {2006})}\BibitemShut {NoStop}%
\bibitem [{\citenamefont {Morozovska}\ \emph {et~al.}(2006)\citenamefont
  {Morozovska}, \citenamefont {Eliseev},\ and\ \citenamefont
  {Glinchuk}}]{morozovska2006ferroelectricity}%
  \BibitemOpen
  \bibfield  {author} {\bibinfo {author} {\bibfnamefont {Anna~N}\ \bibnamefont
  {Morozovska}}, \bibinfo {author} {\bibfnamefont {Eugene~A}\ \bibnamefont
  {Eliseev}}, \ and\ \bibinfo {author} {\bibfnamefont {Maya~D}\ \bibnamefont
  {Glinchuk}},\ }\bibfield  {title} {\enquote {\bibinfo {title}
  {Ferroelectricity enhancement in confined nanorods: Direct variational
  method},}\ }\href {\doibase https://doi.org/10.1103/PhysRevB.73.214106}
  {\bibfield  {journal} {\bibinfo  {journal} {Phys. Rev. B}\ }\textbf {\bibinfo
  {volume} {73}},\ \bibinfo {pages} {214106} (\bibinfo {year}
  {2006})}\BibitemShut {NoStop}%
\bibitem [{\citenamefont {Hong}\ \emph {et~al.}(2010)\citenamefont {Hong},
  \citenamefont {Catalan}, \citenamefont {Fang}, \citenamefont {Artacho},\ and\
  \citenamefont {Scott}}]{hong2010topology}%
  \BibitemOpen
  \bibfield  {author} {\bibinfo {author} {\bibfnamefont {Jiawang}\ \bibnamefont
  {Hong}}, \bibinfo {author} {\bibfnamefont {G}~\bibnamefont {Catalan}},
  \bibinfo {author} {\bibfnamefont {DN}~\bibnamefont {Fang}}, \bibinfo {author}
  {\bibfnamefont {Emilio}\ \bibnamefont {Artacho}}, \ and\ \bibinfo {author}
  {\bibfnamefont {JF}~\bibnamefont {Scott}},\ }\bibfield  {title} {\enquote
  {\bibinfo {title} {Topology of the polarization field in ferroelectric
  nanowires from first principles},}\ }\href {\doibase
  https://doi.org/10.1103/PhysRevB.81.172101} {\bibfield  {journal} {\bibinfo
  {journal} {Phys. Rev. B}\ }\textbf {\bibinfo {volume} {81}},\ \bibinfo
  {pages} {172101} (\bibinfo {year} {2010})}\BibitemShut {NoStop}%
\bibitem [{\citenamefont {Junquera}\ \emph {et~al.}(2023)\citenamefont
  {Junquera}, \citenamefont {Nahas}, \citenamefont {Prokhorenko}, \citenamefont
  {Bellaiche}, \citenamefont {\'I\~niguez}, \citenamefont {Schlom},
  \citenamefont {Chen}, \citenamefont {Salahuddin}, \citenamefont {Muller},
  \citenamefont {Martin},\ and\ \citenamefont
  {Ramesh}}]{junquera2023topologicaly}%
  \BibitemOpen
  \bibfield  {author} {\bibinfo {author} {\bibfnamefont {Javier}\ \bibnamefont
  {Junquera}}, \bibinfo {author} {\bibfnamefont {Yousra}\ \bibnamefont
  {Nahas}}, \bibinfo {author} {\bibfnamefont {Sergei}\ \bibnamefont
  {Prokhorenko}}, \bibinfo {author} {\bibfnamefont {Laurent}\ \bibnamefont
  {Bellaiche}}, \bibinfo {author} {\bibfnamefont {Jorge}\ \bibnamefont
  {\'I\~niguez}}, \bibinfo {author} {\bibfnamefont {Darrell~G.}\ \bibnamefont
  {Schlom}}, \bibinfo {author} {\bibfnamefont {Long-Qing}\ \bibnamefont
  {Chen}}, \bibinfo {author} {\bibfnamefont {Sayeef}\ \bibnamefont
  {Salahuddin}}, \bibinfo {author} {\bibfnamefont {David~A.}\ \bibnamefont
  {Muller}}, \bibinfo {author} {\bibfnamefont {Lane~W.}\ \bibnamefont
  {Martin}}, \ and\ \bibinfo {author} {\bibfnamefont {R.}~\bibnamefont
  {Ramesh}},\ }\bibfield  {title} {\enquote {\bibinfo {title} {Topological
  phases in polar oxide nanostructures},}\ }\href {\doibase
  10.1103/RevModPhys.95.025001} {\bibfield  {journal} {\bibinfo  {journal}
  {Rev. Mod. Phys.}\ }\textbf {\bibinfo {volume} {95}},\ \bibinfo {pages}
  {025001} (\bibinfo {year} {2023})}\BibitemShut {NoStop}%
\bibitem [{\citenamefont {Nahas}\ \emph {et~al.}(2015)\citenamefont {Nahas},
  \citenamefont {Prokhorenko}, \citenamefont {Louis}, \citenamefont {Gui},
  \citenamefont {Kornev},\ and\ \citenamefont
  {Bellaiche}}]{nahas2015discovery}%
  \BibitemOpen
  \bibfield  {author} {\bibinfo {author} {\bibfnamefont {Y}~\bibnamefont
  {Nahas}}, \bibinfo {author} {\bibfnamefont {S}~\bibnamefont {Prokhorenko}},
  \bibinfo {author} {\bibfnamefont {L}~\bibnamefont {Louis}}, \bibinfo {author}
  {\bibfnamefont {Z}~\bibnamefont {Gui}}, \bibinfo {author} {\bibfnamefont
  {Igor}\ \bibnamefont {Kornev}}, \ and\ \bibinfo {author} {\bibfnamefont
  {Laurent}\ \bibnamefont {Bellaiche}},\ }\bibfield  {title} {\enquote
  {\bibinfo {title} {Discovery of stable skyrmionic state in ferroelectric
  nanocomposites},}\ }\href {\doibase https://doi.org/10.1038/ncomms9542}
  {\bibfield  {journal} {\bibinfo  {journal} {Nat. Commun.}\ }\textbf {\bibinfo
  {volume} {6}},\ \bibinfo {pages} {1--6} (\bibinfo {year} {2015})}\BibitemShut
  {NoStop}%
\bibitem [{\citenamefont {Pereira~Gon{\c{c}}alves}\ \emph
  {et~al.}(2019)\citenamefont {Pereira~Gon{\c{c}}alves}, \citenamefont
  {Escorihuela-Sayalero}, \citenamefont {Garca-Fern{\'a}ndez}, \citenamefont
  {Junquera},\ and\ \citenamefont {{\'I}{\~n}iguez}}]{pereira2019theoretical}%
  \BibitemOpen
  \bibfield  {author} {\bibinfo {author} {\bibfnamefont {Mauro~Ant{\'o}nio}\
  \bibnamefont {Pereira~Gon{\c{c}}alves}}, \bibinfo {author} {\bibfnamefont
  {Carlos}\ \bibnamefont {Escorihuela-Sayalero}}, \bibinfo {author}
  {\bibfnamefont {Pablo}\ \bibnamefont {Garca-Fern{\'a}ndez}}, \bibinfo
  {author} {\bibfnamefont {Javier}\ \bibnamefont {Junquera}}, \ and\ \bibinfo
  {author} {\bibfnamefont {Jorge}\ \bibnamefont {{\'I}{\~n}iguez}},\ }\bibfield
   {title} {\enquote {\bibinfo {title} {Theoretical guidelines to create and
  tune electric skyrmion bubbles},}\ }\href {\doibase
  https://doi.org/10.1126/sciadv.aau7023} {\bibfield  {journal} {\bibinfo
  {journal} {Sci. Adv.}\ }\textbf {\bibinfo {volume} {5}},\ \bibinfo {pages}
  {eaau7023} (\bibinfo {year} {2019})}\BibitemShut {NoStop}%
\bibitem [{\citenamefont {Das}\ \emph {et~al.}(2019)\citenamefont {Das},
  \citenamefont {Tang}, \citenamefont {Hong}, \citenamefont {Gon{\c{c}}alves},
  \citenamefont {McCarter}, \citenamefont {Klewe}, \citenamefont {Nguyen},
  \citenamefont {G{\'o}mez-Ortiz}, \citenamefont {Shafer}, \citenamefont
  {Arenholz} \emph {et~al.}}]{das2019observation}%
  \BibitemOpen
  \bibfield  {author} {\bibinfo {author} {\bibfnamefont {S}~\bibnamefont
  {Das}}, \bibinfo {author} {\bibfnamefont {YL}~\bibnamefont {Tang}}, \bibinfo
  {author} {\bibfnamefont {Z}~\bibnamefont {Hong}}, \bibinfo {author}
  {\bibfnamefont {MAP}\ \bibnamefont {Gon{\c{c}}alves}}, \bibinfo {author}
  {\bibfnamefont {MR}~\bibnamefont {McCarter}}, \bibinfo {author}
  {\bibfnamefont {C}~\bibnamefont {Klewe}}, \bibinfo {author} {\bibfnamefont
  {KX}~\bibnamefont {Nguyen}}, \bibinfo {author} {\bibfnamefont
  {F}~\bibnamefont {G{\'o}mez-Ortiz}}, \bibinfo {author} {\bibfnamefont
  {P}~\bibnamefont {Shafer}}, \bibinfo {author} {\bibfnamefont {E}~\bibnamefont
  {Arenholz}},  \emph {et~al.},\ }\bibfield  {title} {\enquote {\bibinfo
  {title} {Observation of room-temperature polar skyrmions},}\ }\href {\doibase
  https://doi.org/10.1038/s41586-019-1092-8} {\bibfield  {journal} {\bibinfo
  {journal} {Nature}\ }\textbf {\bibinfo {volume} {568}},\ \bibinfo {pages}
  {368--372} (\bibinfo {year} {2019})}\BibitemShut {NoStop}%
\bibitem [{\citenamefont {Han}\ \emph {et~al.}(2022)\citenamefont {Han},
  \citenamefont {Addiego}, \citenamefont {Prokhorenko}, \citenamefont {Wang},
  \citenamefont {Fu}, \citenamefont {Nahas}, \citenamefont {Yan}, \citenamefont
  {Cai}, \citenamefont {Wei}, \citenamefont {Fang} \emph
  {et~al.}}]{han2022high}%
  \BibitemOpen
  \bibfield  {author} {\bibinfo {author} {\bibfnamefont {Lu}~\bibnamefont
  {Han}}, \bibinfo {author} {\bibfnamefont {Christopher}\ \bibnamefont
  {Addiego}}, \bibinfo {author} {\bibfnamefont {Sergei}\ \bibnamefont
  {Prokhorenko}}, \bibinfo {author} {\bibfnamefont {Meiyu}\ \bibnamefont
  {Wang}}, \bibinfo {author} {\bibfnamefont {Hanyu}\ \bibnamefont {Fu}},
  \bibinfo {author} {\bibfnamefont {Yousra}\ \bibnamefont {Nahas}}, \bibinfo
  {author} {\bibfnamefont {Xingxu}\ \bibnamefont {Yan}}, \bibinfo {author}
  {\bibfnamefont {Songhua}\ \bibnamefont {Cai}}, \bibinfo {author}
  {\bibfnamefont {Tianqi}\ \bibnamefont {Wei}}, \bibinfo {author}
  {\bibfnamefont {Yanhan}\ \bibnamefont {Fang}},  \emph {et~al.},\ }\bibfield
  {title} {\enquote {\bibinfo {title} {High-density switchable skyrmion-like
  polar nanodomains integrated on silicon},}\ }\href {\doibase
  https://doi.org/10.1038/s41586-021-04338-w} {\bibfield  {journal} {\bibinfo
  {journal} {Nature}\ }\textbf {\bibinfo {volume} {603}},\ \bibinfo {pages}
  {63--67} (\bibinfo {year} {2022})}\BibitemShut {NoStop}%
\bibitem [{\citenamefont {Shao}\ \emph {et~al.}(2023)\citenamefont {Shao},
  \citenamefont {Das}, \citenamefont {Hong}, \citenamefont {Xu}, \citenamefont
  {Chandrika}, \citenamefont {G{\'o}mez-Ortiz}, \citenamefont
  {Garc{\'\i}a-Fern{\'a}ndez}, \citenamefont {Chen}, \citenamefont {Hwang},
  \citenamefont {Junquera} \emph {et~al.}}]{shao2023emergent}%
  \BibitemOpen
  \bibfield  {author} {\bibinfo {author} {\bibfnamefont {Yu-Tsun}\ \bibnamefont
  {Shao}}, \bibinfo {author} {\bibfnamefont {Sujit}\ \bibnamefont {Das}},
  \bibinfo {author} {\bibfnamefont {Zijian}\ \bibnamefont {Hong}}, \bibinfo
  {author} {\bibfnamefont {Ruijuan}\ \bibnamefont {Xu}}, \bibinfo {author}
  {\bibfnamefont {Swathi}\ \bibnamefont {Chandrika}}, \bibinfo {author}
  {\bibfnamefont {Fernando}\ \bibnamefont {G{\'o}mez-Ortiz}}, \bibinfo {author}
  {\bibfnamefont {Pablo}\ \bibnamefont {Garc{\'\i}a-Fern{\'a}ndez}}, \bibinfo
  {author} {\bibfnamefont {Long-Qing}\ \bibnamefont {Chen}}, \bibinfo {author}
  {\bibfnamefont {Harold~Y}\ \bibnamefont {Hwang}}, \bibinfo {author}
  {\bibfnamefont {Javier}\ \bibnamefont {Junquera}},  \emph {et~al.},\
  }\bibfield  {title} {\enquote {\bibinfo {title} {Emergent chirality in a
  polar meron to skyrmion phase transition},}\ }\href {\doibase
  https://doi.org/10.1038/s41467-023-36950-x} {\bibfield  {journal} {\bibinfo
  {journal} {Nat. Commun.}\ }\textbf {\bibinfo {volume} {14}},\ \bibinfo
  {pages} {1355} (\bibinfo {year} {2023})}\BibitemShut {NoStop}%
\bibitem [{\citenamefont {Wang}\ \emph {et~al.}(2020)\citenamefont {Wang},
  \citenamefont {Feng}, \citenamefont {Zhu}, \citenamefont {Tang},
  \citenamefont {Yang}, \citenamefont {Zou}, \citenamefont {Geng},
  \citenamefont {Han}, \citenamefont {Guo}, \citenamefont {Wu} \emph
  {et~al.}}]{wang2020polar}%
  \BibitemOpen
  \bibfield  {author} {\bibinfo {author} {\bibfnamefont {YJ}~\bibnamefont
  {Wang}}, \bibinfo {author} {\bibfnamefont {YP}~\bibnamefont {Feng}}, \bibinfo
  {author} {\bibfnamefont {YL}~\bibnamefont {Zhu}}, \bibinfo {author}
  {\bibfnamefont {YL}~\bibnamefont {Tang}}, \bibinfo {author} {\bibfnamefont
  {LX}~\bibnamefont {Yang}}, \bibinfo {author} {\bibfnamefont {MJ}~\bibnamefont
  {Zou}}, \bibinfo {author} {\bibfnamefont {WR}~\bibnamefont {Geng}}, \bibinfo
  {author} {\bibfnamefont {MJ}~\bibnamefont {Han}}, \bibinfo {author}
  {\bibfnamefont {XW}~\bibnamefont {Guo}}, \bibinfo {author} {\bibfnamefont
  {B}~\bibnamefont {Wu}},  \emph {et~al.},\ }\bibfield  {title} {\enquote
  {\bibinfo {title} {Polar meron lattice in strained oxide ferroelectrics},}\
  }\href {\doibase https://doi.org/10.1038/s41563-020-0694-8} {\bibfield
  {journal} {\bibinfo  {journal} {Nat. Mater.}\ }\textbf {\bibinfo {volume}
  {19}},\ \bibinfo {pages} {881--886} (\bibinfo {year} {2020})}\BibitemShut
  {NoStop}%
\bibitem [{\citenamefont {Li}\ and\ \citenamefont {Wu}(2017)}]{li2017binary}%
  \BibitemOpen
  \bibfield  {author} {\bibinfo {author} {\bibfnamefont {Lei}\ \bibnamefont
  {Li}}\ and\ \bibinfo {author} {\bibfnamefont {Menghao}\ \bibnamefont {Wu}},\
  }\bibfield  {title} {\enquote {\bibinfo {title} {Binary compound bilayer and
  multilayer with vertical polarizations: Two-dimensional ferroelectrics,
  multiferroics, and nanogenerators},}\ }\href
  {https://pubs.acs.org/doi/abs/10.1021/acsnano.7b02756} {\bibfield  {journal}
  {\bibinfo  {journal} {ACS Nano}\ }\textbf {\bibinfo {volume} {11}},\ \bibinfo
  {pages} {6382--6388} (\bibinfo {year} {2017})}\BibitemShut {NoStop}%
\bibitem [{\citenamefont {Bennett}\ and\ \citenamefont
  {Remez}(2022)}]{bennett2022electrically}%
  \BibitemOpen
  \bibfield  {author} {\bibinfo {author} {\bibfnamefont {Daniel}\ \bibnamefont
  {Bennett}}\ and\ \bibinfo {author} {\bibfnamefont {Benjamin}\ \bibnamefont
  {Remez}},\ }\bibfield  {title} {\enquote {\bibinfo {title} {On electrically
  tunable stacking domains and ferroelectricity in moir{\'e} superlattices},}\
  }\href {\doibase https://doi.org/10.1038/s41699-021-00281-6} {\bibfield
  {journal} {\bibinfo  {journal} {npj 2D Mater. Appl.}\ }\textbf {\bibinfo
  {volume} {6}},\ \bibinfo {pages} {1--6} (\bibinfo {year} {2022})}\BibitemShut
  {NoStop}%
\bibitem [{\citenamefont {Bennett}(2020)}]{bennett2022theory}%
  \BibitemOpen
  \bibfield  {author} {\bibinfo {author} {\bibfnamefont {Daniel}\ \bibnamefont
  {Bennett}},\ }\bibfield  {title} {\enquote {\bibinfo {title} {Theory of polar
  domains in moiré heterostructures},}\ }\href {\doibase
  https://doi.org/10.1103/PhysRevB.105.235445} {\bibfield  {journal} {\bibinfo
  {journal} {Phys. Rev. B}\ }\textbf {\bibinfo {volume} {105}},\ \bibinfo
  {pages} {235445} (\bibinfo {year} {2020})}\BibitemShut {NoStop}%
\bibitem [{\citenamefont {Zheng}\ \emph {et~al.}(2020)\citenamefont {Zheng},
  \citenamefont {Ma}, \citenamefont {Bi}, \citenamefont {de~la Barrera},
  \citenamefont {Liu}, \citenamefont {Mao}, \citenamefont {Zhang},
  \citenamefont {Kiper}, \citenamefont {Watanabe}, \citenamefont {Taniguchi}
  \emph {et~al.}}]{zheng2020unconventional}%
  \BibitemOpen
  \bibfield  {author} {\bibinfo {author} {\bibfnamefont {Zhiren}\ \bibnamefont
  {Zheng}}, \bibinfo {author} {\bibfnamefont {Qiong}\ \bibnamefont {Ma}},
  \bibinfo {author} {\bibfnamefont {Zhen}\ \bibnamefont {Bi}}, \bibinfo
  {author} {\bibfnamefont {Sergio}\ \bibnamefont {de~la Barrera}}, \bibinfo
  {author} {\bibfnamefont {Ming-Hao}\ \bibnamefont {Liu}}, \bibinfo {author}
  {\bibfnamefont {Nannan}\ \bibnamefont {Mao}}, \bibinfo {author}
  {\bibfnamefont {Yang}\ \bibnamefont {Zhang}}, \bibinfo {author}
  {\bibfnamefont {Natasha}\ \bibnamefont {Kiper}}, \bibinfo {author}
  {\bibfnamefont {Kenji}\ \bibnamefont {Watanabe}}, \bibinfo {author}
  {\bibfnamefont {Takashi}\ \bibnamefont {Taniguchi}},  \emph {et~al.},\
  }\bibfield  {title} {\enquote {\bibinfo {title} {Unconventional
  ferroelectricity in moir{\'e} heterostructures},}\ }\href {\doibase
  https://doi.org/10.1038/s41586-020-2970-9} {\bibfield  {journal} {\bibinfo
  {journal} {Nature}\ }\textbf {\bibinfo {volume} {588}},\ \bibinfo {pages}
  {71} (\bibinfo {year} {2020})}\BibitemShut {NoStop}%
\bibitem [{\citenamefont {Yasuda}\ \emph {et~al.}(2021)\citenamefont {Yasuda},
  \citenamefont {Wang}, \citenamefont {Watanabe}, \citenamefont {Taniguchi},\
  and\ \citenamefont {Jarillo-Herrero}}]{yasuda2021stacking}%
  \BibitemOpen
  \bibfield  {author} {\bibinfo {author} {\bibfnamefont {Kenji}\ \bibnamefont
  {Yasuda}}, \bibinfo {author} {\bibfnamefont {Xirui}\ \bibnamefont {Wang}},
  \bibinfo {author} {\bibfnamefont {Kenji}\ \bibnamefont {Watanabe}}, \bibinfo
  {author} {\bibfnamefont {Takashi}\ \bibnamefont {Taniguchi}}, \ and\ \bibinfo
  {author} {\bibfnamefont {Pablo}\ \bibnamefont {Jarillo-Herrero}},\ }\bibfield
   {title} {\enquote {\bibinfo {title} {Stacking-engineered ferroelectricity in
  bilayer boron nitride},}\ }\href {https://doi.org/10.1126/science.abd3230}
  {\bibfield  {journal} {\bibinfo  {journal} {Science}\ }\textbf {\bibinfo
  {volume} {372}},\ \bibinfo {pages} {1458} (\bibinfo {year}
  {2021})}\BibitemShut {NoStop}%
\bibitem [{\citenamefont {Stern}\ \emph {et~al.}(2021)\citenamefont {Stern},
  \citenamefont {Waschitz}, \citenamefont {Cao}, \citenamefont {Nevo},
  \citenamefont {Watanabe}, \citenamefont {Taniguchi}, \citenamefont {Sela},
  \citenamefont {Urbakh}, \citenamefont {Hod},\ and\ \citenamefont
  {Shalom}}]{stern2020interfacial}%
  \BibitemOpen
  \bibfield  {author} {\bibinfo {author} {\bibfnamefont {M.~Vizner}\
  \bibnamefont {Stern}}, \bibinfo {author} {\bibfnamefont {Y.}~\bibnamefont
  {Waschitz}}, \bibinfo {author} {\bibfnamefont {W.}~\bibnamefont {Cao}},
  \bibinfo {author} {\bibfnamefont {I.}~\bibnamefont {Nevo}}, \bibinfo {author}
  {\bibfnamefont {K.}~\bibnamefont {Watanabe}}, \bibinfo {author}
  {\bibfnamefont {T.}~\bibnamefont {Taniguchi}}, \bibinfo {author}
  {\bibfnamefont {E.}~\bibnamefont {Sela}}, \bibinfo {author} {\bibfnamefont
  {M.}~\bibnamefont {Urbakh}}, \bibinfo {author} {\bibfnamefont
  {O.}~\bibnamefont {Hod}}, \ and\ \bibinfo {author} {\bibfnamefont {M.~Ben}\
  \bibnamefont {Shalom}},\ }\bibfield  {title} {\enquote {\bibinfo {title}
  {Interfacial ferroelectricity by van der {W}aals sliding},}\ }\href {\doibase
  10.1126/science.abe8177} {\bibfield  {journal} {\bibinfo  {journal}
  {Science}\ }\textbf {\bibinfo {volume} {372}},\ \bibinfo {pages} {1462}
  (\bibinfo {year} {2021})}\BibitemShut {NoStop}%
\bibitem [{\citenamefont {Bennett}\ \emph {et~al.}(2023)\citenamefont
  {Bennett}, \citenamefont {Chaudhary}, \citenamefont {Slager}, \citenamefont
  {Bousquet},\ and\ \citenamefont {Ghosez}}]{bennett2023polar}%
  \BibitemOpen
  \bibfield  {author} {\bibinfo {author} {\bibfnamefont {Daniel}\ \bibnamefont
  {Bennett}}, \bibinfo {author} {\bibfnamefont {Gaurav}\ \bibnamefont
  {Chaudhary}}, \bibinfo {author} {\bibfnamefont {Robert-Jan}\ \bibnamefont
  {Slager}}, \bibinfo {author} {\bibfnamefont {Eric}\ \bibnamefont {Bousquet}},
  \ and\ \bibinfo {author} {\bibfnamefont {Philippe}\ \bibnamefont {Ghosez}},\
  }\bibfield  {title} {\enquote {\bibinfo {title} {Polar meron-antimeron
  networks in strained and twisted bilayers},}\ }\href {\doibase
  https://doi.org/10.1038/s41467-023-37337-8} {\bibfield  {journal} {\bibinfo
  {journal} {Nat. Commun.}\ }\textbf {\bibinfo {volume} {14}},\ \bibinfo
  {pages} {1629} (\bibinfo {year} {2023})}\BibitemShut {NoStop}%
\bibitem [{\citenamefont {Shen}\ \emph {et~al.}(2022)\citenamefont {Shen},
  \citenamefont {Dong}, \citenamefont {Qi}, \citenamefont {Zhang},
  \citenamefont {Zhu}, \citenamefont {Wu},\ and\ \citenamefont
  {Li}}]{shen2022observation}%
  \BibitemOpen
  \bibfield  {author} {\bibinfo {author} {\bibfnamefont {Jiaying}\ \bibnamefont
  {Shen}}, \bibinfo {author} {\bibfnamefont {Zhengang}\ \bibnamefont {Dong}},
  \bibinfo {author} {\bibfnamefont {MingQun}\ \bibnamefont {Qi}}, \bibinfo
  {author} {\bibfnamefont {Yang}\ \bibnamefont {Zhang}}, \bibinfo {author}
  {\bibfnamefont {Chao}\ \bibnamefont {Zhu}}, \bibinfo {author} {\bibfnamefont
  {Zhenping}\ \bibnamefont {Wu}}, \ and\ \bibinfo {author} {\bibfnamefont
  {Danfeng}\ \bibnamefont {Li}},\ }\bibfield  {title} {\enquote {\bibinfo
  {title} {Observation of moir{\'e} patterns in twisted stacks of bilayer
  perovskite oxide nanomembranes with various lattice symmetries},}\ }\href
  {\doibase https://doi.org/10.1021/acsami.2c14746} {\bibfield  {journal}
  {\bibinfo  {journal} {ACS Appl. Mater. Interfaces}\ }\textbf {\bibinfo
  {volume} {14}},\ \bibinfo {pages} {50386--50392} (\bibinfo {year}
  {2022})}\BibitemShut {NoStop}%
\bibitem [{\citenamefont {Sanchez-Santolino}\ \emph {et~al.}(2023)\citenamefont
  {Sanchez-Santolino}, \citenamefont {Rouco}, \citenamefont {Puebla},
  \citenamefont {Aramberri}, \citenamefont {Zamora}, \citenamefont {Cuellar},
  \citenamefont {Munuera}, \citenamefont {Mompean}, \citenamefont
  {Garcia-Hernandez}, \citenamefont {Castellanos-Gomez} \emph
  {et~al.}}]{sanchez20232d}%
  \BibitemOpen
  \bibfield  {author} {\bibinfo {author} {\bibfnamefont {Gabriel}\ \bibnamefont
  {Sanchez-Santolino}}, \bibinfo {author} {\bibfnamefont {Victor}\ \bibnamefont
  {Rouco}}, \bibinfo {author} {\bibfnamefont {Sergio}\ \bibnamefont {Puebla}},
  \bibinfo {author} {\bibfnamefont {Hugo}\ \bibnamefont {Aramberri}}, \bibinfo
  {author} {\bibfnamefont {Victor}\ \bibnamefont {Zamora}}, \bibinfo {author}
  {\bibfnamefont {Fabian~A}\ \bibnamefont {Cuellar}}, \bibinfo {author}
  {\bibfnamefont {Carmen}\ \bibnamefont {Munuera}}, \bibinfo {author}
  {\bibfnamefont {Federico}\ \bibnamefont {Mompean}}, \bibinfo {author}
  {\bibfnamefont {Mar}\ \bibnamefont {Garcia-Hernandez}}, \bibinfo {author}
  {\bibfnamefont {Aandres}\ \bibnamefont {Castellanos-Gomez}},  \emph
  {et~al.},\ }\bibfield  {title} {\enquote {\bibinfo {title} {{A 2D
  ferroelectric vortex lattice in twisted BaTiO$_3$ freestanding layers}},}\
  }\href {https://doi.org/10.48550/arXiv.2301.04438} {\bibfield  {journal}
  {\bibinfo  {journal} {arXiv:2301.04438}\ } (\bibinfo {year}
  {2023})}\BibitemShut {NoStop}%
\bibitem [{\citenamefont {Artyukhov}\ \emph {et~al.}(2020)\citenamefont
  {Artyukhov}, \citenamefont {Gupta}, \citenamefont {Kutana},\ and\
  \citenamefont {Yakobson}}]{artyukhov2020flexoelectricity}%
  \BibitemOpen
  \bibfield  {author} {\bibinfo {author} {\bibfnamefont {Vasilii~I}\
  \bibnamefont {Artyukhov}}, \bibinfo {author} {\bibfnamefont {Sunny}\
  \bibnamefont {Gupta}}, \bibinfo {author} {\bibfnamefont {Alex}\ \bibnamefont
  {Kutana}}, \ and\ \bibinfo {author} {\bibfnamefont {Boris~I}\ \bibnamefont
  {Yakobson}},\ }\bibfield  {title} {\enquote {\bibinfo {title}
  {Flexoelectricity and charge separation in carbon nanotubes},}\ }\href
  {\doibase https://doi.org/10.1021/acs.nanolett.9b05345} {\bibfield  {journal}
  {\bibinfo  {journal} {Nano Lett.}\ }\textbf {\bibinfo {volume} {20}},\
  \bibinfo {pages} {3240--3246} (\bibinfo {year} {2020})}\BibitemShut {NoStop}%
\bibitem [{\citenamefont {Springolo}\ \emph {et~al.}(2021)\citenamefont
  {Springolo}, \citenamefont {Royo},\ and\ \citenamefont
  {Stengel}}]{springolo2021direct}%
  \BibitemOpen
  \bibfield  {author} {\bibinfo {author} {\bibfnamefont {Matteo}\ \bibnamefont
  {Springolo}}, \bibinfo {author} {\bibfnamefont {Miquel}\ \bibnamefont
  {Royo}}, \ and\ \bibinfo {author} {\bibfnamefont {Massimiliano}\ \bibnamefont
  {Stengel}},\ }\bibfield  {title} {\enquote {\bibinfo {title} {Direct and
  converse flexoelectricity in two-dimensional materials},}\ }\href {\doibase
  https://doi.org/10.1103/PhysRevLett.127.216801} {\bibfield  {journal}
  {\bibinfo  {journal} {Phys. Rev. Lett.}\ }\textbf {\bibinfo {volume} {127}},\
  \bibinfo {pages} {216801} (\bibinfo {year} {2021})}\BibitemShut {NoStop}%
\bibitem [{\citenamefont {Bennett}(2021)}]{bennett2021flexoelectric}%
  \BibitemOpen
  \bibfield  {author} {\bibinfo {author} {\bibfnamefont {Daniel}\ \bibnamefont
  {Bennett}},\ }\bibfield  {title} {\enquote {\bibinfo {title}
  {Flexoelectric-like radial polarization of single-walled nanotubes from
  first-principles},}\ }\href {\doibase
  https://doi.org/10.1088/2516-1075/aba095} {\bibfield  {journal} {\bibinfo
  {journal} {Electron. Struct.}\ }\textbf {\bibinfo {volume} {3}},\ \bibinfo
  {pages} {015001} (\bibinfo {year} {2021})}\BibitemShut {NoStop}%
\bibitem [{\citenamefont {Vanderbilt}\ and\ \citenamefont
  {King-Smith}(1993)}]{vanderbilt1993electric}%
  \BibitemOpen
  \bibfield  {author} {\bibinfo {author} {\bibfnamefont {David}\ \bibnamefont
  {Vanderbilt}}\ and\ \bibinfo {author} {\bibfnamefont {RD}~\bibnamefont
  {King-Smith}},\ }\bibfield  {title} {\enquote {\bibinfo {title} {Electric
  polarization as a bulk quantity and its relation to surface charge},}\ }\href
  {\doibase https://doi.org/10.1103/PhysRevB.48.4442} {\bibfield  {journal}
  {\bibinfo  {journal} {Phys. Rev. B}\ }\textbf {\bibinfo {volume} {48}},\
  \bibinfo {pages} {4442} (\bibinfo {year} {1993})}\BibitemShut {NoStop}%
\bibitem [{\citenamefont {King-Smith}\ and\ \citenamefont
  {Vanderbilt}(1993)}]{king1993theory}%
  \BibitemOpen
  \bibfield  {author} {\bibinfo {author} {\bibfnamefont {RD}~\bibnamefont
  {King-Smith}}\ and\ \bibinfo {author} {\bibfnamefont {David}\ \bibnamefont
  {Vanderbilt}},\ }\bibfield  {title} {\enquote {\bibinfo {title} {Theory of
  polarization of crystalline solids},}\ }\href {\doibase
  https://doi.org/10.1103/PhysRevB.47.1651} {\bibfield  {journal} {\bibinfo
  {journal} {Phys. Rev. B}\ }\textbf {\bibinfo {volume} {47}},\ \bibinfo
  {pages} {1651} (\bibinfo {year} {1993})}\BibitemShut {NoStop}%
\bibitem [{\citenamefont {Resta}(1994)}]{resta1994macroscopic}%
  \BibitemOpen
  \bibfield  {author} {\bibinfo {author} {\bibfnamefont {Raffaele}\
  \bibnamefont {Resta}},\ }\bibfield  {title} {\enquote {\bibinfo {title}
  {Macroscopic polarization in crystalline dielectrics: the geometric phase
  approach},}\ }\href {\doibase 10.1103/RevModPhys.66.899} {\bibfield
  {journal} {\bibinfo  {journal} {Rev. Mod. Phys.}\ }\textbf {\bibinfo {volume}
  {66}},\ \bibinfo {pages} {899--915} (\bibinfo {year} {1994})}\BibitemShut
  {NoStop}%
\bibitem [{\citenamefont {Nakagawa}\ \emph {et~al.}(2020)\citenamefont
  {Nakagawa}, \citenamefont {Slager}, \citenamefont {Higashikawa},\ and\
  \citenamefont {Oka}}]{Wannier_Nakagawa}%
  \BibitemOpen
  \bibfield  {author} {\bibinfo {author} {\bibfnamefont {Masaya}\ \bibnamefont
  {Nakagawa}}, \bibinfo {author} {\bibfnamefont {Robert-Jan}\ \bibnamefont
  {Slager}}, \bibinfo {author} {\bibfnamefont {Sho}\ \bibnamefont
  {Higashikawa}}, \ and\ \bibinfo {author} {\bibfnamefont {Takashi}\
  \bibnamefont {Oka}},\ }\bibfield  {title} {\enquote {\bibinfo {title}
  {Wannier representation of floquet topological states},}\ }\href {\doibase
  10.1103/PhysRevB.101.075108} {\bibfield  {journal} {\bibinfo  {journal}
  {Phys. Rev. B}\ }\textbf {\bibinfo {volume} {101}},\ \bibinfo {pages}
  {075108} (\bibinfo {year} {2020})}\BibitemShut {NoStop}%
\bibitem [{\citenamefont {Vanderbilt}(2018)}]{vanderbilt2018berry}%
  \BibitemOpen
  \bibfield  {author} {\bibinfo {author} {\bibfnamefont {David}\ \bibnamefont
  {Vanderbilt}},\ }\href {\doibase https://doi.org/10.1017/9781316662205}
  {\emph {\bibinfo {title} {Berry phases in electronic structure theory:
  electric polarization, orbital magnetization and topological insulators}}}\
  (\bibinfo  {publisher} {Cambridge University Press},\ \bibinfo {year}
  {2018})\BibitemShut {NoStop}%
\bibitem [{\citenamefont {Van~Mechelen}\ \emph {et~al.}(2022)\citenamefont
  {Van~Mechelen}, \citenamefont {Bharadwaj}, \citenamefont {Jacob},\ and\
  \citenamefont {Slager}}]{vanMechelen_optical}%
  \BibitemOpen
  \bibfield  {author} {\bibinfo {author} {\bibfnamefont {Todd}\ \bibnamefont
  {Van~Mechelen}}, \bibinfo {author} {\bibfnamefont {Sathwik}\ \bibnamefont
  {Bharadwaj}}, \bibinfo {author} {\bibfnamefont {Zubin}\ \bibnamefont
  {Jacob}}, \ and\ \bibinfo {author} {\bibfnamefont {Robert-Jan}\ \bibnamefont
  {Slager}},\ }\bibfield  {title} {\enquote {\bibinfo {title} {Optical
  ${N}$-insulators: Topological obstructions to optical {W}annier functions in
  the atomistic susceptibility tensor},}\ }\href {\doibase
  10.1103/PhysRevResearch.4.023011} {\bibfield  {journal} {\bibinfo  {journal}
  {Phys. Rev. Res.}\ }\textbf {\bibinfo {volume} {4}},\ \bibinfo {pages}
  {023011} (\bibinfo {year} {2022})}\BibitemShut {NoStop}%
\bibitem [{\citenamefont {Wu}\ \emph {et~al.}(2006)\citenamefont {Wu},
  \citenamefont {Di{\'e}guez}, \citenamefont {Rabe},\ and\ \citenamefont
  {Vanderbilt}}]{wu2006wannier}%
  \BibitemOpen
  \bibfield  {author} {\bibinfo {author} {\bibfnamefont {Xifan}\ \bibnamefont
  {Wu}}, \bibinfo {author} {\bibfnamefont {Oswaldo}\ \bibnamefont
  {Di{\'e}guez}}, \bibinfo {author} {\bibfnamefont {Karin~M}\ \bibnamefont
  {Rabe}}, \ and\ \bibinfo {author} {\bibfnamefont {David}\ \bibnamefont
  {Vanderbilt}},\ }\bibfield  {title} {\enquote {\bibinfo {title}
  {Wannier-based definition of layer polarizations in perovskite
  superlattices},}\ }\href {\doibase
  https://doi.org/10.1103/PhysRevLett.97.107602} {\bibfield  {journal}
  {\bibinfo  {journal} {Phys. Rev. Lett.}\ }\textbf {\bibinfo {volume} {97}},\
  \bibinfo {pages} {107602} (\bibinfo {year} {2006})}\BibitemShut {NoStop}%
\bibitem [{\citenamefont {Meyer}\ and\ \citenamefont
  {Vanderbilt}(2002)}]{meyer2002ab}%
  \BibitemOpen
  \bibfield  {author} {\bibinfo {author} {\bibfnamefont {B}~\bibnamefont
  {Meyer}}\ and\ \bibinfo {author} {\bibfnamefont {David}\ \bibnamefont
  {Vanderbilt}},\ }\bibfield  {title} {\enquote {\bibinfo {title} {{Ab initio
  study of ferroelectric domain walls in PbTiO$_3$}},}\ }\href {\doibase
  https://doi.org/10.1103/PhysRevB.65.104111} {\bibfield  {journal} {\bibinfo
  {journal} {Phys. Rev. B}\ }\textbf {\bibinfo {volume} {65}},\ \bibinfo
  {pages} {104111} (\bibinfo {year} {2002})}\BibitemShut {NoStop}%
\bibitem [{\citenamefont {Stengel}\ \emph {et~al.}(2011)\citenamefont
  {Stengel}, \citenamefont {Aguado-Puente}, \citenamefont {Spaldin},\ and\
  \citenamefont {Junquera}}]{stengel2011band}%
  \BibitemOpen
  \bibfield  {author} {\bibinfo {author} {\bibfnamefont {Massimiliano}\
  \bibnamefont {Stengel}}, \bibinfo {author} {\bibfnamefont {Pablo}\
  \bibnamefont {Aguado-Puente}}, \bibinfo {author} {\bibfnamefont {Nicola~A}\
  \bibnamefont {Spaldin}}, \ and\ \bibinfo {author} {\bibfnamefont {Javier}\
  \bibnamefont {Junquera}},\ }\bibfield  {title} {\enquote {\bibinfo {title}
  {Band alignment at metal/ferroelectric interfaces: Insights and artifacts
  from first principles},}\ }\href {\doibase
  https://doi.org/10.1103/PhysRevB.83.235112} {\bibfield  {journal} {\bibinfo
  {journal} {Phys. Rev. B}\ }\textbf {\bibinfo {volume} {83}},\ \bibinfo
  {pages} {235112} (\bibinfo {year} {2011})}\BibitemShut {NoStop}%
\bibitem [{\citenamefont {Qi}\ and\ \citenamefont {Zhang}(2011)}]{Rmp1}%
  \BibitemOpen
  \bibfield  {author} {\bibinfo {author} {\bibfnamefont {Xiao-Liang}\
  \bibnamefont {Qi}}\ and\ \bibinfo {author} {\bibfnamefont {Shou-Cheng}\
  \bibnamefont {Zhang}},\ }\bibfield  {title} {\enquote {\bibinfo {title}
  {Topological insulators and superconductors},}\ }\href {\doibase
  10.1103/RevModPhys.83.1057} {\bibfield  {journal} {\bibinfo  {journal} {Rev.
  Mod. Phys.}\ }\textbf {\bibinfo {volume} {83}},\ \bibinfo {pages}
  {1057--1110} (\bibinfo {year} {2011})}\BibitemShut {NoStop}%
\bibitem [{\citenamefont {Hasan}\ and\ \citenamefont {Kane}(2010)}]{Rmp2}%
  \BibitemOpen
  \bibfield  {author} {\bibinfo {author} {\bibfnamefont {M.~Z.}\ \bibnamefont
  {Hasan}}\ and\ \bibinfo {author} {\bibfnamefont {C.~L.}\ \bibnamefont
  {Kane}},\ }\bibfield  {title} {\enquote {\bibinfo {title} {Colloquium:
  Topological insulators},}\ }\href {\doibase 10.1103/RevModPhys.82.3045}
  {\bibfield  {journal} {\bibinfo  {journal} {Rev. Mod. Phys.}\ }\textbf
  {\bibinfo {volume} {82}},\ \bibinfo {pages} {3045--3067} (\bibinfo {year}
  {2010})}\BibitemShut {NoStop}%
\bibitem [{\citenamefont {Fu}(2011)}]{Clas1}%
  \BibitemOpen
  \bibfield  {author} {\bibinfo {author} {\bibfnamefont {Liang}\ \bibnamefont
  {Fu}},\ }\bibfield  {title} {\enquote {\bibinfo {title} {Topological
  crystalline insulators},}\ }\href {\doibase 10.1103/PhysRevLett.106.106802}
  {\bibfield  {journal} {\bibinfo  {journal} {Phys. Rev. Lett.}\ }\textbf
  {\bibinfo {volume} {106}},\ \bibinfo {pages} {106802} (\bibinfo {year}
  {2011})}\BibitemShut {NoStop}%
\bibitem [{\citenamefont {Slager}\ \emph {et~al.}(2013)\citenamefont {Slager},
  \citenamefont {Mesaros}, \citenamefont {Juri{\v{c}}i{\'c}},\ and\
  \citenamefont {Zaanen}}]{Clas2}%
  \BibitemOpen
  \bibfield  {author} {\bibinfo {author} {\bibfnamefont {Robert-Jan}\
  \bibnamefont {Slager}}, \bibinfo {author} {\bibfnamefont {Andrej}\
  \bibnamefont {Mesaros}}, \bibinfo {author} {\bibfnamefont {Vladimir}\
  \bibnamefont {Juri{\v{c}}i{\'c}}}, \ and\ \bibinfo {author} {\bibfnamefont
  {Jan}\ \bibnamefont {Zaanen}},\ }\bibfield  {title} {\enquote {\bibinfo
  {title} {The space group classification of topological band-insulators},}\
  }\href {\doibase https://doi.org/10.1038/nphys2513} {\bibfield  {journal}
  {\bibinfo  {journal} {Nat. Phys.}\ }\textbf {\bibinfo {volume} {9}},\
  \bibinfo {pages} {98--102} (\bibinfo {year} {2013})}\BibitemShut {NoStop}%
\bibitem [{\citenamefont {\"Unal}\ \emph {et~al.}(2020)\citenamefont {\"Unal},
  \citenamefont {Bouhon},\ and\ \citenamefont {Slager}}]{Unal_quenched_Euler}%
  \BibitemOpen
  \bibfield  {author} {\bibinfo {author} {\bibfnamefont {F.~Nur}\ \bibnamefont
  {\"Unal}}, \bibinfo {author} {\bibfnamefont {Adrien}\ \bibnamefont {Bouhon}},
  \ and\ \bibinfo {author} {\bibfnamefont {Robert-Jan}\ \bibnamefont
  {Slager}},\ }\bibfield  {title} {\enquote {\bibinfo {title} {Topological
  {E}uler class as a dynamical observable in optical lattices},}\ }\href
  {\doibase 10.1103/PhysRevLett.125.053601} {\bibfield  {journal} {\bibinfo
  {journal} {Phys. Rev. Lett.}\ }\textbf {\bibinfo {volume} {125}},\ \bibinfo
  {pages} {053601} (\bibinfo {year} {2020})}\BibitemShut {NoStop}%
\bibitem [{\citenamefont {Po}\ \emph {et~al.}(2017)\citenamefont {Po},
  \citenamefont {Vishwanath},\ and\ \citenamefont {Watanabe}}]{Clas4}%
  \BibitemOpen
  \bibfield  {author} {\bibinfo {author} {\bibfnamefont {Hoi~Chun}\
  \bibnamefont {Po}}, \bibinfo {author} {\bibfnamefont {Ashvin}\ \bibnamefont
  {Vishwanath}}, \ and\ \bibinfo {author} {\bibfnamefont {Haruki}\ \bibnamefont
  {Watanabe}},\ }\bibfield  {title} {\enquote {\bibinfo {title} {Symmetry-based
  indicators of band topology in the 230 space groups},}\ }\href {\doibase
  https://doi.org/10.1038/s41467-017-00133-2} {\bibfield  {journal} {\bibinfo
  {journal} {Nat. Commun.}\ }\textbf {\bibinfo {volume} {8}},\ \bibinfo {pages}
  {50} (\bibinfo {year} {2017})}\BibitemShut {NoStop}%
\bibitem [{\citenamefont {Bradlyn}\ \emph {et~al.}(2017)\citenamefont
  {Bradlyn}, \citenamefont {Elcoro}, \citenamefont {Cano}, \citenamefont
  {Vergniory}, \citenamefont {Wang}, \citenamefont {Felser}, \citenamefont
  {Aroyo},\ and\ \citenamefont {Bernevig}}]{Clas5}%
  \BibitemOpen
  \bibfield  {author} {\bibinfo {author} {\bibfnamefont {Barry}\ \bibnamefont
  {Bradlyn}}, \bibinfo {author} {\bibfnamefont {Luis}\ \bibnamefont {Elcoro}},
  \bibinfo {author} {\bibfnamefont {Jennifer}\ \bibnamefont {Cano}}, \bibinfo
  {author} {\bibfnamefont {Maia~G}\ \bibnamefont {Vergniory}}, \bibinfo
  {author} {\bibfnamefont {Zhijun}\ \bibnamefont {Wang}}, \bibinfo {author}
  {\bibfnamefont {Claudia}\ \bibnamefont {Felser}}, \bibinfo {author}
  {\bibfnamefont {Mois~I}\ \bibnamefont {Aroyo}}, \ and\ \bibinfo {author}
  {\bibfnamefont {B~Andrei}\ \bibnamefont {Bernevig}},\ }\bibfield  {title}
  {\enquote {\bibinfo {title} {Topological quantum chemistry},}\ }\href
  {\doibase https://doi.org/10.1038/nature23268} {\bibfield  {journal}
  {\bibinfo  {journal} {Nature}\ }\textbf {\bibinfo {volume} {547}},\ \bibinfo
  {pages} {298--305} (\bibinfo {year} {2017})}\BibitemShut {NoStop}%
\bibitem [{\citenamefont {Kruthoff}\ \emph {et~al.}(2017)\citenamefont
  {Kruthoff}, \citenamefont {De~Boer}, \citenamefont {Van~Wezel}, \citenamefont
  {Kane},\ and\ \citenamefont {Slager}}]{Clas3}%
  \BibitemOpen
  \bibfield  {author} {\bibinfo {author} {\bibfnamefont {Jorrit}\ \bibnamefont
  {Kruthoff}}, \bibinfo {author} {\bibfnamefont {Jan}\ \bibnamefont {De~Boer}},
  \bibinfo {author} {\bibfnamefont {Jasper}\ \bibnamefont {Van~Wezel}},
  \bibinfo {author} {\bibfnamefont {Charles~L}\ \bibnamefont {Kane}}, \ and\
  \bibinfo {author} {\bibfnamefont {Robert-Jan}\ \bibnamefont {Slager}},\
  }\bibfield  {title} {\enquote {\bibinfo {title} {Topological classification
  of crystalline insulators through band structure combinatorics},}\ }\href
  {\doibase https://doi.org/10.1103/PhysRevX.7.041069} {\bibfield  {journal}
  {\bibinfo  {journal} {Phys. Rev. X}\ }\textbf {\bibinfo {volume} {7}},\
  \bibinfo {pages} {041069} (\bibinfo {year} {2017})}\BibitemShut {NoStop}%
\bibitem [{\citenamefont {Shiozaki}\ and\ \citenamefont
  {Sato}(2014)}]{Shiozaki14}%
  \BibitemOpen
  \bibfield  {author} {\bibinfo {author} {\bibfnamefont {Ken}\ \bibnamefont
  {Shiozaki}}\ and\ \bibinfo {author} {\bibfnamefont {Masatoshi}\ \bibnamefont
  {Sato}},\ }\bibfield  {title} {\enquote {\bibinfo {title} {Topology of
  crystalline insulators and superconductors},}\ }\href {\doibase
  https://doi.org/10.1103/PhysRevB.90.165114} {\bibfield  {journal} {\bibinfo
  {journal} {Phys. Rev. B}\ }\textbf {\bibinfo {volume} {90}},\ \bibinfo
  {pages} {165114} (\bibinfo {year} {2014})}\BibitemShut {NoStop}%
\bibitem [{\citenamefont {Bouhon}\ \emph
  {et~al.}(2020{\natexlab{a}})\citenamefont {Bouhon}, \citenamefont {Wu},
  \citenamefont {Slager}, \citenamefont {Weng}, \citenamefont {Yazyev},\ and\
  \citenamefont {Bzdusek}}]{bouhon2019nonabelian}%
  \BibitemOpen
  \bibfield  {author} {\bibinfo {author} {\bibfnamefont {Adrien}\ \bibnamefont
  {Bouhon}}, \bibinfo {author} {\bibfnamefont {QuanSheng}\ \bibnamefont {Wu}},
  \bibinfo {author} {\bibfnamefont {Robert-Jan}\ \bibnamefont {Slager}},
  \bibinfo {author} {\bibfnamefont {Hongming}\ \bibnamefont {Weng}}, \bibinfo
  {author} {\bibfnamefont {Oleg~V}\ \bibnamefont {Yazyev}}, \ and\ \bibinfo
  {author} {\bibfnamefont {Tomas}\ \bibnamefont {Bzdusek}},\ }\bibfield
  {title} {\enquote {\bibinfo {title} {Non-{A}belian reciprocal braiding of
  {W}eyl points and its manifestation in {Z}r{T}e},}\ }\href {\doibase
  https://doi.org/10.1038/s41567-020-0967-9} {\bibfield  {journal} {\bibinfo
  {journal} {Nat. Phys.}\ }\textbf {\bibinfo {volume} {16}},\ \bibinfo {pages}
  {1137--1143} (\bibinfo {year} {2020}{\natexlab{a}})}\BibitemShut {NoStop}%
\bibitem [{\citenamefont {Bouhon}\ \emph
  {et~al.}(2020{\natexlab{b}})\citenamefont {Bouhon}, \citenamefont {Bzdusek},\
  and\ \citenamefont {Slager}}]{bouhon2020geometric}%
  \BibitemOpen
  \bibfield  {author} {\bibinfo {author} {\bibfnamefont {Adrien}\ \bibnamefont
  {Bouhon}}, \bibinfo {author} {\bibfnamefont {Tomas}\ \bibnamefont {Bzdusek}},
  \ and\ \bibinfo {author} {\bibfnamefont {Robert-Jan}\ \bibnamefont
  {Slager}},\ }\bibfield  {title} {\enquote {\bibinfo {title} {Geometric
  approach to fragile topology beyond symmetry indicators},}\ }\href {\doibase
  10.1103/PhysRevB.102.115135} {\bibfield  {journal} {\bibinfo  {journal}
  {Phys. Rev. B}\ }\textbf {\bibinfo {volume} {102}},\ \bibinfo {pages}
  {115135} (\bibinfo {year} {2020}{\natexlab{b}})}\BibitemShut {NoStop}%
\bibitem [{\citenamefont {Song}\ \emph {et~al.}(2018)\citenamefont {Song},
  \citenamefont {Zhang}, \citenamefont {Fang},\ and\ \citenamefont
  {Fang}}]{Song_2018}%
  \BibitemOpen
  \bibfield  {author} {\bibinfo {author} {\bibfnamefont {Zhida}\ \bibnamefont
  {Song}}, \bibinfo {author} {\bibfnamefont {Tiantian}\ \bibnamefont {Zhang}},
  \bibinfo {author} {\bibfnamefont {Zhong}\ \bibnamefont {Fang}}, \ and\
  \bibinfo {author} {\bibfnamefont {Chen}\ \bibnamefont {Fang}},\ }\bibfield
  {title} {\enquote {\bibinfo {title} {Quantitative mappings between symmetry
  and topology in solids},}\ }\href {\doibase
  https://doi.org/10.1038/s41467-018-06010-w} {\bibfield  {journal} {\bibinfo
  {journal} {Nat. Commun.}\ }\textbf {\bibinfo {volume} {9}},\ \bibinfo {pages}
  {3530} (\bibinfo {year} {2018})}\BibitemShut {NoStop}%
\bibitem [{\citenamefont {Coh}\ and\ \citenamefont
  {Vanderbilt}(2009)}]{Sinisa2009}%
  \BibitemOpen
  \bibfield  {author} {\bibinfo {author} {\bibfnamefont {Sinisa}\ \bibnamefont
  {Coh}}\ and\ \bibinfo {author} {\bibfnamefont {David}\ \bibnamefont
  {Vanderbilt}},\ }\bibfield  {title} {\enquote {\bibinfo {title} {Electric
  polarization in a {C}hern insulator},}\ }\href {\doibase
  10.1103/PhysRevLett.102.107603} {\bibfield  {journal} {\bibinfo  {journal}
  {Phys. Rev. Lett.}\ }\textbf {\bibinfo {volume} {102}},\ \bibinfo {pages}
  {107603} (\bibinfo {year} {2009})}\BibitemShut {NoStop}%
\bibitem [{\citenamefont {Song}\ \emph {et~al.}(2021)\citenamefont {Song},
  \citenamefont {He}, \citenamefont {Vishwanath},\ and\ \citenamefont
  {Wang}}]{Song2021}%
  \BibitemOpen
  \bibfield  {author} {\bibinfo {author} {\bibfnamefont {Xue-Yang}\
  \bibnamefont {Song}}, \bibinfo {author} {\bibfnamefont {Yin-Chen}\
  \bibnamefont {He}}, \bibinfo {author} {\bibfnamefont {Ashvin}\ \bibnamefont
  {Vishwanath}}, \ and\ \bibinfo {author} {\bibfnamefont {Chong}\ \bibnamefont
  {Wang}},\ }\bibfield  {title} {\enquote {\bibinfo {title} {Electric
  polarization as a nonquantized topological response and boundary {L}uttinger
  theorem},}\ }\href {\doibase 10.1103/PhysRevResearch.3.023011} {\bibfield
  {journal} {\bibinfo  {journal} {Phys. Rev. Research}\ }\textbf {\bibinfo
  {volume} {3}},\ \bibinfo {pages} {023011} (\bibinfo {year}
  {2021})}\BibitemShut {NoStop}%
\bibitem [{\citenamefont {Alexandradinata}\ and\ \citenamefont
  {Bernevig}(2016)}]{Alex_BerryPhase}%
  \BibitemOpen
  \bibfield  {author} {\bibinfo {author} {\bibfnamefont {A}~\bibnamefont
  {Alexandradinata}}\ and\ \bibinfo {author} {\bibfnamefont {B.~A.}\
  \bibnamefont {Bernevig}},\ }\bibfield  {title} {\enquote {\bibinfo {title}
  {{Berry-phase description of topological crystalline insulators}},}\ }\href
  {\doibase 10.1103/PhysRevB.93.205104} {\bibfield  {journal} {\bibinfo
  {journal} {Phys. Rev. B}\ }\textbf {\bibinfo {volume} {93}},\ \bibinfo
  {pages} {205104} (\bibinfo {year} {2016})}\BibitemShut {NoStop}%
\bibitem [{\citenamefont {Bouhon}\ \emph {et~al.}(2019)\citenamefont {Bouhon},
  \citenamefont {Black-Schaffer},\ and\ \citenamefont
  {Slager}}]{bouhon2019wilson}%
  \BibitemOpen
  \bibfield  {author} {\bibinfo {author} {\bibfnamefont {Adrien}\ \bibnamefont
  {Bouhon}}, \bibinfo {author} {\bibfnamefont {Annica~M}\ \bibnamefont
  {Black-Schaffer}}, \ and\ \bibinfo {author} {\bibfnamefont {Robert-Jan}\
  \bibnamefont {Slager}},\ }\bibfield  {title} {\enquote {\bibinfo {title}
  {{Wilson loop approach to fragile topology of split elementary band
  representations and topological crystalline insulators with time-reversal
  symmetry}},}\ }\href {\doibase 10.1103/PhysRevB.100.195135} {\bibfield
  {journal} {\bibinfo  {journal} {Phys. Rev. B}\ }\textbf {\bibinfo {volume}
  {100}},\ \bibinfo {pages} {195135} (\bibinfo {year} {2019})}\BibitemShut
  {NoStop}%
\bibitem [{\citenamefont {Liang}\ \emph {et~al.}(2023)\citenamefont {Liang},
  \citenamefont {Zheng}, \citenamefont {Frauenheim},\ and\ \citenamefont
  {Zhao}}]{liang2023ferroelectric}%
  \BibitemOpen
  \bibfield  {author} {\bibinfo {author} {\bibfnamefont {Yan}\ \bibnamefont
  {Liang}}, \bibinfo {author} {\bibfnamefont {Fulu}\ \bibnamefont {Zheng}},
  \bibinfo {author} {\bibfnamefont {Thomas}\ \bibnamefont {Frauenheim}}, \ and\
  \bibinfo {author} {\bibfnamefont {Pei}\ \bibnamefont {Zhao}},\ }\bibfield
  {title} {\enquote {\bibinfo {title} {Ferroelectric antiferromagnetic quantum
  anomalous {H}all insulator in two dimensional van der {W}aals materials},}\
  }\href {https://doi.org/10.48550/arXiv.2302.05091} {\bibfield  {journal}
  {\bibinfo  {journal} {arXiv:2302.05091}\ } (\bibinfo {year}
  {2023})}\BibitemShut {NoStop}%
\bibitem [{\citenamefont {Szab\'o}\ and\ \citenamefont
  {Schneider}(2018)}]{Attila2018}%
  \BibitemOpen
  \bibfield  {author} {\bibinfo {author} {\bibfnamefont {Attila}\ \bibnamefont
  {Szab\'o}}\ and\ \bibinfo {author} {\bibfnamefont {Ulrich}\ \bibnamefont
  {Schneider}},\ }\bibfield  {title} {\enquote {\bibinfo {title} {Non-power-law
  universality in one-dimensional quasicrystals},}\ }\href {\doibase
  10.1103/PhysRevB.98.134201} {\bibfield  {journal} {\bibinfo  {journal} {Phys.
  Rev. B}\ }\textbf {\bibinfo {volume} {98}},\ \bibinfo {pages} {134201}
  (\bibinfo {year} {2018})}\BibitemShut {NoStop}%
\bibitem [{\citenamefont {Borgnia}\ \emph {et~al.}(2022)\citenamefont
  {Borgnia}, \citenamefont {Vishwanath},\ and\ \citenamefont {Slager}}]{qp1}%
  \BibitemOpen
  \bibfield  {author} {\bibinfo {author} {\bibfnamefont {Dan~S.}\ \bibnamefont
  {Borgnia}}, \bibinfo {author} {\bibfnamefont {Ashvin}\ \bibnamefont
  {Vishwanath}}, \ and\ \bibinfo {author} {\bibfnamefont {Robert-Jan}\
  \bibnamefont {Slager}},\ }\bibfield  {title} {\enquote {\bibinfo {title}
  {Rational approximations of quasiperiodicity via projected green's
  functions},}\ }\href {\doibase 10.1103/PhysRevB.106.054204} {\bibfield
  {journal} {\bibinfo  {journal} {Phys. Rev. B}\ }\textbf {\bibinfo {volume}
  {106}},\ \bibinfo {pages} {054204} (\bibinfo {year} {2022})}\BibitemShut
  {NoStop}%
\bibitem [{\citenamefont {Borgnia}\ and\ \citenamefont {Slager}(2023)}]{qp2}%
  \BibitemOpen
  \bibfield  {author} {\bibinfo {author} {\bibfnamefont {Dan~S.}\ \bibnamefont
  {Borgnia}}\ and\ \bibinfo {author} {\bibfnamefont {Robert-Jan}\ \bibnamefont
  {Slager}},\ }\bibfield  {title} {\enquote {\bibinfo {title} {Localization as
  a consequence of quasiperiodic bulk-bulk correspondence},}\ }\href {\doibase
  10.1103/PhysRevB.107.085111} {\bibfield  {journal} {\bibinfo  {journal}
  {Phys. Rev. B}\ }\textbf {\bibinfo {volume} {107}},\ \bibinfo {pages}
  {085111} (\bibinfo {year} {2023})}\BibitemShut {NoStop}%
\bibitem [{\citenamefont {Aubry}\ and\ \citenamefont
  {Andr{\'e}}(1980)}]{aubry1980analyticity}%
  \BibitemOpen
  \bibfield  {author} {\bibinfo {author} {\bibfnamefont {Serge}\ \bibnamefont
  {Aubry}}\ and\ \bibinfo {author} {\bibfnamefont {Gilles}\ \bibnamefont
  {Andr{\'e}}},\ }\bibfield  {title} {\enquote {\bibinfo {title} {Analyticity
  breaking and anderson localization in incommensurate lattices},}\ }\href@noop
  {} {\bibfield  {journal} {\bibinfo  {journal} {Ann. Israel Phys. Soc}\
  }\textbf {\bibinfo {volume} {3}},\ \bibinfo {pages} {18} (\bibinfo {year}
  {1980})}\BibitemShut {NoStop}%
\bibitem [{\citenamefont {Bistritzer}\ and\ \citenamefont
  {MacDonald}(2011)}]{Bistritzer2011}%
  \BibitemOpen
  \bibfield  {author} {\bibinfo {author} {\bibfnamefont {Rafi}\ \bibnamefont
  {Bistritzer}}\ and\ \bibinfo {author} {\bibfnamefont {Allan~H.}\ \bibnamefont
  {MacDonald}},\ }\bibfield  {title} {\enquote {\bibinfo {title} {Moiré bands
  in twisted double-layer graphene},}\ }\href {\doibase
  10.1073/pnas.1108174108} {\bibfield  {journal} {\bibinfo  {journal}
  {Proceedings of the National Academy of Sciences}\ }\textbf {\bibinfo
  {volume} {108}},\ \bibinfo {pages} {12233--12237} (\bibinfo {year} {2011})},\
  \Eprint
  {http://arxiv.org/abs/https://www.pnas.org/doi/pdf/10.1073/pnas.1108174108}
  {https://www.pnas.org/doi/pdf/10.1073/pnas.1108174108} \BibitemShut {NoStop}%
\bibitem [{\citenamefont {Balents}(2019)}]{Balents2019}%
  \BibitemOpen
  \bibfield  {author} {\bibinfo {author} {\bibfnamefont {L.}~\bibnamefont
  {Balents}},\ }\bibfield  {title} {\enquote {\bibinfo {title} {{General
  continuum model for twisted bilayer graphene and arbitrary smooth
  deformations}},}\ }\href {\doibase 10.21468/SciPostPhys.7.4.048} {\bibfield
  {journal} {\bibinfo  {journal} {SciPost Phys.}\ }\textbf {\bibinfo {volume}
  {7}},\ \bibinfo {pages} {48} (\bibinfo {year} {2019})}\BibitemShut {NoStop}%
\bibitem [{\citenamefont {Carr}\ \emph {et~al.}(2017)\citenamefont {Carr},
  \citenamefont {Massatt}, \citenamefont {Fang}, \citenamefont {Cazeaux},
  \citenamefont {Luskin},\ and\ \citenamefont {Kaxiras}}]{carr2017twistronics}%
  \BibitemOpen
  \bibfield  {author} {\bibinfo {author} {\bibfnamefont {Stephen}\ \bibnamefont
  {Carr}}, \bibinfo {author} {\bibfnamefont {Daniel}\ \bibnamefont {Massatt}},
  \bibinfo {author} {\bibfnamefont {Shiang}\ \bibnamefont {Fang}}, \bibinfo
  {author} {\bibfnamefont {Paul}\ \bibnamefont {Cazeaux}}, \bibinfo {author}
  {\bibfnamefont {Mitchell}\ \bibnamefont {Luskin}}, \ and\ \bibinfo {author}
  {\bibfnamefont {Efthimios}\ \bibnamefont {Kaxiras}},\ }\bibfield  {title}
  {\enquote {\bibinfo {title} {Twistronics: Manipulating the electronic
  properties of two-dimensional layered structures through their twist
  angle},}\ }\href {\doibase https://doi.org/10.1103/PhysRevB.95.075420}
  {\bibfield  {journal} {\bibinfo  {journal} {Phys. Rev. B}\ }\textbf {\bibinfo
  {volume} {95}},\ \bibinfo {pages} {075420} (\bibinfo {year}
  {2017})}\BibitemShut {NoStop}%
\bibitem [{\citenamefont {Cazeaux}\ \emph {et~al.}(2017)\citenamefont
  {Cazeaux}, \citenamefont {Luskin},\ and\ \citenamefont
  {Tadmor}}]{cazeaux2017analysis}%
  \BibitemOpen
  \bibfield  {author} {\bibinfo {author} {\bibfnamefont {Paul}\ \bibnamefont
  {Cazeaux}}, \bibinfo {author} {\bibfnamefont {Mitchell}\ \bibnamefont
  {Luskin}}, \ and\ \bibinfo {author} {\bibfnamefont {Ellad~B}\ \bibnamefont
  {Tadmor}},\ }\bibfield  {title} {\enquote {\bibinfo {title} {Analysis of
  rippling in incommensurate one-dimensional coupled chains},}\ }\href
  {\doibase https://doi.org/10.1137/16M1076198} {\bibfield  {journal} {\bibinfo
   {journal} {Multiscale Model. Simul.}\ }\textbf {\bibinfo {volume} {15}},\
  \bibinfo {pages} {56--73} (\bibinfo {year} {2017})}\BibitemShut {NoStop}%
\bibitem [{\citenamefont {Massatt}\ \emph {et~al.}(2017)\citenamefont
  {Massatt}, \citenamefont {Luskin},\ and\ \citenamefont
  {Ortner}}]{massatt2017electronic}%
  \BibitemOpen
  \bibfield  {author} {\bibinfo {author} {\bibfnamefont {Daniel}\ \bibnamefont
  {Massatt}}, \bibinfo {author} {\bibfnamefont {Mitchell}\ \bibnamefont
  {Luskin}}, \ and\ \bibinfo {author} {\bibfnamefont {Christoph}\ \bibnamefont
  {Ortner}},\ }\bibfield  {title} {\enquote {\bibinfo {title} {Electronic
  density of states for incommensurate layers},}\ }\href {\doibase
  https://doi.org/10.1137/16M1088363} {\bibfield  {journal} {\bibinfo
  {journal} {Multiscale Model. Simul.}\ }\textbf {\bibinfo {volume} {15}},\
  \bibinfo {pages} {476--499} (\bibinfo {year} {2017})}\BibitemShut {NoStop}%
\bibitem [{\citenamefont {Carr}\ \emph {et~al.}(2018)\citenamefont {Carr},
  \citenamefont {Massatt}, \citenamefont {Torrisi}, \citenamefont {Cazeaux},
  \citenamefont {Luskin},\ and\ \citenamefont {Kaxiras}}]{carr2018relaxation}%
  \BibitemOpen
  \bibfield  {author} {\bibinfo {author} {\bibfnamefont {Stephen}\ \bibnamefont
  {Carr}}, \bibinfo {author} {\bibfnamefont {Daniel}\ \bibnamefont {Massatt}},
  \bibinfo {author} {\bibfnamefont {Steven~B}\ \bibnamefont {Torrisi}},
  \bibinfo {author} {\bibfnamefont {Paul}\ \bibnamefont {Cazeaux}}, \bibinfo
  {author} {\bibfnamefont {Mitchell}\ \bibnamefont {Luskin}}, \ and\ \bibinfo
  {author} {\bibfnamefont {Efthimios}\ \bibnamefont {Kaxiras}},\ }\bibfield
  {title} {\enquote {\bibinfo {title} {Relaxation and domain formation in
  incommensurate two-dimensional heterostructures},}\ }\href {\doibase
  https://doi.org/10.1103/PhysRevB.98.224102} {\bibfield  {journal} {\bibinfo
  {journal} {Phys. Rev. B}\ }\textbf {\bibinfo {volume} {98}},\ \bibinfo
  {pages} {224102} (\bibinfo {year} {2018})}\BibitemShut {NoStop}%
\bibitem [{\citenamefont {Cohen}(1992)}]{cohen1992origin}%
  \BibitemOpen
  \bibfield  {author} {\bibinfo {author} {\bibfnamefont {Ronald~E}\
  \bibnamefont {Cohen}},\ }\bibfield  {title} {\enquote {\bibinfo {title}
  {Origin of ferroelectricity in perovskite oxides},}\ }\href {\doibase
  https://doi.org/10.1038/358136a0} {\bibfield  {journal} {\bibinfo  {journal}
  {Nature}\ }\textbf {\bibinfo {volume} {358}},\ \bibinfo {pages} {136--138}
  (\bibinfo {year} {1992})}\BibitemShut {NoStop}%
\bibitem [{\citenamefont {Sluka}\ \emph {et~al.}(2013)\citenamefont {Sluka},
  \citenamefont {Tagantsev}, \citenamefont {Bednyakov},\ and\ \citenamefont
  {Setter}}]{sluka2013free}%
  \BibitemOpen
  \bibfield  {author} {\bibinfo {author} {\bibfnamefont {Tomas}\ \bibnamefont
  {Sluka}}, \bibinfo {author} {\bibfnamefont {Alexander~K}\ \bibnamefont
  {Tagantsev}}, \bibinfo {author} {\bibfnamefont {Petr}\ \bibnamefont
  {Bednyakov}}, \ and\ \bibinfo {author} {\bibfnamefont {Nava}\ \bibnamefont
  {Setter}},\ }\bibfield  {title} {\enquote {\bibinfo {title} {Free-electron
  gas at charged domain walls in insulating {BaTiO$_3$}},}\ }\href {\doibase
  https://doi.org/10.1038/ncomms2839} {\bibfield  {journal} {\bibinfo
  {journal} {Nat. Commun.}\ }\textbf {\bibinfo {volume} {4}},\ \bibinfo {pages}
  {1808} (\bibinfo {year} {2013})}\BibitemShut {NoStop}%
\bibitem [{\citenamefont {Bednyakov}\ \emph {et~al.}(2015)\citenamefont
  {Bednyakov}, \citenamefont {Sluka}, \citenamefont {Tagantsev}, \citenamefont
  {Damjanovic},\ and\ \citenamefont {Setter}}]{bednyakov2015formation}%
  \BibitemOpen
  \bibfield  {author} {\bibinfo {author} {\bibfnamefont {Petr~S}\ \bibnamefont
  {Bednyakov}}, \bibinfo {author} {\bibfnamefont {Tomas}\ \bibnamefont
  {Sluka}}, \bibinfo {author} {\bibfnamefont {Alexander~K}\ \bibnamefont
  {Tagantsev}}, \bibinfo {author} {\bibfnamefont {Dragan}\ \bibnamefont
  {Damjanovic}}, \ and\ \bibinfo {author} {\bibfnamefont {Nava}\ \bibnamefont
  {Setter}},\ }\bibfield  {title} {\enquote {\bibinfo {title} {Formation of
  charged ferroelectric domain walls with controlled periodicity},}\ }\href
  {\doibase https://doi.org/10.1038/srep15819} {\bibfield  {journal} {\bibinfo
  {journal} {Sci. Rep.}\ }\textbf {\bibinfo {volume} {5}},\ \bibinfo {pages}
  {15819} (\bibinfo {year} {2015})}\BibitemShut {NoStop}%
\bibitem [{\citenamefont {Engelke}\ \emph {et~al.}(2023)\citenamefont
  {Engelke}, \citenamefont {Yoo}, \citenamefont {Carr}, \citenamefont {Xu},
  \citenamefont {Cazeaux}, \citenamefont {Allen}, \citenamefont {Valdivia},
  \citenamefont {Luskin}, \citenamefont {Kaxiras}, \citenamefont {Kim},
  \citenamefont {Han},\ and\ \citenamefont {Kim}}]{engelke2023non}%
  \BibitemOpen
  \bibfield  {author} {\bibinfo {author} {\bibfnamefont {Rebecca}\ \bibnamefont
  {Engelke}}, \bibinfo {author} {\bibfnamefont {Hyobin}\ \bibnamefont {Yoo}},
  \bibinfo {author} {\bibfnamefont {Stephen}\ \bibnamefont {Carr}}, \bibinfo
  {author} {\bibfnamefont {Kevin}\ \bibnamefont {Xu}}, \bibinfo {author}
  {\bibfnamefont {Paul}\ \bibnamefont {Cazeaux}}, \bibinfo {author}
  {\bibfnamefont {Richard}\ \bibnamefont {Allen}}, \bibinfo {author}
  {\bibfnamefont {Andres~Mier}\ \bibnamefont {Valdivia}}, \bibinfo {author}
  {\bibfnamefont {Mitchell}\ \bibnamefont {Luskin}}, \bibinfo {author}
  {\bibfnamefont {Efthimios}\ \bibnamefont {Kaxiras}}, \bibinfo {author}
  {\bibfnamefont {Minhyong}\ \bibnamefont {Kim}}, \bibinfo {author}
  {\bibfnamefont {Jung~Hoon}\ \bibnamefont {Han}}, \ and\ \bibinfo {author}
  {\bibfnamefont {Philip}\ \bibnamefont {Kim}},\ }\bibfield  {title} {\enquote
  {\bibinfo {title} {Topological nature of dislocation networks in
  two-dimensional moir\'e materials},}\ }\href {\doibase
  10.1103/PhysRevB.107.125413} {\bibfield  {journal} {\bibinfo  {journal}
  {Phys. Rev. B}\ }\textbf {\bibinfo {volume} {107}},\ \bibinfo {pages}
  {125413} (\bibinfo {year} {2023})}\BibitemShut {NoStop}%
\bibitem [{\citenamefont {Resta}(1998)}]{PhysRevLett.80.1800}%
  \BibitemOpen
  \bibfield  {author} {\bibinfo {author} {\bibfnamefont {Raffaele}\
  \bibnamefont {Resta}},\ }\bibfield  {title} {\enquote {\bibinfo {title}
  {Quantum-mechanical position operator in extended systems},}\ }\href
  {\doibase 10.1103/PhysRevLett.80.1800} {\bibfield  {journal} {\bibinfo
  {journal} {Phys. Rev. Lett.}\ }\textbf {\bibinfo {volume} {80}},\ \bibinfo
  {pages} {1800--1803} (\bibinfo {year} {1998})}\BibitemShut {NoStop}%
\bibitem [{\citenamefont {Wu}\ \emph {et~al.}(2005)\citenamefont {Wu},
  \citenamefont {Vanderbilt},\ and\ \citenamefont {Hamann}}]{Wu2005}%
  \BibitemOpen
  \bibfield  {author} {\bibinfo {author} {\bibfnamefont {Xifan}\ \bibnamefont
  {Wu}}, \bibinfo {author} {\bibfnamefont {David}\ \bibnamefont {Vanderbilt}},
  \ and\ \bibinfo {author} {\bibfnamefont {D.~R.}\ \bibnamefont {Hamann}},\
  }\bibfield  {title} {\enquote {\bibinfo {title} {Systematic treatment of
  displacements, strains, and electric fields in density-functional
  perturbation theory},}\ }\href {\doibase 10.1103/PhysRevB.72.035105}
  {\bibfield  {journal} {\bibinfo  {journal} {Phys. Rev. B}\ }\textbf {\bibinfo
  {volume} {72}},\ \bibinfo {pages} {035105} (\bibinfo {year}
  {2005})}\BibitemShut {NoStop}%
\bibitem [{\citenamefont {Gonze}\ and\ \citenamefont
  {Lee}(1997)}]{gonze1997dynamical}%
  \BibitemOpen
  \bibfield  {author} {\bibinfo {author} {\bibfnamefont {Xavier}\ \bibnamefont
  {Gonze}}\ and\ \bibinfo {author} {\bibfnamefont {Changyol}\ \bibnamefont
  {Lee}},\ }\bibfield  {title} {\enquote {\bibinfo {title} {Dynamical matrices,
  {B}orn effective charges, dielectric permittivity tensors, and interatomic
  force constants from density-functional perturbation theory},}\ }\href
  {\doibase 10.1103/PhysRevB.55.10355} {\bibfield  {journal} {\bibinfo
  {journal} {Phys. Rev. B}\ }\textbf {\bibinfo {volume} {55}},\ \bibinfo
  {pages} {10355} (\bibinfo {year} {1997})}\BibitemShut {NoStop}%
\bibitem [{\citenamefont {Vanderbilt}(2000)}]{Vanderbilt2000}%
  \BibitemOpen
  \bibfield  {author} {\bibinfo {author} {\bibfnamefont {D}~\bibnamefont
  {Vanderbilt}},\ }\bibfield  {title} {\enquote {\bibinfo {title} {Berry-phase
  theory of proper piezoelectric response},}\ }\href {\doibase
  https://doi.org/10.1016/S0022-3697(99)00273-5} {\bibfield  {journal}
  {\bibinfo  {journal} {J. Phys. Chem. Solids}\ }\textbf {\bibinfo {volume}
  {61}},\ \bibinfo {pages} {147--151} (\bibinfo {year} {2000})}\BibitemShut
  {NoStop}%
\bibitem [{\citenamefont {Bennett}\ \emph {et~al.}(2022)\citenamefont
  {Bennett}, \citenamefont {Tanner}, \citenamefont {Ghosez}, \citenamefont
  {Janolin},\ and\ \citenamefont {Bousquet}}]{bennett2022generalized}%
  \BibitemOpen
  \bibfield  {author} {\bibinfo {author} {\bibfnamefont {Daniel}\ \bibnamefont
  {Bennett}}, \bibinfo {author} {\bibfnamefont {Daniel}\ \bibnamefont
  {Tanner}}, \bibinfo {author} {\bibfnamefont {Philippe}\ \bibnamefont
  {Ghosez}}, \bibinfo {author} {\bibfnamefont {Pierre-Eymeric}\ \bibnamefont
  {Janolin}}, \ and\ \bibinfo {author} {\bibfnamefont {Eric}\ \bibnamefont
  {Bousquet}},\ }\bibfield  {title} {\enquote {\bibinfo {title} {Generalized
  relation between electromechanical responses at fixed voltage and fixed
  electric field},}\ }\href {\doibase 10.1103/PhysRevB.106.174105} {\bibfield
  {journal} {\bibinfo  {journal} {Phys. Rev. B}\ }\textbf {\bibinfo {volume}
  {106}},\ \bibinfo {pages} {174105} (\bibinfo {year} {2022})}\BibitemShut
  {NoStop}%
\bibitem [{\citenamefont {Ghosez}\ \emph {et~al.}(1998)\citenamefont {Ghosez},
  \citenamefont {Michenaud},\ and\ \citenamefont
  {Gonze}}]{ghosez1998dynamical}%
  \BibitemOpen
  \bibfield  {author} {\bibinfo {author} {\bibfnamefont {Ph}~\bibnamefont
  {Ghosez}}, \bibinfo {author} {\bibfnamefont {J-P}\ \bibnamefont {Michenaud}},
  \ and\ \bibinfo {author} {\bibfnamefont {Xavier}\ \bibnamefont {Gonze}},\
  }\bibfield  {title} {\enquote {\bibinfo {title} {{Dynamical atomic charges:
  The case of ABO$_3$ compounds}},}\ }\href {\doibase 10.1103/PhysRevB.58.6224}
  {\bibfield  {journal} {\bibinfo  {journal} {Phys. Rev. B}\ }\textbf {\bibinfo
  {volume} {58}},\ \bibinfo {pages} {6224} (\bibinfo {year}
  {1998})}\BibitemShut {NoStop}%
\bibitem [{\citenamefont {Ghosez}\ and\ \citenamefont
  {Gonze}(2000)}]{ghosez2000band}%
  \BibitemOpen
  \bibfield  {author} {\bibinfo {author} {\bibfnamefont {Philippe}\
  \bibnamefont {Ghosez}}\ and\ \bibinfo {author} {\bibfnamefont {Xavier}\
  \bibnamefont {Gonze}},\ }\bibfield  {title} {\enquote {\bibinfo {title}
  {Band-by-band decompositions of the {B}orn effective charges},}\ }\href
  {\doibase https://doi.org/10.1088/0953-8984/12/43/308} {\bibfield  {journal}
  {\bibinfo  {journal} {J. Phys.: Condens. Matter}\ }\textbf {\bibinfo {volume}
  {12}},\ \bibinfo {pages} {9179} (\bibinfo {year} {2000})}\BibitemShut
  {NoStop}%
\bibitem [{\citenamefont {Sai}\ \emph {et~al.}(2002)\citenamefont {Sai},
  \citenamefont {Rabe},\ and\ \citenamefont {Vanderbilt}}]{sai2002theory}%
  \BibitemOpen
  \bibfield  {author} {\bibinfo {author} {\bibfnamefont {Na}~\bibnamefont
  {Sai}}, \bibinfo {author} {\bibfnamefont {Karin~M}\ \bibnamefont {Rabe}}, \
  and\ \bibinfo {author} {\bibfnamefont {David}\ \bibnamefont {Vanderbilt}},\
  }\bibfield  {title} {\enquote {\bibinfo {title} {Theory of structural
  response to macroscopic electric fields in ferroelectric systems},}\ }\href
  {\doibase https://doi.org/10.1103/PhysRevB.66.104108} {\bibfield  {journal}
  {\bibinfo  {journal} {Phys. Rev. B}\ }\textbf {\bibinfo {volume} {66}},\
  \bibinfo {pages} {104108} (\bibinfo {year} {2002})}\BibitemShut {NoStop}%
\bibitem [{\citenamefont {Marzari}\ \emph {et~al.}(2012)\citenamefont
  {Marzari}, \citenamefont {Mostofi}, \citenamefont {Yates}, \citenamefont
  {Souza},\ and\ \citenamefont {Vanderbilt}}]{marzari2012maximally}%
  \BibitemOpen
  \bibfield  {author} {\bibinfo {author} {\bibfnamefont {Nicola}\ \bibnamefont
  {Marzari}}, \bibinfo {author} {\bibfnamefont {Arash~A.}\ \bibnamefont
  {Mostofi}}, \bibinfo {author} {\bibfnamefont {Jonathan~R.}\ \bibnamefont
  {Yates}}, \bibinfo {author} {\bibfnamefont {Ivo}\ \bibnamefont {Souza}}, \
  and\ \bibinfo {author} {\bibfnamefont {David}\ \bibnamefont {Vanderbilt}},\
  }\bibfield  {title} {\enquote {\bibinfo {title} {Maximally localized
  {W}annier functions: {T}heory and applications},}\ }\href {\doibase
  10.1103/RevModPhys.84.1419} {\bibfield  {journal} {\bibinfo  {journal} {Rev.
  Mod. Phys.}\ }\textbf {\bibinfo {volume} {84}},\ \bibinfo {pages}
  {1419--1475} (\bibinfo {year} {2012})}\BibitemShut {NoStop}%
\bibitem [{\citenamefont {Gresch}\ \emph {et~al.}(2017)\citenamefont {Gresch},
  \citenamefont {Aut\`es}, \citenamefont {Yazyev}, \citenamefont {Troyer},
  \citenamefont {Vanderbilt}, \citenamefont {Bernevig},\ and\ \citenamefont
  {Soluyanov}}]{gresch2017z2pack}%
  \BibitemOpen
  \bibfield  {author} {\bibinfo {author} {\bibfnamefont {Dominik}\ \bibnamefont
  {Gresch}}, \bibinfo {author} {\bibfnamefont {Gabriel}\ \bibnamefont
  {Aut\`es}}, \bibinfo {author} {\bibfnamefont {Oleg~V.}\ \bibnamefont
  {Yazyev}}, \bibinfo {author} {\bibfnamefont {Matthias}\ \bibnamefont
  {Troyer}}, \bibinfo {author} {\bibfnamefont {David}\ \bibnamefont
  {Vanderbilt}}, \bibinfo {author} {\bibfnamefont {B.~Andrei}\ \bibnamefont
  {Bernevig}}, \ and\ \bibinfo {author} {\bibfnamefont {Alexey~A.}\
  \bibnamefont {Soluyanov}},\ }\bibfield  {title} {\enquote {\bibinfo {title}
  {Z2pack: Numerical implementation of hybrid {W}annier centers for identifying
  topological materials},}\ }\href {\doibase 10.1103/PhysRevB.95.075146}
  {\bibfield  {journal} {\bibinfo  {journal} {Phys. Rev. B}\ }\textbf {\bibinfo
  {volume} {95}},\ \bibinfo {pages} {075146} (\bibinfo {year}
  {2017})}\BibitemShut {NoStop}%
\bibitem [{\citenamefont {Pizzi}\ \emph {et~al.}(2020)\citenamefont {Pizzi},
  \citenamefont {Vitale}, \citenamefont {Arita}, \citenamefont {Bl{\"u}gel},
  \citenamefont {Freimuth}, \citenamefont {G{\'e}ranton}, \citenamefont
  {Gibertini}, \citenamefont {Gresch}, \citenamefont {Johnson}, \citenamefont
  {Koretsune} \emph {et~al.}}]{pizzi2020wannier90}%
  \BibitemOpen
  \bibfield  {author} {\bibinfo {author} {\bibfnamefont {Giovanni}\
  \bibnamefont {Pizzi}}, \bibinfo {author} {\bibfnamefont {Valerio}\
  \bibnamefont {Vitale}}, \bibinfo {author} {\bibfnamefont {Ryotaro}\
  \bibnamefont {Arita}}, \bibinfo {author} {\bibfnamefont {Stefan}\
  \bibnamefont {Bl{\"u}gel}}, \bibinfo {author} {\bibfnamefont {Frank}\
  \bibnamefont {Freimuth}}, \bibinfo {author} {\bibfnamefont {Guillaume}\
  \bibnamefont {G{\'e}ranton}}, \bibinfo {author} {\bibfnamefont {Marco}\
  \bibnamefont {Gibertini}}, \bibinfo {author} {\bibfnamefont {Dominik}\
  \bibnamefont {Gresch}}, \bibinfo {author} {\bibfnamefont {Charles}\
  \bibnamefont {Johnson}}, \bibinfo {author} {\bibfnamefont {Takashi}\
  \bibnamefont {Koretsune}},  \emph {et~al.},\ }\bibfield  {title} {\enquote
  {\bibinfo {title} {Wannier90 as a community code: new features and
  applications},}\ }\href {\doibase https://doi.org/10.1088/1361-648X/ab51ff}
  {\bibfield  {journal} {\bibinfo  {journal} {J. Phys.: Condens. Matter}\
  }\textbf {\bibinfo {volume} {32}},\ \bibinfo {pages} {165902} (\bibinfo
  {year} {2020})}\BibitemShut {NoStop}%
\bibitem [{\citenamefont {Carr}\ \emph
  {et~al.}(2019{\natexlab{a}})\citenamefont {Carr}, \citenamefont {Fang},
  \citenamefont {Zhu},\ and\ \citenamefont {Kaxiras}}]{carr2019exact}%
  \BibitemOpen
  \bibfield  {author} {\bibinfo {author} {\bibfnamefont {Stephen}\ \bibnamefont
  {Carr}}, \bibinfo {author} {\bibfnamefont {Shiang}\ \bibnamefont {Fang}},
  \bibinfo {author} {\bibfnamefont {Ziyan}\ \bibnamefont {Zhu}}, \ and\
  \bibinfo {author} {\bibfnamefont {Efthimios}\ \bibnamefont {Kaxiras}},\
  }\bibfield  {title} {\enquote {\bibinfo {title} {Exact continuum model for
  low-energy electronic states of twisted bilayer graphene},}\ }\href {\doibase
  10.1103/PhysRevResearch.1.013001} {\bibfield  {journal} {\bibinfo  {journal}
  {Phys. Rev. Res.}\ }\textbf {\bibinfo {volume} {1}},\ \bibinfo {pages}
  {013001} (\bibinfo {year} {2019}{\natexlab{a}})}\BibitemShut {NoStop}%
\bibitem [{\citenamefont {Carr}\ \emph
  {et~al.}(2019{\natexlab{b}})\citenamefont {Carr}, \citenamefont {Fang},
  \citenamefont {Po}, \citenamefont {Vishwanath},\ and\ \citenamefont
  {Kaxiras}}]{carr2019derivation}%
  \BibitemOpen
  \bibfield  {author} {\bibinfo {author} {\bibfnamefont {Stephen}\ \bibnamefont
  {Carr}}, \bibinfo {author} {\bibfnamefont {Shiang}\ \bibnamefont {Fang}},
  \bibinfo {author} {\bibfnamefont {Hoi~Chun}\ \bibnamefont {Po}}, \bibinfo
  {author} {\bibfnamefont {Ashvin}\ \bibnamefont {Vishwanath}}, \ and\ \bibinfo
  {author} {\bibfnamefont {Efthimios}\ \bibnamefont {Kaxiras}},\ }\bibfield
  {title} {\enquote {\bibinfo {title} {Derivation of {W}annier orbitals and
  minimal-basis tight-binding {H}amiltonians for twisted bilayer graphene:
  First-principles approach},}\ }\href {\doibase
  10.1103/PhysRevResearch.1.033072} {\bibfield  {journal} {\bibinfo  {journal}
  {Phys. Rev. Res.}\ }\textbf {\bibinfo {volume} {1}},\ \bibinfo {pages}
  {033072} (\bibinfo {year} {2019}{\natexlab{b}})}\BibitemShut {NoStop}%
\bibitem [{\citenamefont {Sternheimer}(1954)}]{sternheimer1954electronic}%
  \BibitemOpen
  \bibfield  {author} {\bibinfo {author} {\bibfnamefont {R.~M.}\ \bibnamefont
  {Sternheimer}},\ }\bibfield  {title} {\enquote {\bibinfo {title} {Electronic
  polarizabilities of ions from the {H}artree-{F}ock wave functions},}\ }\href
  {\doibase 10.1103/PhysRev.96.951} {\bibfield  {journal} {\bibinfo  {journal}
  {Phys. Rev.}\ }\textbf {\bibinfo {volume} {96}},\ \bibinfo {pages} {951--968}
  (\bibinfo {year} {1954})}\BibitemShut {NoStop}%
\bibitem [{\citenamefont {Gonze}(1995)}]{gonze1995adiabatic}%
  \BibitemOpen
  \bibfield  {author} {\bibinfo {author} {\bibfnamefont {Xavier}\ \bibnamefont
  {Gonze}},\ }\bibfield  {title} {\enquote {\bibinfo {title} {Adiabatic
  density-functional perturbation theory},}\ }\href {\doibase
  10.1103/PhysRevA.52.1096} {\bibfield  {journal} {\bibinfo  {journal} {Phys.
  Rev. A}\ }\textbf {\bibinfo {volume} {52}},\ \bibinfo {pages} {1096--1114}
  (\bibinfo {year} {1995})}\BibitemShut {NoStop}%
\bibitem [{\citenamefont {Baroni}\ \emph {et~al.}(2001)\citenamefont {Baroni},
  \citenamefont {de~Gironcoli}, \citenamefont {Dal~Corso},\ and\ \citenamefont
  {Giannozzi}}]{baroni2001phonons}%
  \BibitemOpen
  \bibfield  {author} {\bibinfo {author} {\bibfnamefont {Stefano}\ \bibnamefont
  {Baroni}}, \bibinfo {author} {\bibfnamefont {Stefano}\ \bibnamefont
  {de~Gironcoli}}, \bibinfo {author} {\bibfnamefont {Andrea}\ \bibnamefont
  {Dal~Corso}}, \ and\ \bibinfo {author} {\bibfnamefont {Paolo}\ \bibnamefont
  {Giannozzi}},\ }\bibfield  {title} {\enquote {\bibinfo {title} {Phonons and
  related crystal properties from density-functional perturbation theory},}\
  }\href {\doibase 10.1103/RevModPhys.73.515} {\bibfield  {journal} {\bibinfo
  {journal} {Rev. Mod. Phys.}\ }\textbf {\bibinfo {volume} {73}},\ \bibinfo
  {pages} {515--562} (\bibinfo {year} {2001})}\BibitemShut {NoStop}%
\bibitem [{\citenamefont {Gonze}\ \emph {et~al.}(2009)\citenamefont {Gonze},
  \citenamefont {Amadon}, \citenamefont {Anglade}, \citenamefont {Beuken},
  \citenamefont {Bottin}, \citenamefont {Boulanger}, \citenamefont {Bruneval},
  \citenamefont {Caliste}, \citenamefont {Caracas}, \citenamefont
  {C{\^o}t{\'e}} \emph {et~al.}}]{gonze2009abinit}%
  \BibitemOpen
  \bibfield  {author} {\bibinfo {author} {\bibfnamefont {Xavier}\ \bibnamefont
  {Gonze}}, \bibinfo {author} {\bibfnamefont {Bernard}\ \bibnamefont {Amadon}},
  \bibinfo {author} {\bibfnamefont {P-M}\ \bibnamefont {Anglade}}, \bibinfo
  {author} {\bibfnamefont {J-M}\ \bibnamefont {Beuken}}, \bibinfo {author}
  {\bibfnamefont {Fran{\c{c}}ois}\ \bibnamefont {Bottin}}, \bibinfo {author}
  {\bibfnamefont {Paul}\ \bibnamefont {Boulanger}}, \bibinfo {author}
  {\bibfnamefont {Fabien}\ \bibnamefont {Bruneval}}, \bibinfo {author}
  {\bibfnamefont {Damien}\ \bibnamefont {Caliste}}, \bibinfo {author}
  {\bibfnamefont {Razvan}\ \bibnamefont {Caracas}}, \bibinfo {author}
  {\bibfnamefont {Michel}\ \bibnamefont {C{\^o}t{\'e}}},  \emph {et~al.},\
  }\bibfield  {title} {\enquote {\bibinfo {title} {Abinit: First-principles
  approach to material and nanosystem properties},}\ }\href {\doibase
  https://doi.org/10.1016/j.cpc.2009.07.007} {\bibfield  {journal} {\bibinfo
  {journal} {Comput. Phys. Commun.}\ }\textbf {\bibinfo {volume} {180}},\
  \bibinfo {pages} {2582} (\bibinfo {year} {2009})}\BibitemShut {NoStop}%
\bibitem [{\citenamefont {Garc{\'\i}a}\ \emph {et~al.}(2018)\citenamefont
  {Garc{\'\i}a}, \citenamefont {Verstraete}, \citenamefont {Pouillon},\ and\
  \citenamefont {Junquera}}]{psml}%
  \BibitemOpen
  \bibfield  {author} {\bibinfo {author} {\bibfnamefont {Alberto}\ \bibnamefont
  {Garc{\'\i}a}}, \bibinfo {author} {\bibfnamefont {Matthieu~J}\ \bibnamefont
  {Verstraete}}, \bibinfo {author} {\bibfnamefont {Yann}\ \bibnamefont
  {Pouillon}}, \ and\ \bibinfo {author} {\bibfnamefont {Javier}\ \bibnamefont
  {Junquera}},\ }\bibfield  {title} {\enquote {\bibinfo {title} {The {PSML}
  format and library for norm-conserving pseudopotential data curation and
  interoperability},}\ }\href {\doibase
  https://doi.org/10.1016/j.cpc.2018.02.011} {\bibfield  {journal} {\bibinfo
  {journal} {Comput. Phys. Commun.}\ }\textbf {\bibinfo {volume} {227}},\
  \bibinfo {pages} {51} (\bibinfo {year} {2018})}\BibitemShut {NoStop}%
\bibitem [{\citenamefont {Hamann}(2013)}]{norm_conserving}%
  \BibitemOpen
  \bibfield  {author} {\bibinfo {author} {\bibfnamefont {DR}~\bibnamefont
  {Hamann}},\ }\bibfield  {title} {\enquote {\bibinfo {title} {Optimized
  norm-conserving {V}anderbilt pseudopotentials},}\ }\href {\doibase
  https://doi.org/10.1103/PhysRevB.88.085117} {\bibfield  {journal} {\bibinfo
  {journal} {Phys. Rev. B}\ }\textbf {\bibinfo {volume} {88}},\ \bibinfo
  {pages} {085117} (\bibinfo {year} {2013})}\BibitemShut {NoStop}%
\bibitem [{\citenamefont {Van~Setten}\ \emph {et~al.}(2018)\citenamefont
  {Van~Setten}, \citenamefont {Giantomassi}, \citenamefont {Bousquet},
  \citenamefont {Verstraete}, \citenamefont {Hamann}, \citenamefont {Gonze},\
  and\ \citenamefont {Rignanese}}]{pseudodojo}%
  \BibitemOpen
  \bibfield  {author} {\bibinfo {author} {\bibfnamefont {MJ}~\bibnamefont
  {Van~Setten}}, \bibinfo {author} {\bibfnamefont {Matteo}\ \bibnamefont
  {Giantomassi}}, \bibinfo {author} {\bibfnamefont {Eric}\ \bibnamefont
  {Bousquet}}, \bibinfo {author} {\bibfnamefont {Matthieu~J}\ \bibnamefont
  {Verstraete}}, \bibinfo {author} {\bibfnamefont {Don~R}\ \bibnamefont
  {Hamann}}, \bibinfo {author} {\bibfnamefont {Xavier}\ \bibnamefont {Gonze}},
  \ and\ \bibinfo {author} {\bibfnamefont {G-M}\ \bibnamefont {Rignanese}},\
  }\bibfield  {title} {\enquote {\bibinfo {title} {The {P}seudo{D}ojo:
  {T}raining and grading a 85 element optimized norm-conserving pseudopotential
  table},}\ }\href {\doibase https://doi.org/10.1016/j.cpc.2018.01.012}
  {\bibfield  {journal} {\bibinfo  {journal} {Comput. Phys. Commun.}\ }\textbf
  {\bibinfo {volume} {226}},\ \bibinfo {pages} {39} (\bibinfo {year}
  {2018})}\BibitemShut {NoStop}%
\bibitem [{\citenamefont {Monkhorst}\ and\ \citenamefont {Pack}(1976)}]{mp}%
  \BibitemOpen
  \bibfield  {author} {\bibinfo {author} {\bibfnamefont {Hendrik~J}\
  \bibnamefont {Monkhorst}}\ and\ \bibinfo {author} {\bibfnamefont {James~D}\
  \bibnamefont {Pack}},\ }\bibfield  {title} {\enquote {\bibinfo {title}
  {Special points for {B}rillouin-zone integrations},}\ }\href {\doibase
  https://doi.org/10.1103/PhysRevB.13.5188} {\bibfield  {journal} {\bibinfo
  {journal} {Phys. Rev. B}\ }\textbf {\bibinfo {volume} {13}},\ \bibinfo
  {pages} {5188} (\bibinfo {year} {1976})}\BibitemShut {NoStop}%
\bibitem [{\citenamefont {Zhang}\ and\ \citenamefont
  {Yang}(1998)}]{zhang1998comment}%
  \BibitemOpen
  \bibfield  {author} {\bibinfo {author} {\bibfnamefont {Yingkai}\ \bibnamefont
  {Zhang}}\ and\ \bibinfo {author} {\bibfnamefont {Weitao}\ \bibnamefont
  {Yang}},\ }\bibfield  {title} {\enquote {\bibinfo {title} {Comment on
  “{G}eneralized gradient approximation made simple”},}\ }\href {\doibase
  https://doi.org/10.1103/PhysRevLett.80.890} {\bibfield  {journal} {\bibinfo
  {journal} {Phys. Rev. Lett.}\ }\textbf {\bibinfo {volume} {80}},\ \bibinfo
  {pages} {890} (\bibinfo {year} {1998})}\BibitemShut {NoStop}%
\bibitem [{\citenamefont {Becke}\ and\ \citenamefont
  {Johnson}(2006)}]{becke2006simple}%
  \BibitemOpen
  \bibfield  {author} {\bibinfo {author} {\bibfnamefont {Axel~D}\ \bibnamefont
  {Becke}}\ and\ \bibinfo {author} {\bibfnamefont {Erin~R}\ \bibnamefont
  {Johnson}},\ }\bibfield  {title} {\enquote {\bibinfo {title} {A simple
  effective potential for exchange},}\ }\href {\doibase
  https://doi.org/10.1063/1.2213970} {\bibfield  {journal} {\bibinfo  {journal}
  {J. Chem. Phys.}\ }\textbf {\bibinfo {volume} {124}},\ \bibinfo {pages}
  {221101} (\bibinfo {year} {2006})}\BibitemShut {NoStop}%
\bibitem [{\citenamefont {Ghorashi}\ \emph {et~al.}(2022)\citenamefont
  {Ghorashi}, \citenamefont {Dunbrack}, \citenamefont {Sun}, \citenamefont
  {Du},\ and\ \citenamefont {Cano}}]{ghorashi2022topological}%
  \BibitemOpen
  \bibfield  {author} {\bibinfo {author} {\bibfnamefont {Sayed Ali~Akbar}\
  \bibnamefont {Ghorashi}}, \bibinfo {author} {\bibfnamefont {Aaron}\
  \bibnamefont {Dunbrack}}, \bibinfo {author} {\bibfnamefont {Jiacheng}\
  \bibnamefont {Sun}}, \bibinfo {author} {\bibfnamefont {Xu}~\bibnamefont
  {Du}}, \ and\ \bibinfo {author} {\bibfnamefont {Jennifer}\ \bibnamefont
  {Cano}},\ }\bibfield  {title} {\enquote {\bibinfo {title} {Topological and
  stacked flat bands in bilayer graphene with a superlattice potential},}\
  }\href {https://doi.org/10.48550/arXiv.2206.13501} {\bibfield  {journal}
  {\bibinfo  {journal} {arXiv:2206.13501}\ } (\bibinfo {year}
  {2022})}\BibitemShut {NoStop}%
\bibitem [{\citenamefont {Slager}\ \emph {et~al.}(2015)\citenamefont {Slager},
  \citenamefont {Rademaker}, \citenamefont {Zaanen},\ and\ \citenamefont
  {Balents}}]{Slager_bbc}%
  \BibitemOpen
  \bibfield  {author} {\bibinfo {author} {\bibfnamefont {Robert-Jan}\
  \bibnamefont {Slager}}, \bibinfo {author} {\bibfnamefont {Louk}\ \bibnamefont
  {Rademaker}}, \bibinfo {author} {\bibfnamefont {Jan}\ \bibnamefont {Zaanen}},
  \ and\ \bibinfo {author} {\bibfnamefont {Leon}\ \bibnamefont {Balents}},\
  }\bibfield  {title} {\enquote {\bibinfo {title} {Impurity-bound states and
  {G}reen's function zeros as local signatures of topology},}\ }\href {\doibase
  10.1103/PhysRevB.92.085126} {\bibfield  {journal} {\bibinfo  {journal} {Phys.
  Rev. B}\ }\textbf {\bibinfo {volume} {92}},\ \bibinfo {pages} {085126}
  (\bibinfo {year} {2015})}\BibitemShut {NoStop}%
\bibitem [{\citenamefont {Rhim}\ \emph {et~al.}(2018)\citenamefont {Rhim},
  \citenamefont {Bardarson},\ and\ \citenamefont
  {Slager}}]{Slager2018_unifiedbbc}%
  \BibitemOpen
  \bibfield  {author} {\bibinfo {author} {\bibfnamefont {Jun-Won}\ \bibnamefont
  {Rhim}}, \bibinfo {author} {\bibfnamefont {Jens~H.}\ \bibnamefont
  {Bardarson}}, \ and\ \bibinfo {author} {\bibfnamefont {Robert-Jan}\
  \bibnamefont {Slager}},\ }\bibfield  {title} {\enquote {\bibinfo {title}
  {Unified bulk-boundary correspondence for band insulators},}\ }\href
  {\doibase 10.1103/PhysRevB.97.115143} {\bibfield  {journal} {\bibinfo
  {journal} {Phys. Rev. B}\ }\textbf {\bibinfo {volume} {97}},\ \bibinfo
  {pages} {115143} (\bibinfo {year} {2018})}\BibitemShut {NoStop}%
\bibitem [{\citenamefont {Hatsugai}(1993)}]{Hatsugai93}%
  \BibitemOpen
  \bibfield  {author} {\bibinfo {author} {\bibfnamefont {Yasuhiro}\
  \bibnamefont {Hatsugai}},\ }\bibfield  {title} {\enquote {\bibinfo {title}
  {Chern number and edge states in the integer quantum {H}all effect},}\ }\href
  {\doibase 10.1103/PhysRevLett.71.3697} {\bibfield  {journal} {\bibinfo
  {journal} {Phys. Rev. Lett.}\ }\textbf {\bibinfo {volume} {71}},\ \bibinfo
  {pages} {3697--3700} (\bibinfo {year} {1993})}\BibitemShut {NoStop}%
\end{thebibliography}

%

\end{document}